# Activation cross-sections of proton induced reactions on $^{nat}$Hf in the 38-65 MeV energy range: production of $^{172}$Lu and of $^{169}$Yb

F. Tárkányi[1], A. Hermanne[2], F. Ditrói[1*], S. Takács[1], A.V. Ignatyuk[3]

[1] *Institute for Nuclear Research, Hungarian Academy of Sciences (ATOMKI), Debrecen, Hungary*

[2] *Cyclotron Laboratory, Vrije Universiteit Brussel (VUB), Brussels, Belgium*

*3 Institute of Physics and Power Engineering (IPPE), Obninsk 249020, Russia*

## Abstract

In the frame of a systematical study of light ion induced nuclear reactions on hafnium, activation cross sections for proton induced reactions were investigated. Excitation functions were measured in the 38 - 65 MeV energy range for the $^{nat}$Hf(p,xn)$^{180g,177,176,175,173}$Ta, $^{nat}$Hf(p,x)$^{180m,179m,175,173,172,171}$Hf, $^{177g,173,172,171,170,169}$Lu and $^{nat}$Hf(p,x)$^{169}$Yb reactions by using the activation method, combining stacked foil irradiation and off line gamma ray spectroscopy. The experimental results are compared with earlier results in the overlapping energy range, and with the theoretical predictions of the ALICE IPPE and EMPIRE theoretical codes and of the TALYS code reported in the TENDL-2015 and TENDL-2017 libraries. The production routes of $^{172}$Lu (and its parent $^{172}$Hf) and of $^{169}$Yb are reviewed.

* Corresponding author: ditroi@atomki.hu

## 1. Introduction

A systematic study of proton and deuteron induced reactions up 50 MeV deuteron and 65- 70 MeV proton is in progress to investigate the production routes of medical radioisotopes (diagnostic and therapeutic) and testing the prediction possibilities of theoretical nuclear reaction model codes. Activation cross section on hafnium are of interest for production of the medically relevant radionuclides $^{167}$Tm, $^{169}$Yb, $^{177}$Lu, $^{179}$Lu $^{177}$Ta, $^{179}$Ta (X-ray fluorescent source) and for $^{172}$Hf/$^{172}$Lu and $^{178}$W/$^{178}$Ta generators. The $^{179}$Ta is used in medicine and industry for multiphoton absorptiometry techniques.

We already published activation cross sections on proton induced reaction up to 36 MeV [1], on deuteron induced reactions up to 50 MeV [2] and for alpha induced reactions up to 35 MeV [3]. In this work we extend the energy range of the proton induced reactions up to 65 MeV.

In the energy range up to 200 MeV the following experimental activation cross section of proton induced reaction are available:

Norman et al 1982 [4]: $^{180}$Hf (p,n)$^{180m,180g}$Ta reactions up to 9 MeV;

Batij et al. 1986 [5]: $^{180}$Hf (p,n)$^{180m,180g}$Ta and $^{178}$Hf(p,n)$^{178}$Ta reactions up to 9.2 MeV;

Medvedev et al. 2008 [6]: $^{nat}$Hf (p,x) $^{169,170,171,172,177}$Lu, $^{167}$Tm, $^{170,171}$Hf reactions at 108 and 195 MeV;

Siiskonen et al. 2009 [7]: $^{nat}$Hf(p,x)$^{177}$Lu, $^{175,176,177,178}$Ta reactions up to 17 MeV;

Takacs et al. 2011 [8]: $^{nat}$Hf(p,x)$^{173,174,175,176,177,178m,180g}$Ta, $^{173,175,179m2,180m}$Hf and $^{172g,173,177g}$Lu reactions up to 36 MeV;.

Shahid et al. 2014 [9]: $^{nat}$Hf(p,x)$^{173,175,176,177,178m,180g}$Ta, $^{173,175,179m2,180m}$Hf, and $^{172g,173,177g}$Lu reactions up to 45 MeV;

Murakami et al. 2014 [10]: $^{nat}$Hf(p,x) $^{175,176,177,178,179}$Ta $^{175}$Hf reactions up to 14 MeV.

No experimental data were presented earlier for the largest part (45-65 MeV) of the presently investigated energy range (38-65 MeV).

## 2. Experiment and data evaluation

An activation method based on stacked foil irradiation, followed by off-line gamma-ray spectroscopy was used.

The stack consisted of a sequence of Al, Dy, Al, Hf, Al, Ti, Be, Ti targets repeated 25 times and bombarded for 3600 s with a 65 MeV proton beam of 25 nA at Louvain la Neuve. From this irradiation the activation cross sections for reactions on beryllium and dysprosium were already published in [11, 12], where the experimental parameters, methods of data evaluation and additional details can be found on monitoring of the number of the incident protons by re-measurement of the excitation functions of monitor reactions on simultaneously irradiated Ti and Al foils. The main experimental parameters and the methods of data evaluation for the present study are summarized in Table 1. The used decay data are collected in Table 2.
Two important factors should be mentioned:
Due to the large number of irradiated samples during the reserved beam time and the limited detector capacity, the elapsed time before the first series of measurements was long (more than 20 hours for the hafnium samples) and hence the statistics for shorter-lived isotopes was poor, or they were even not detected.
During the calculation of the cross sections all listed gamma-lines were used as the first attempt. The gamma-lines were investigated for possible interferences as described in our previous work [8], interference-free lines were selected if possible, and corrections were made when interference-free gamma-lines were not available.

Table 1. Main parameters of the experiment and the methods of data evaluations.

| Experiment | | Data evaluation | |
|---|---|---|---|
| Incident particle | Proton | Gamma spectra evaluation | Genie 2000 [13], Forgamma [14] |
| Method | Stacked foil | Determination of beam intensity | Faraday cup (preliminary)<br>Fitted monitor reaction (final) [15] |
| Target stack and thicknesses | Al(49.6 μm), Dy(22.1 μm), Al(98 μm), Hf(10.7 μm), Al(49.6 μm), Ti(10.9 μm), Be(285 μm), Ti(10.9 μm) block<br>Repeated 25 times | Decay data | NUDAT 2.6 [16] |
| Number of target foils | 25 | Reaction Q-values | Q-value calculator [17] |
| Accelerator | Cyclone 90 cyclotron of the Université Catholique in Louvain la Neuve (LLN) Belgium | Determination of beam energy | Andersen (preliminary) [18]<br>Fitted monitor reaction (final) [19] |
| Primary energy | 65 MeV | Uncertainty of energy | Cumulative effects of possible uncertainties |
| Irradiation time | 60 min | Cross sections | Isotopic cross section |
| Beam current | 25 nA | Uncertainty of cross sections | Sum in quadrature of all individual contributions [20] |
| Monitor reaction, [recommended values] | $^{nat}Ti(p,x)^{48}V$<br>$^{27}Al(p,x)^{22,24}Na$<br>Reactions [19] | Yield | Physical yield [21, 22] |
| Monitor target and thickness | $^{nat}Ti$ 10.9 μm<br>$^{27}Al$ 98 μm | Theory | ALICE-IPPE [23], EMPIRE [24], TALYS (TENDL 2015, TENDL2017) [25] |
| detector | HPGe | | |
| γ-spectra measurements | 3 series | | |
| Cooling times (h) | 21.5-31.6<br>72.8-95.5<br>360.6-554.6 | | |

Table 2 Decay characteristics of the investigated reaction products and Q-values of reactions for their productions.

| **Nuclide**<br>(***level energy***)<br>**Decay mode** | **Half-life** | $E_\gamma$**(keV)** | $I_\gamma$**(%)** | **Contributing reaction** | **Q-value (keV)** |
|---|---|---|---|---|---|
| $^{180}$**Ta**<br>ε: 85 %<br>$\beta^-$: 15 % | 8.154 h | 93.324<br>103.6 | 4.51<br>0.87 | $^{180}$Hf(p,n) | -1628.82 |
| $^{178}$**Ta**<br>ε: 100 % | 2.36 h | 213.440<br>325.562<br>331.613<br>426.383 | 81.4<br>94.1<br>31.19<br>97.0 | $^{178}$Hf(p,n)<br>$^{179}$Hf(p,2n)<br>$^{180}$Hf(p,3n) | -2619.3<br>-8718.3<br>-16106.3 |
| $^{177}$**Ta**<br>EC: 100 % | 56.56 h | 112.9<br>208.4 | 7.2<br>0.94 | $^{177}$Hf(p,n)<br>$^{178}$Hf(p,2n)<br>$^{179}$Hf(p,3n)<br>$^{180}$Hf(p,4n) | -1948.35<br>-9574.29<br>-15673.27<br>-23061.03 |
| $^{176}$**Ta**<br>ε: 100 % | 8.09 h | 201.84<br>521.6<br>710.50<br>1159.30 | 5.7<br>2.4<br>5.4<br>24.7 | $^{176}$Hf(p,n)<br>$^{177}$Hf(p,2n)<br>$^{178}$Hf(p,3n)<br>$^{179}$Hf(p,4n)<br>$^{180}$Hf(p,5n) | -3993.3<br>-10368.9<br>-17994.8<br>-25022.9<br>-31481.6 |
| $^{175}$**Ta**<br>ε: 100 % | 10.5 h | 207.4<br>266.9<br>348.5<br>998.3 4 | 14.0<br>10.8<br>12.0<br>2.6 | $^{176}$Hf(p,2n)<br>$^{177}$Hf(p,3n)<br>$^{178}$Hf(p,4n)<br>$^{179}$Hf(p,5n)<br>$^{180}$Hf(p,6n) | -11021.3<br>-17397.0<br>-25022.9<br>-31121.9<br>-38509.6 |
| $^{173}$**Ta**<br>ε: 100 % | 3.14 h | 160.4<br>172.2<br>180.6<br>1030.0<br>1208.2 | 4.9<br>17.5<br>2.22<br>1.42<br>2.7 | $^{174}$Hf(p,n)<br>$^{176}$Hf(p,3n)<br>$^{177}$Hf(p,4n)<br>$^{178}$Hf(p,5n)<br>$^{179}$Hf(p,6n)<br>$^{180}$Hf(p,7n) | -4886.1<br>-19760.5<br>-26136.2<br>-33762.1<br>-39861.1<br>-47248.8 |
| $^{180m}$**Hf**<br>(*1141.7 keV*)<br>IT: 100 % | 5.53 h | 215.426<br>332.274<br>443.162<br>500.697 | 81.6<br>94<br>81.7<br>14.2 | $^{180}$Hf(p,p) | 0. |
| $^{179m}$**Hf**<br>(*1105.74 keV*)<br>IT: 100 % | 25.05 d | 122.70<br>146.15<br>169.78<br>192.66<br>217.04<br>236.48<br>268.85<br>315.93 | 27.7<br>27.1<br>19.4<br>21.5<br>9.0<br>18.8<br>11.3<br>20.3 | $^{179}$Hf(p,p)<br>$^{180}$Hf(p,pn) | 0.<br>-38509.6 |
| | | 362.55<br>409.72<br>453.59 | 39.6<br>21.5<br>68 | | |
| **$^{175}$Hf**<br>ε: 100 % | 70 d | 343.40<br>433.0 | 84<br>1.44 | $^{176}$Hf(p,pn)<br>$^{177}$Hf(p,p2n)<br>$^{178}$Hf(p,p3n)<br>$^{179}$Hf(p,p4n)<br>$^{180}$Hf(p,p5n)<br>$^{175}$Ta decay | -8165.98<br>-14541.59<br>-22167.53<br>-28266.52<br>-35654.28<br>-11021.3 |
| **$^{173}$Hf**<br>ε: 100 % | 23.6 h | 123.675<br>139.635<br>296.974<br>311.239 | 83<br>12.7<br>33.9<br>10.7 | $^{174}$Hf(p,pn)<br>$^{176}$Hf(p,p3n)<br>$^{177}$Hf(p,p4n)<br>$^{178}$Hf(p,p5n)<br>$^{179}$Hf(p,p6n)<br>$^{180}$Hf(p,p7n)<br>$^{173}$Ta decay | -8504.0<br>-23378.5<br>-29754.1<br>-37380.0<br>-43479.0<br>-50866.8<br>-12301.6 |
| **$^{172}$Hf**<br>ε: 100 % | 1.87 y | 114.061<br>122.916<br>125.812<br>127.91 | 2.6<br>1.14<br>11.3<br>1.46 | $^{174}$Hf(p,p2n)<br>$^{176}$Hf(p,p4n)<br>$^{177}$Hf(p,p5n)<br>$^{178}$Hf(p,p6n)<br>$^{179}$Hf(p,p7n)<br>$^{180}$Hf(p,p8n)<br>$^{172}$Ta decay | -15584.9<br>-30459.4<br>-36835.0<br>-44460.9<br>-50559.9<br>-57947.7<br>-21439. |
| **$^{171}$Hf**<br>ε: 100 % | 12.2 h | | | $^{174}$Hf(p,p3n)<br>$^{176}$Hf(p,p5n)<br>$^{177}$Hf(p,p6n)<br>$^{178}$Hf(p,p7n)<br>$^{179}$Hf(p,p8n)<br>$^{180}$Hf(p,p9n)<br>$^{171}$Ta decay | -24627.1<br>-39501.6<br>-45877.2<br>-53503.1<br>-59602.1<br>-66989.9<br>-29120.5 |
| **$^{170}$Hf**<br>ε: 100 % | 16.01 h | 120.19<br>164.71<br>208.1<br>572.9 | 15<br>26<br>2.7<br>15 | $^{174}$Hf(p,p4n)<br>$^{176}$Hf(p,p6n)<br>$^{177}$Hf(p,p7n)<br>$^{178}$Hf(p,p8n)<br>$^{179}$Hf(p,p9n)<br>$^{170}$Ta decay | -31875.9<br>-50971.2<br>-53126.0<br>-60751.9<br>-66850.9<br>-38774.4 |
| **$^{179}$Lu**<br>$\beta^-$: 100 % | 4.59 h | 214.33 | 12 | $^{180}$Hf(p,2p) | -8009.4 |
| **$^{177g}$Lu**<br>$\beta^-$: 100 % | 6.647 d | 112.9498<br>208.3662 | 6.17<br>10.36 | $^{178}$Hf(p,2p)<br>$^{179}$Hf(p,2pn)<br>$^{180}$Hf(p,2p2n)<br>$^{177m}$Lu decay | -7340.4<br>-13439.39<br>-20827.14 |
| **$^{177m}$Lu**<br>(*970.175 keV*)<br>$\beta^-$: 78.6%<br>IT: 21.4 % | 160.44 d | 105.3589<br>112.9498<br>128.5027<br>153.2842<br>174.3988<br>204.1050<br>208.3662<br>228.4838 | 12.4<br>21.9<br>15.6<br>17.0<br>12.7<br>13.9<br>57.4<br>37.1 | $^{178}$Hf(p,2p)<br>$^{179}$Hf(p,2pn)<br>$^{180}$Hf(p,2p2n) | -7340.4<br>-13439.39<br>-20827.14 |
| | | 281.7868<br>327.6829<br>378.5036<br>418.5388 | 14.2<br>18.1<br>29.9<br>21.3 | | |
| **$^{174m}$Lu**<br>IT: 99.38 %<br>ε: 0.62 % | 142 d | 111.762 | 0.298 | $^{176}$Hf(p,2pn)<br>$^{177}$Hf(p,2p2n)<br>$^{178}$Hf(p,2p3n)<br>$^{179}$Hf(p,2p4n)<br>$^{180}$Hf(p,2p5n) | -14366.41<br>-20742.03<br>-28367.97<br>-34466.96<br>-41854.71 |
| **$^{174g}$Lu**<br>ε: 100 % | 3.31 y | 1241.847 | 5.14 | $^{176}$Hf(p,2pn)<br>$^{177}$Hf(p,2p2n)<br>$^{178}$Hf(p,2p3n)<br>$^{179}$Hf(p,2p4n)<br>$^{180}$Hf(p,2p5n) | -14366.41<br>-20742.03<br>-28367.97<br>-34466.96<br>-41854.71 |
| **$^{173}$Lu**<br>ε: 100 % | 1.37 y | 171.393<br>272.105 | 2.90<br>21.2 | $^{174}$Hf(p,2p)<br>$^{176}$Hf(p,2p2n)<br>$^{177}$Hf(p,2p3n)<br>$^{178}$Hf(p,2p4n)<br>$^{179}$Hf(p,2p5n)<br>$^{180}$Hf(p,2p6n)<br>$^{173}$Hf decay | -6252.54<br>-21127.02<br>-27502.63<br>-35128.57<br>-41227.56<br>-48615.31<br>-8504.0 |
| **$^{172g}$Lu**<br>ε: 100 % | 6.70 d | 181.525<br>810.064<br>900.724<br>912.079<br>1093.63 | 20.6<br>16.6<br>29.8<br>15.3<br>63 | $^{174}$Hf(p,2pn)<br>$^{176}$Hf(p,2p3n)<br>$^{177}$Hf(p,2p4n)<br>$^{178}$Hf(p,2p5n)<br>$^{179}$Hf(p,2p6n)<br>$^{180}$Hf(p,2p7n)<br>$^{172}$Hf decay | -14468.79<br>-29343.27<br>-35718.89<br>-43344.82<br>-49443.82<br>-56831.57<br>-15584.9 |
| **$^{171}$Lu**<br>ε: 100 % | 8.24 d | 667.422<br>739.793<br>780.711<br>839.961 | 11.1<br>47.9<br>4.37<br>3.05 | $^{174}$Hf(p,2p2n)<br>$^{176}$Hf(p,2p4n)<br>$^{177}$Hf(p,2p5n)<br>$^{178}$Hf(p,2p6n)<br>$^{179}$Hf(p,2p7n)<br>$^{180}$Hf(p,2p8n)<br>$^{171}$Hf decay | -21447.69<br>-36322.18<br>-42697.79<br>-50323.73<br>-56422.72<br>-63810.47<br>-24627.1 |
| **$^{170}$Lu**<br>ε: 100 % | 2.012 d | 938.75<br>985.10<br>987.25<br>1003.20<br>1225.65 | 1.57<br>5.4<br>1.65<br>3.44<br>4.83 | $^{174}$Hf(p,2p3n)<br>$^{176}$Hf(p,2p5n)<br>$^{177}$Hf(p,2p6n)<br>$^{178}$Hf(p,2p7n)<br>$^{179}$Hf(p,2p8n)<br>$^{180}$Hf(p,2p9n)<br>$^{170}$Hf decay | -30041.2<br>-44915.7<br>-51291.3<br>-58917.2<br>-65016.2<br>-72404.0<br>-31875.9 |
| **$^{169}$Lu**<br>ε: 100 % | 34.06 h | 191.217<br>889.753<br>960.622<br>1449.74 | 18.7<br>4.85<br>21.2<br>9.0 | $^{174}$Hf(p,2p4n)<br>$^{176}$Hf(p,2p6n)<br>$^{177}$Hf(p,2p7n)<br>$^{178}$Hf(p,2p8n)<br>$^{179}$Hf(p,2p9n)<br>$^{169}$Hf decay | -37334.16<br>-52208.65<br>-58584.27<br>-66210.2<br>-72309.2<br>-41484.2 |
| **$^{169}$Yb**<br>ε: 100 % | 32.018 d | 109.77924<br>130.52293 | 17.39<br>11.38 | $^{174}$Hf(p,3p3n)<br>$^{176}$Hf(p,3p5n) | -34258.81<br>-49133.3 |
| | | 177.21307<br>197.95675<br>307.73586 | 22.28<br>35.93<br>10.05 | $^{177}$Hf(p,3p6n)<br>$^{178}$Hf(p,3p7n)<br>$^{179}$Hf(p,3p8n)<br>$^{180}$Hf(p,3p9n)<br>$^{169}$Lu decay | -55508.91<br>-63134.85<br>-69233.84<br>-76621.59<br>-37334.16 |

Abundance of isotopes of natural Hf -%: $^{174}$Hf-0.162, $^{176}$Hf-5.206, $^{177}$Hf-18.60, $^{178}$Hf-27.30, $^{179}$Hf-13.63, $^{180}$Hf-35.10

The *Q*-values shown in Table 2 refer to formation of the ground state. Decrease Q-values for isomeric states with level energy of the isomer

For emission of clusters increase Q-value by : pn→d +2.2 MeV, p2n→t +8.5 MeV, 2pn→$^{3}$He +7.7 MeV, 2p2n→$\alpha$ +28.3 MeV

## 3. Theoretical calculations with model codes

The cross sections of the investigated reactions were calculated using the pre-compound model codes ALICE-IPPE [23] (Dityuk et al., 1998) and EMPIRE-II [24] (Herman, 2008)]. The theoretical curves were determined using one recommended input data-set [26] (Belgya, 2006) without any optimization or adjustment of parameters to the individual reactions or stable target isotopes. Independent data for isomers with ALICE-IPPE code were obtained by using the isomeric ratios calculated with EMPIRE-II.

The experimental data were also compared with cross section data reported in the TENDL-2015 and TENDL-2017 databases, calculated by using the TALYS code with global parameters [25, 27]. The reaction cross sections for the investigated radionuclides on the individual target isotopes were summed according to the abundance of the target isotopes, if cumulative cross sections had to be calculated the contributing reactions have also been added.

# 4. Results

## 4.1 Cross sections

The measured cross sections for the $^{nat}Hf(p,xn)^{180g,177,176,175,173}Ta$, $^{nat}Hf(p,x)^{180m,179m,175,173,172,171}Hf$, $^{179,177g,173,172,171,170,169}Lu$ and $^{nat}Hf(p,x)^{169}Yb$ reactions are shown in Table 3-6 and Figures 1-21. The figures also show the theoretical results calculated with the ALICE-IPPE and the EMPIRE codes and the values available in the TENDL-2015 and TENDL-2017 on-line libraries. Due to the experimental circumstances (stacked foil technique, large dose at EOB, limited detector capacity) no cross section data were obtained for most short-lived activation products. For some short-lived radio-products the counting statistics were low due to the long cooling time before the first series of gamma spectra measurements, and for very long-lived radio-products due to limited detector capacity. Naturally occurring hafnium is composed of 6 stable isotopes ($^{174}Hf$-0.162 %, $^{176}Hf$-5.206 %, $^{177}Hf$-18.60 %, $^{178}Hf$-27.30 %, $^{179}Hf$-13.63, $^{180}Hf$-35.10 %). The relevant contributing reactions are collected in Table. 2.

Table 3. Experimental cross sections for the $^{nat}Hf(p,xn)^{180g,178g,177g,176,175g}Ta$ reactions.

| | | $^{180g}$Ta | | $^{178g}$Ta | | $^{177g}$Ta | | $^{176}$Ta | | $^{175g}$Ta | |
|---|---|---|---|---|---|---|---|---|---|---|---|
| **E** | **ΔE** | σ | **Δ** σ | σ | **Δ** σ | σ | **Δ** σ | σ | **Δ** σ | σ | **Δ** σ |
| **MeV** | | **mb** | | | | | | | | | |
| 64.57 | 0.20 | | | | | 78.82 | 10.27 | 120.81 | 14.61 | 307.21 | 36.69 |
| 63.65 | 0.24 | | | 6.39 | 2.98 | 82.99 | 11.29 | 145.77 | 17.68 | 313.48 | 37.41 |
| 62.73 | 0.29 | | | | | 86.65 | 10.58 | 139.61 | 17.19 | 329.09 | 39.29 |
| 61.8 | 0.33 | | | | | 90.23 | 11.99 | 149.04 | 18.39 | 347.71 | 41.51 |
| 60.87 | 0.38 | | | | | 79.50 | 10.95 | 150.24 | 18.08 | 351.40 | 41.95 |
| 59.91 | 0.42 | | | | | 83.58 | 10.77 | 160.25 | 19.53 | 356.91 | 42.59 |
| 58.94 | 0.47 | | | | | 87.86 | 11.59 | 154.41 | 20.47 | 372.40 | 44.46 |
| 57.96 | 0.52 | | | 4.44 | 4.66 | 93.89 | 12.61 | 185.59 | 22.44 | 371.37 | 44.31 |
| 56.96 | 0.57 | | | | | 98.59 | 12.88 | 188.16 | 22.60 | 388.73 | 46.40 |
| 55.96 | 0.61 | | | | | 89.51 | 11.36 | 204.57 | 24.51 | 376.67 | 44.95 |
| 54.93 | 0.66 | | | | | 101.95 | 13.27 | 220.40 | 26.36 | 363.71 | 43.39 |
| 53.89 | 0.71 | | | | | 115.05 | 14.73 | 226.18 | 27.09 | 355.00 | 42.37 |
| 52.83 | 0.76 | | | | | 117.66 | 14.68 | 262.51 | 31.31 | 369.65 | 44.11 |
| 51.76 | 0.81 | | | | | 118.83 | 14.92 | 305.18 | 36.62 | 351.13 | 41.90 |
| 50.67 | 0.87 | | | | | 125.70 | 15.83 | 298.60 | 36.72 | 335.74 | 40.08 |
| 49.55 | 0.92 | | | | | 134.90 | 16.74 | 337.78 | 40.44 | 329.43 | 39.34 |
| 48.41 | 0.98 | | | | | 145.62 | 17.88 | 375.40 | 45.11 | 323.92 | 38.69 |
| 47.25 | 1.03 | | | 17.99 | 6.16 | 145.41 | 17.87 | 357.80 | 42.73 | 304.06 | 36.29 |
| 46.07 | 1.09 | | | | | 172.03 | 21.02 | 385.12 | 46.36 | 335.15 | 40.04 |
| 44.87 | 1.15 | | | | | 194.24 | 23.32 | 370.28 | 44.21 | 347.30 | 41.42 |
| 43.64 | 1.20 | | | | | 213.09 | 25.63 | 335.06 | 40.06 | 354.01 | 42.25 |
| 42.36 | 1.27 | | | 16.73 | 5.25 | 236.41 | 28.45 | 323.29 | 38.60 | 367.58 | 43.85 |
| 41.06 | 1.33 | 7.39 | 1.26 | 13.39 | 3.20 | 283.65 | 33.94 | 299.09 | 35.69 | 384.09 | 45.80 |
| 39.73 | 1.39 | | | 32.05 | 15.93 | 332.91 | 39.92 | 263.98 | 31.63 | 408.75 | 48.78 |
| 38.36 | 1.46 | 8.68 | 1.97 | 35.11 | 5.32 | 350.73 | 42.07 | 253.69 | 30.28 | 408.73 | 48.75 |

Table 4. Experimental cross sections for the $^{nat}Hf(p,xn)^{173}Ta,^{180m,179m,175,173}Hf$ reactions.

| | | $^{173}$Ta | | $^{180m}$Hf | | $^{179m}$Hf | | $^{175}$Hf | | $^{173}$Hf | |
|---|---|---|---|---|---|---|---|---|---|---|---|
| **E** | **ΔE** | σ | **Δ** σ | σ | **Δ** σ | σ | **Δ** σ | σ | **Δ** σ | σ | **Δ** σ |
| **MeV** | | **mb** | | | | | | | | | |
| 64.57 | 0.20 | 357.34 | 45.58 | | | | | 408.29 | 48.67 | 368.63 | 43.95 |
| 63.65 | 0.24 | 248.06 | 30.01 | 1.56 | 0.28 | | | 405.70 | 48.37 | 373.20 | 44.50 |
| 62.73 | 0.29 | 260.84 | 34.91 | | | 0.65 | 0.42 | 428.83 | 51.13 | 369.54 | 44.07 |
| 61.8 | 0.33 | 339.43 | 42.79 | | | | | 454.60 | 54.20 | 373.80 | 44.58 |
| 60.87 | 0.38 | 276.23 | 34.47 | | | | | 520.70 | 62.08 | 366.14 | 43.67 |
| 59.91 | 0.42 | 252.81 | 31.16 | | | 0.47 | 0.46 | 462.04 | 55.07 | 355.31 | 42.36 |
| 58.94 | 0.47 | 290.86 | 39.19 | | | 0.95 | 0.61 | 470.74 | 56.12 | 352.65 | 42.05 |
| 57.96 | 0.52 | 225.37 | 27.39 | 1.10 | 0.60 | | | 490.15 | 58.43 | 355.34 | 42.37 |
| 56.96 | 0.57 | 224.44 | 30.34 | | | | | 489.03 | 58.29 | 341.10 | 40.68 |
| 55.96 | 0.61 | 256.08 | 31.43 | 1.98 | 0.72 | 0.76 | 0.26 | 438.96 | 52.33 | 329.39 | 39.28 |
| 54.93 | 0.66 | 203.27 | 24.82 | | | | | 496.59 | 59.18 | 323.05 | 38.52 |
| 53.89 | 0.71 | 202.24 | 25.63 | | | | | 479.31 | 57.12 | 299.99 | 35.78 |
| 52.83 | 0.76 | 226.22 | 28.44 | | | | | 429.19 | 51.15 | 305.71 | 36.45 |
| 51.76 | 0.81 | 186.14 | 22.39 | 2.16 | 0.63 | | | 395.58 | 47.15 | 288.52 | 34.41 |
| 50.67 | 0.87 | 181.07 | 23.62 | | | | | 453.70 | 54.07 | 267.64 | 31.93 |
| 49.55 | 0.92 | 199.16 | 32.74 | | | 0.44 | 0.46 | 408.35 | 48.68 | 247.59 | 29.54 |
| 48.41 | 0.98 | 165.96 | 30.59 | | | 0.68 | 0.55 | 394.79 | 47.07 | 225.00 | 26.85 |
| 47.25 | 1.03 | 129.53 | 16.68 | | | | | 356.51 | 42.50 | 209.45 | 24.98 |
| 46.07 | 1.09 | | | | | 0.37 | 0.64 | 404.36 | 48.20 | 201.75 | 24.08 |
| 44.87 | 1.15 | 160.42 | 20.03 | | | 0.32 | 0.55 | 402.92 | 48.04 | 182.71 | 21.79 |
| 43.64 | 1.20 | 142.99 | 17.95 | | | | | 390.24 | 46.51 | 168.04 | 20.05 |
| 42.36 | 1.27 | 85.99 | 11.50 | 1.58 | 0.29 | | | 407.20 | 48.54 | 156.42 | 18.67 |
| 41.06 | 1.33 | 58.05 | 7.77 | 1.86 | 0.27 | 0.41 | 0.28 | 434.65 | 51.80 | 146.30 | 17.45 |
| 39.73 | 1.39 | | | | | | | 454.10 | 54.12 | 120.92 | 14.42 |
| 38.36 | 1.46 | 38.72 | 4.82 | 1.86 | 0.29 | | | 454.04 | 54.11 | 126.00 | 15.05 |

Table 5. Experimental cross sections for $^{172,171}Hf$, $^{179,177g,173}Lu$ reactions.

| | | $^{172}Hf$ | | $^{171}Hf$ | | $^{179}Lu$ | | $^{177g}Lu$ | | $^{173}Lu$ | |
|---|---|---|---|---|---|---|---|---|---|---|---|
| **E** | **ΔE** | σ | **Δ** σ | σ | **Δ** σ | σ | **Δ** σ | σ | **Δ** σ | σ | **Δ** σ |
| **MeV** | | **mb** | | | | | | | | | |
| 64.57 | 0.20 | 154.04 | 20.49 | 111.11 | 14.57 | | | | | 458.27 | 55.07 |
| 63.65 | 0.24 | 112.91 | 17.37 | 94.45 | 12.43 | 4.84 | 2.20 | 3.49 | 1.06 | 305.17 | 38.75 |
| 62.73 | 0.29 | 132.63 | 19.61 | 90.60 | 12.29 | | | 3.92 | 0.86 | 307.10 | 37.45 |
| 61.8 | 0.33 | 135.32 | 20.74 | 80.04 | 11.17 | 5.45 | 3.83 | 3.08 | 1.03 | 334.47 | 43.85 |
| 60.87 | 0.38 | 96.39 | 14.72 | 62.53 | 9.39 | | | 1.49 | 0.80 | 318.58 | 40.95 |
| 59.91 | 0.42 | 99.48 | 13.78 | 40.61 | 6.69 | | | 3.90 | 0.84 | 298.63 | 37.20 |
| 58.94 | 0.47 | 108.07 | 15.51 | 48.45 | 8.10 | | | | | 415.20 | 50.93 |
| 57.96 | 0.52 | 94.58 | 14.55 | 29.29 | 6.08 | | | 3.53 | 0.69 | 374.14 | 45.16 |
| 56.96 | 0.57 | 72.21 | 12.29 | 21.65 | 5.72 | | | | | 368.96 | 45.76 |
| 55.96 | 0.61 | 59.74 | 11.99 | 29.81 | 5.79 | | | 2.31 | 0.97 | 326.93 | 39.67 |
| 54.93 | 0.66 | 62.75 | 8.99 | | | | | 3.17 | 0.87 | 294.89 | 35.32 |
| 53.89 | 0.71 | 55.40 | 7.33 | | | | | 2.24 | 0.73 | 350.07 | 41.85 |
| 52.83 | 0.76 | 34.74 | 6.32 | | | | | 3.45 | 0.50 | 228.39 | 27.42 |
| 51.76 | 0.81 | 30.35 | 6.02 | | | | | | | 219.52 | 26.37 |
| 50.67 | 0.87 | 28.96 | 6.50 | | | | | 3.81 | 0.92 | 197.20 | 23.78 |
| 49.55 | 0.92 | 16.32 | 6.19 | | | | | | | 180.33 | 22.07 |
| 48.41 | 0.98 | | | | | | | | | 177.55 | 23.57 |
| 47.25 | 1.03 | 13.31 | 6.40 | | | | | 3.23 | 0.71 | 128.70 | 16.66 |
| 46.07 | 1.09 | | | | | | | 3.45 | 0.84 | 145.91 | 19.79 |
| 44.87 | 1.15 | | | | | | | | | 119.35 | 15.26 |
| 43.64 | 1.20 | | | | | | | 3.44 | 0.81 | 113.40 | 13.81 |
| 42.36 | 1.27 | | | | | | | 3.40 | 2.05 | 75.43 | 9.82 |
| 41.06 | 1.33 | | | | | | | 3.12 | 1.30 | 72.53 | 8.96 |
| 39.73 | 1.39 | | | | | | | 3.23 | 1.16 | 58.01 | 7.37 |
| 38.36 | 1.46 | | | | | | | | | 72.70 | 8.99 |

Table 6. Experimental cross sections for $^{172,171,170,169}$Lu,$^{169}$Yb reactions.

| | | $^{172}$Lu | | $^{171}$Lu | | $^{170}$Lu | | $^{169}$Lu | | $^{169}$Ybu | |
|---|---|---|---|---|---|---|---|---|---|---|---|
| E | **ΔE** | σ | **Δ σ** | σ | **Δ σ** | σ | **Δ σ** | σ | **Δ σ** | σ | **Δ σ** |
| **MeV** | | **mb** | | | | | | | | | |
| 64.57 | 0.20 | 12.20 | 1.48 | 61.03 | 7.34 | 26.46 | 7.31 | 10.22 | 1.37 | 5.38 | 0.90 |
| 63.65 | 0.24 | 12.07 | 1.62 | 50.88 | 6.12 | 41.36 | 8.68 | 7.73 | 1.20 | 5.48 | 0.92 |
| 62.73 | 0.29 | 9.90 | 1.73 | 48.09 | 5.79 | 24.31 | 6.32 | 12.19 | 1.90 | 5.35 | 0.96 |
| 61.8 | 0.33 | 8.81 | 1.66 | 44.96 | 5.42 | 30.71 | 8.86 | 5.68 | 1.13 | 3.49 | 0.87 |
| 60.87 | 0.38 | 9.74 | 1.72 | 38.25 | 4.62 | | | 8.42 | 1.57 | 2.28 | 0.63 |
| 59.91 | 0.42 | 7.40 | 1.60 | 33.88 | 4.08 | 17.33 | 6.39 | 6.74 | 1.34 | 3.40 | 0.60 |
| 58.94 | 0.47 | 2.64 | 2.15 | 25.76 | 3.15 | 21.35 | 6.69 | 1.98 | 1.61 | 2.71 | 0.68 |
| 57.96 | 0.52 | 6.52 | 1.33 | 26.26 | 3.16 | 20.83 | 6.32 | 0.36 | 1.79 | 2.31 | 0.63 |
| 56.96 | 0.57 | 3.40 | 3.63 | 19.41 | 2.41 | 27.02 | 10.12 | | | | |
| 55.96 | 0.61 | 6.34 | 1.57 | 16.74 | 2.86 | | | | | | |
| 54.93 | 0.66 | 6.45 | 0.94 | 13.35 | 1.63 | | | | | | |
| 53.89 | 0.71 | 4.67 | 1.20 | 11.26 | 1.37 | | | | | | |
| 52.83 | 0.76 | | | 9.91 | 1.20 | | | | | | |
| 51.76 | 0.81 | | | 7.87 | 1.01 | | | | | | |
| 50.67 | 0.87 | | | 7.42 | 0.92 | | | | | | |
| 49.55 | 0.92 | | | 6.70 | 0.93 | | | | | | |
| 48.41 | 0.98 | | | 5.74 | 0.82 | | | | | | |
| 47.25 | 1.03 | | | 5.69 | 0.76 | | | | | | |
| 46.07 | 1.09 | | | 5.23 | 0.74 | | | | | | |
| 44.87 | 1.15 | | | 4.31 | 0.69 | | | | | | |
| 43.64 | 1.20 | | | 4.18 | 0.52 | | | | | | |
| 42.36 | 1.27 | | | 3.20 | 0.53 | | | | | | |
| 41.06 | 1.33 | | | 2.27 | 0.36 | | | | | | |
| 39.73 | 1.39 | | | 2.37 | 0.31 | | | | | | |
| 38.36 | 1.46 | | | | | | | | | | |

### *4.1.1 Radioisotopes of tantalum*

The radioisotopes of tantalum are produced directly via (p,xn) reactions.

#### *4.1.1.1 $^{nat}Hf(p,xn)^{180g}Ta$ reaction*

The $^{180}$Ta has two isomeric states. The ground-state, which has 8.15 h half-life and the meta-stable state that has very long ($T_{1/2} > 1.2 \times 10^{15}$ a) half-life. The $^{180m}$Ta decay does not contribute directly to the production of the $^{180g}$Ta state since it populates exclusively the $^{180m}$Hf isomeric state. We got only two cross section data points (Fig. 1), which are slightly above the previous experimental results and also the theory. The predictions of the theoretical codes are acceptable good.

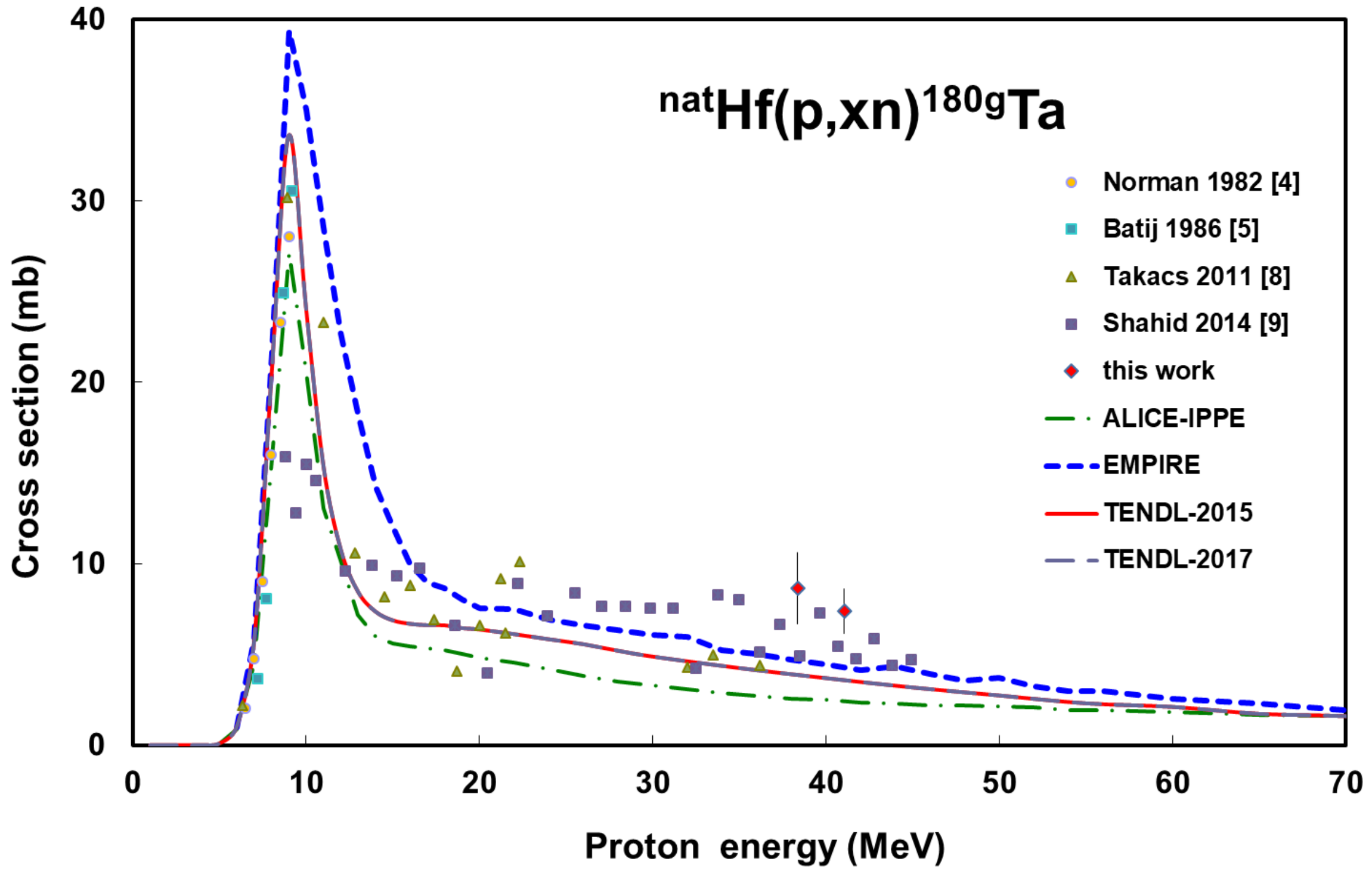

Fig. 1**.** Experimental and theoretical cross sections for the formation of $^{180g}$Ta by the proton bombardment of hafnium

### 4.1.1.2 $^{nat}Hf(p,x)^{178g}Ta$ reaction

The 9.31 min ($J^{\pi}$ = 1+) and 2.36 h ($J^{\pi}$ = $7^-$) states of $^{178}$Ta are decaying independently. Due to long cooling time, only a few points were measured for production of the longer lived (ground-) state (Fig. 2). Our new results are a bit scattered because of the long cooling time, but in good agreement with the previous data. EMPIRE and ALICE give good trend but overestimate the values below 40 MeV.

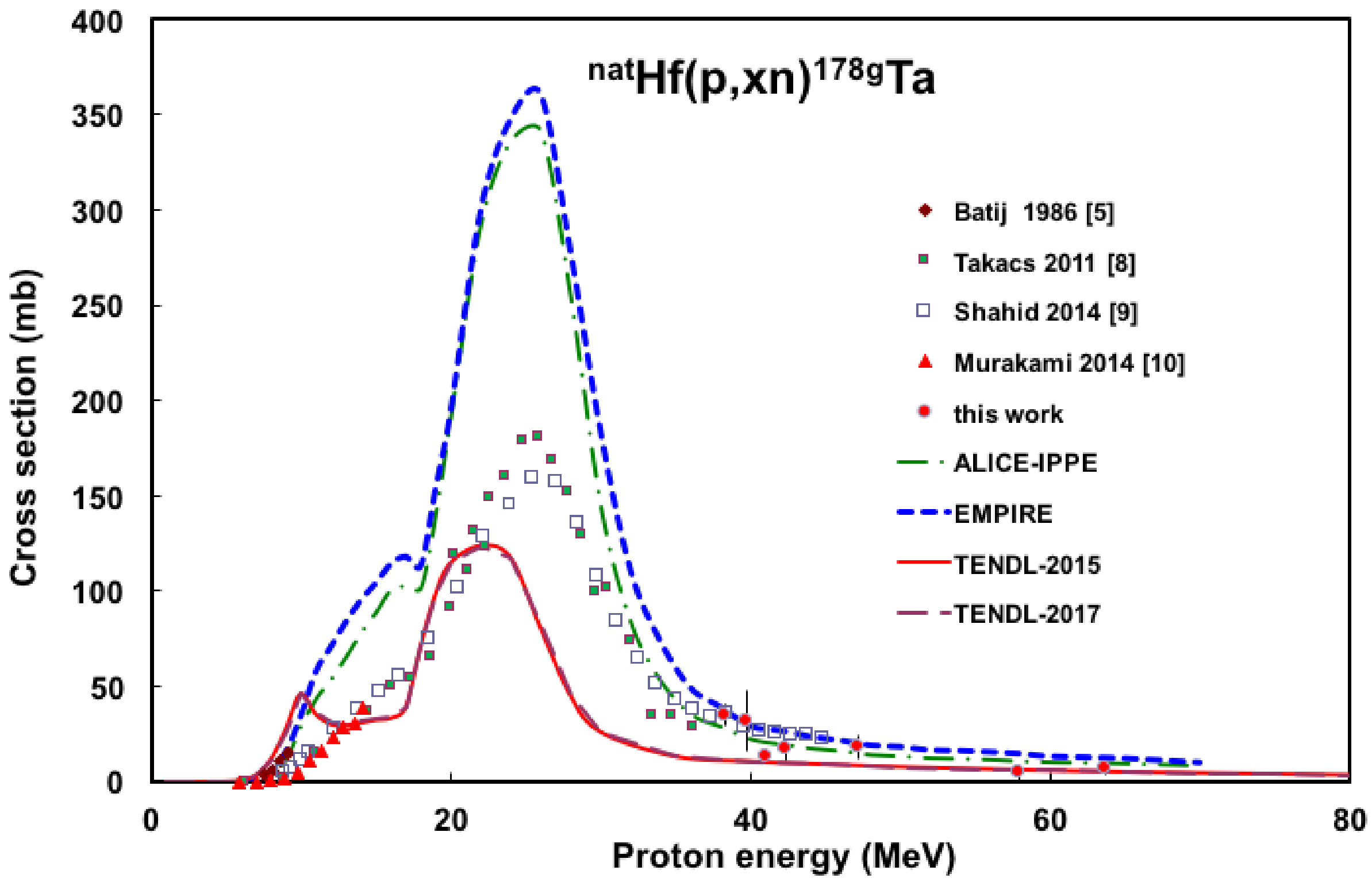

Fig. 2**.** Experimental and theoretical cross sections for the formation of $^{178g}$Ta by the proton bombardment of hafnium.

### 4.1.1.3 $^{nat}Hf(p,x)^{177}Ta$ reaction

Our data for production of $^{177}$Ta (56.56 h) are in good agreement with the earlier experimental data (Fig. 3). The theoretical descriptions are good concerning the magnitude but at high energies a systematic downward energy shift can be observed for the TALYS calculations.

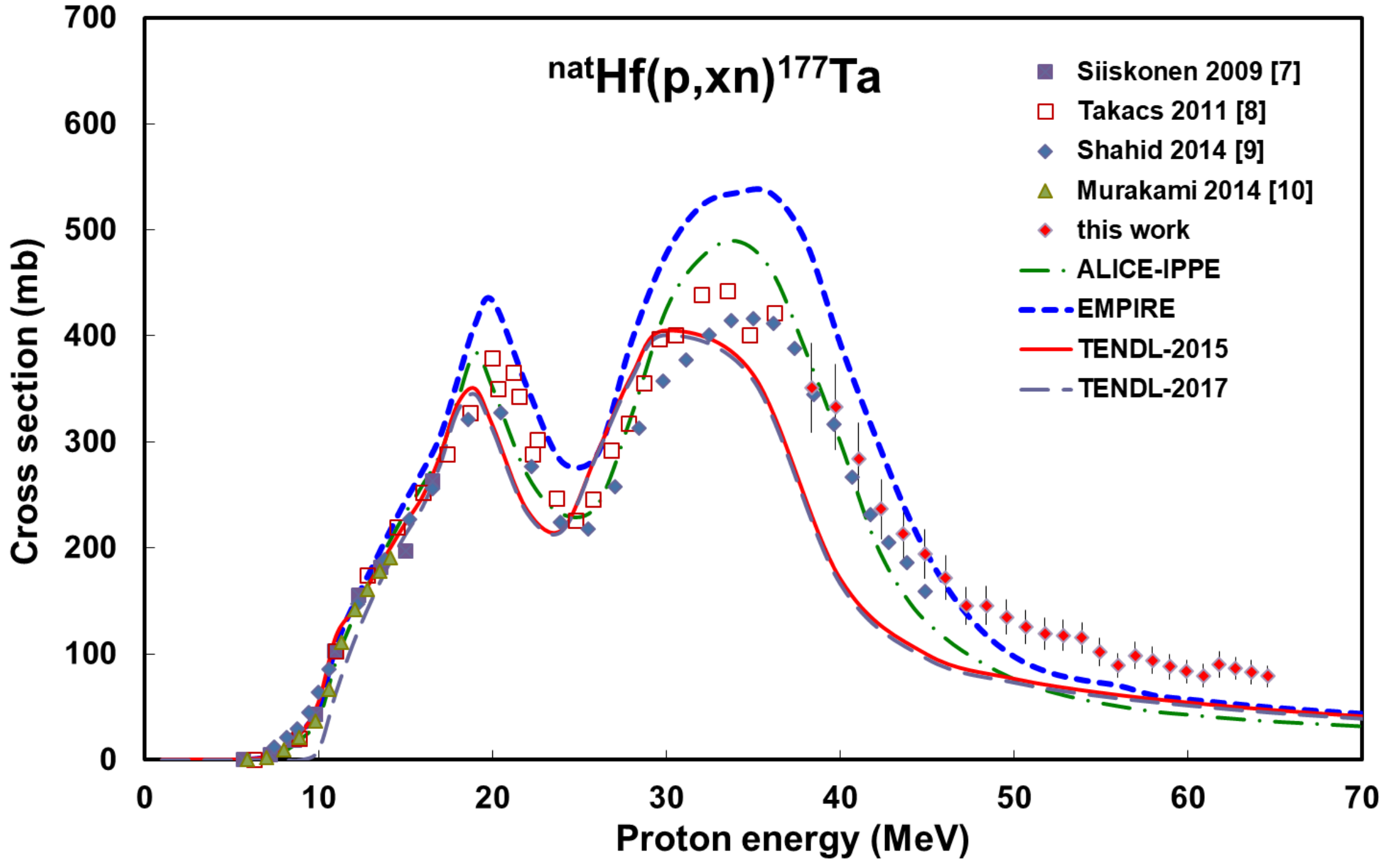

Fig. 3. Experimental and theoretical cross sections for the formation of $^{177}$Ta by the proton bombardment of hafnium.

#### *4.1.1.4 $^{nat}Hf(p,x)^{176}Ta$ reaction*

Our new cross section data for production of $^{176}$Ta (8.09 h) isotope (Fig. 4) fit well in magnitude at 35 MeV to our earlier experimental data [8] The data of Shahid et al. 2014 [9] are systematically higher in the overlapping energy range. The EMPIRE and the ALICE predictions are acceptable, but TENDL data are shifted to lower energy.

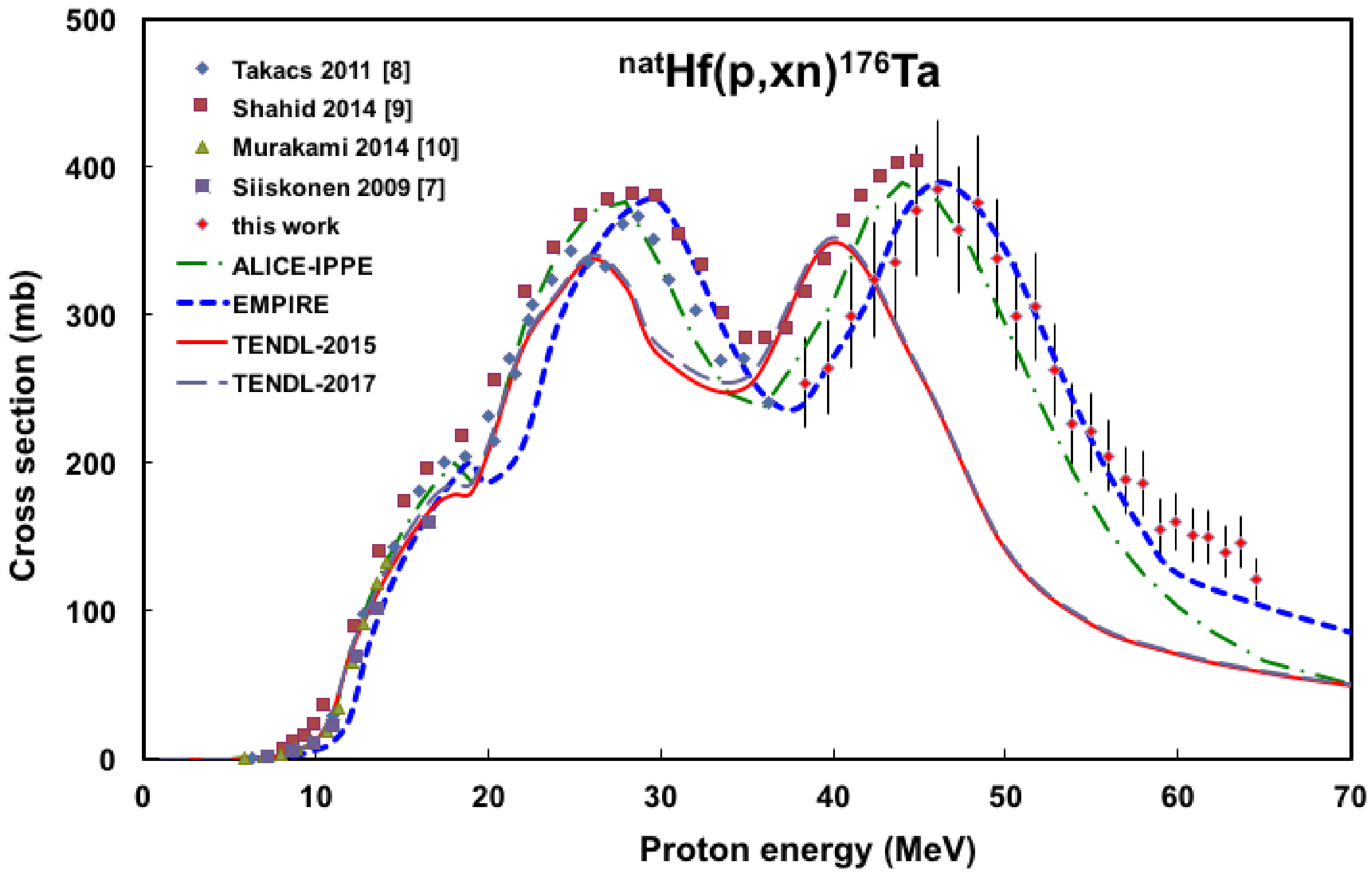

Fig. 4. Experimental and theoretical cross sections for the formation of $^{176}$Ta by the proton bombardment of hafnium.

### 4.1.1.5 $^{nat}Hf(p,x)^{175}Ta$ reaction

The new data for production of $^{175}$Ta (10.5 h) are a little higher comparing to earlier low energy data in the overlapping energy range (Fig. 5). Raising energy shift can be observed for the TALYS predictions.

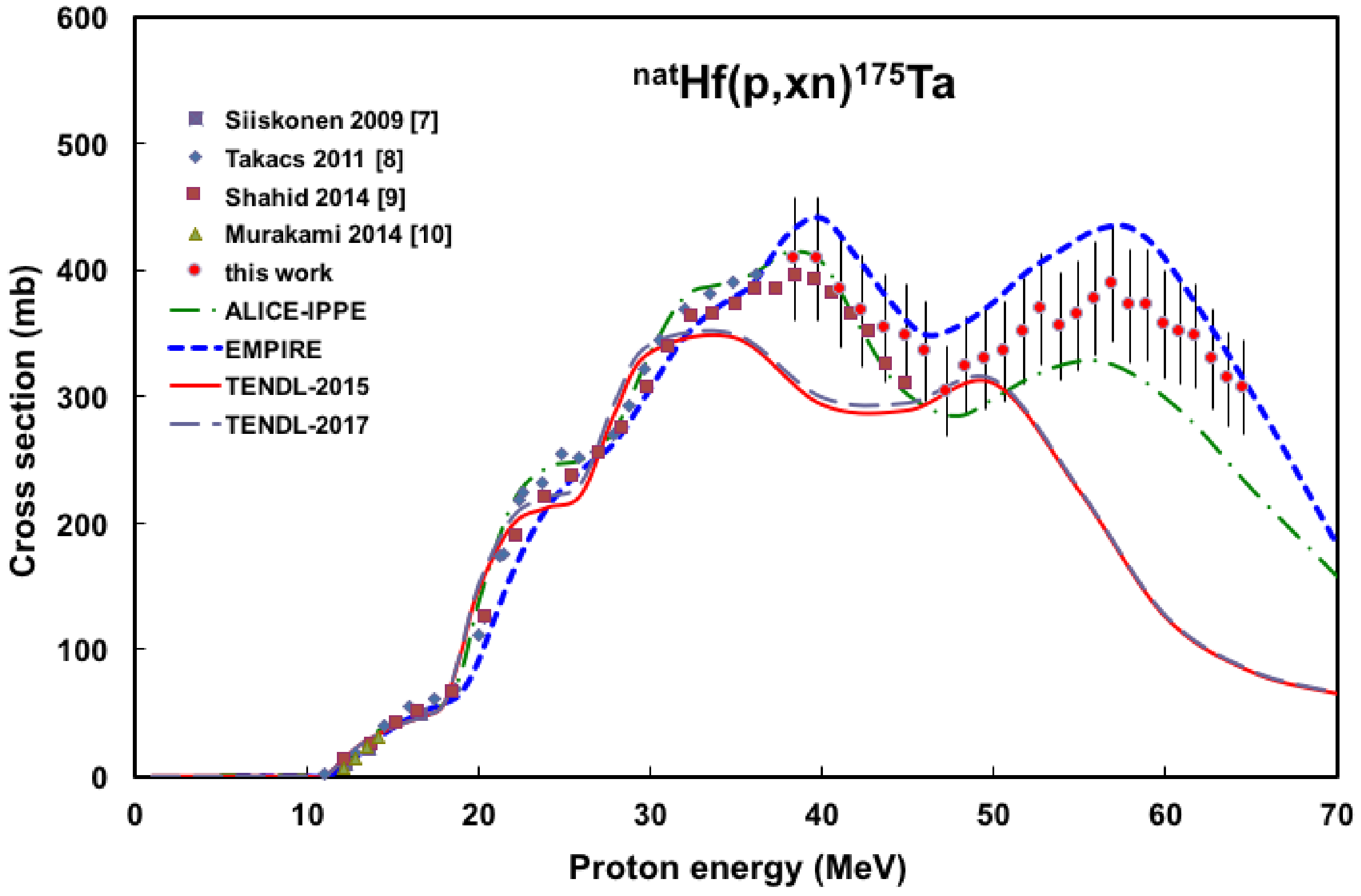

Fig. 5. Experimental and theoretical cross sections for the formation of $^{175}$Ta by the proton bombardment of hafnium.

### 4.1.1.6 ${}^{nat}Hf(p,x){}^{174}Ta$ reaction

No experimental data were obtained in the present experiment due to the short half-life. In Fig. 6 we have collected the earlier experimental data and results of our model calculations. The ALICE and EMPIRE systematically overestimate the experimental data.

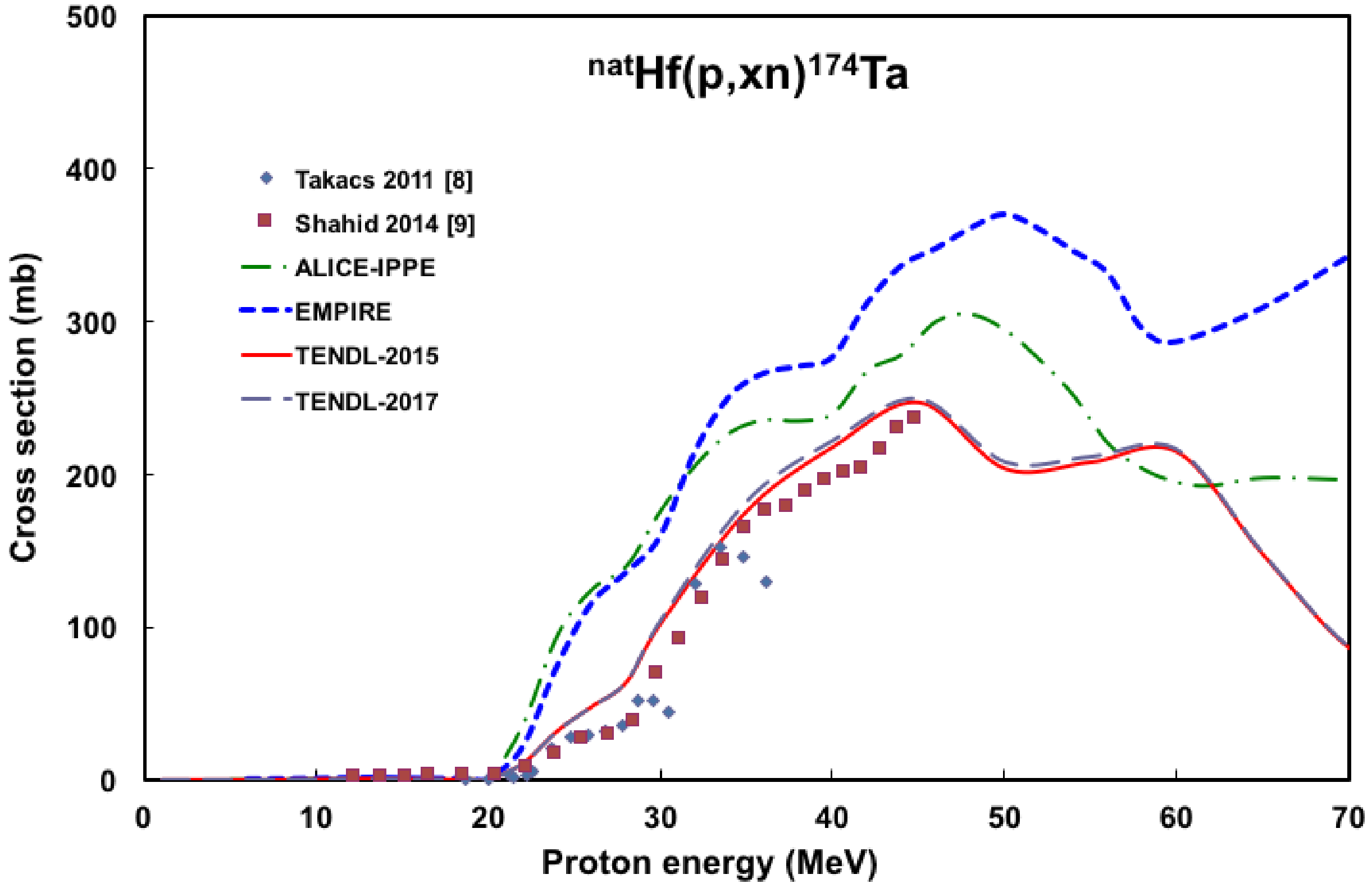

Fig. 6. Experimental and theoretical cross sections for the formation of ${}^{174}Ta$ by the proton bombardment of hafnium.

#### *4.1.1.7 $^{nat}Hf(p,x)^{173}Ta$ reaction*

Our data for $^{173}Ta$ (3.14 h) are in agreement with the earlier experimental data in the overlapping energy range (Fig. 7). Due to the long cooling time (low statistics) our data are scattered. The TENDL data are lower at high energies.

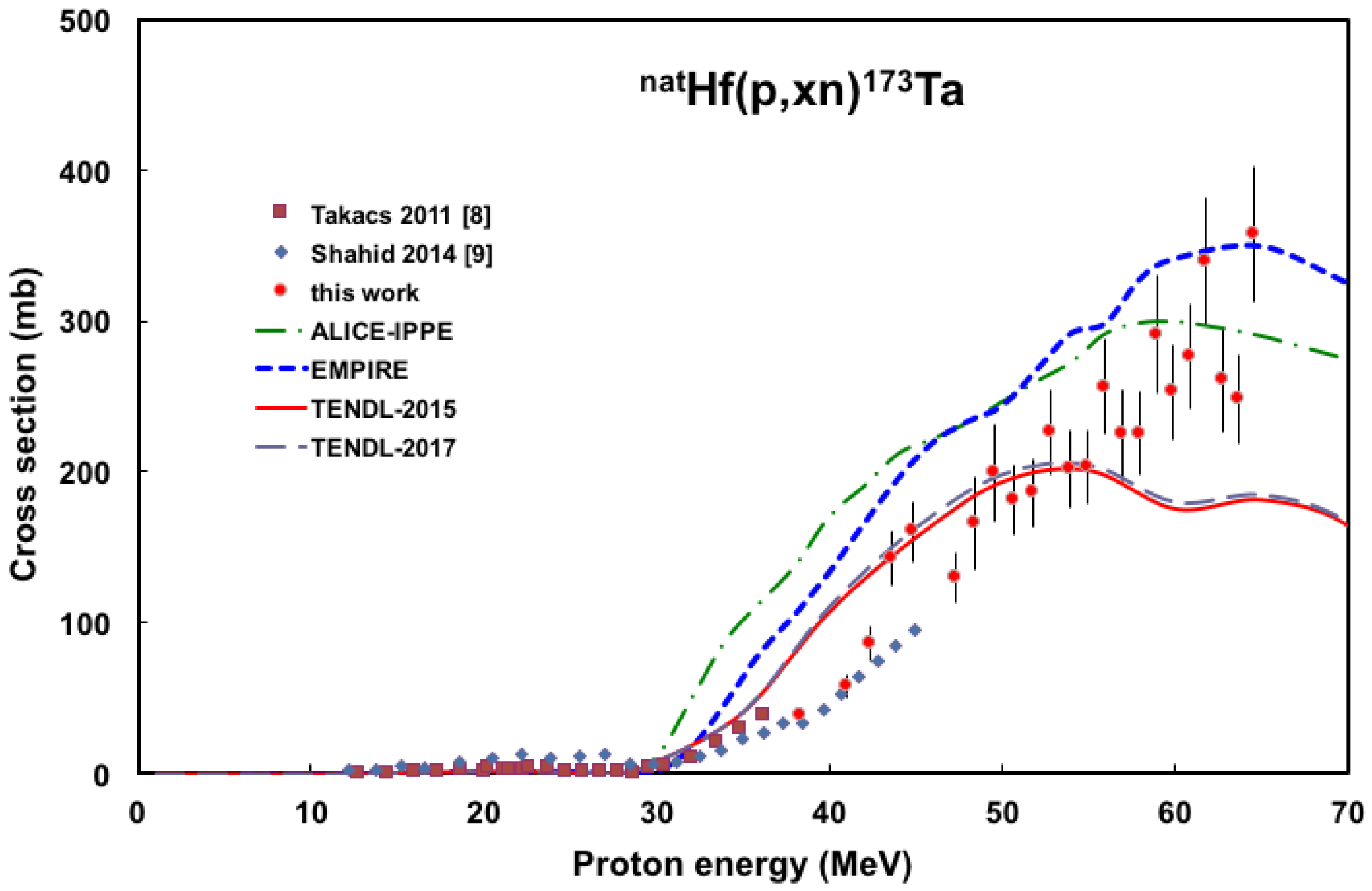

Fig. 7. Experimental and theoretical cross sections for the formation of $^{173}Ta$ by the proton bombardment of hafnium.

### 4.1.2 Radioisotopes of hafnium

The radioisotopes of hafnium are produced directly via reactions where a proton and one or multiple neutrons are emitted complemented with the ($\varepsilon+\beta^+$) decay of isobaric Ta-parents (see section 4.1.1.).

#### 4.1.2.1 $^{nat}Hf(p,x)^{180m}Hf$ reaction

The cross sections for direct production of $^{180m}Hf$ (5.53 h) are shown in Fig. 8. Our new data have large uncertainties due to the needed separation of overlapping gamma-lines. The possible contributors: $^{180}Ta$ (8.15 h) is decaying to $^{180g}Hf$ ground and the 5.7 min $^{170}Lu$ is decaying also to ground state. The TENDL data show large overestimation, the best results are coming from the EMPIRE prediction. The TENDL-2015 and 2017 curves are identical.

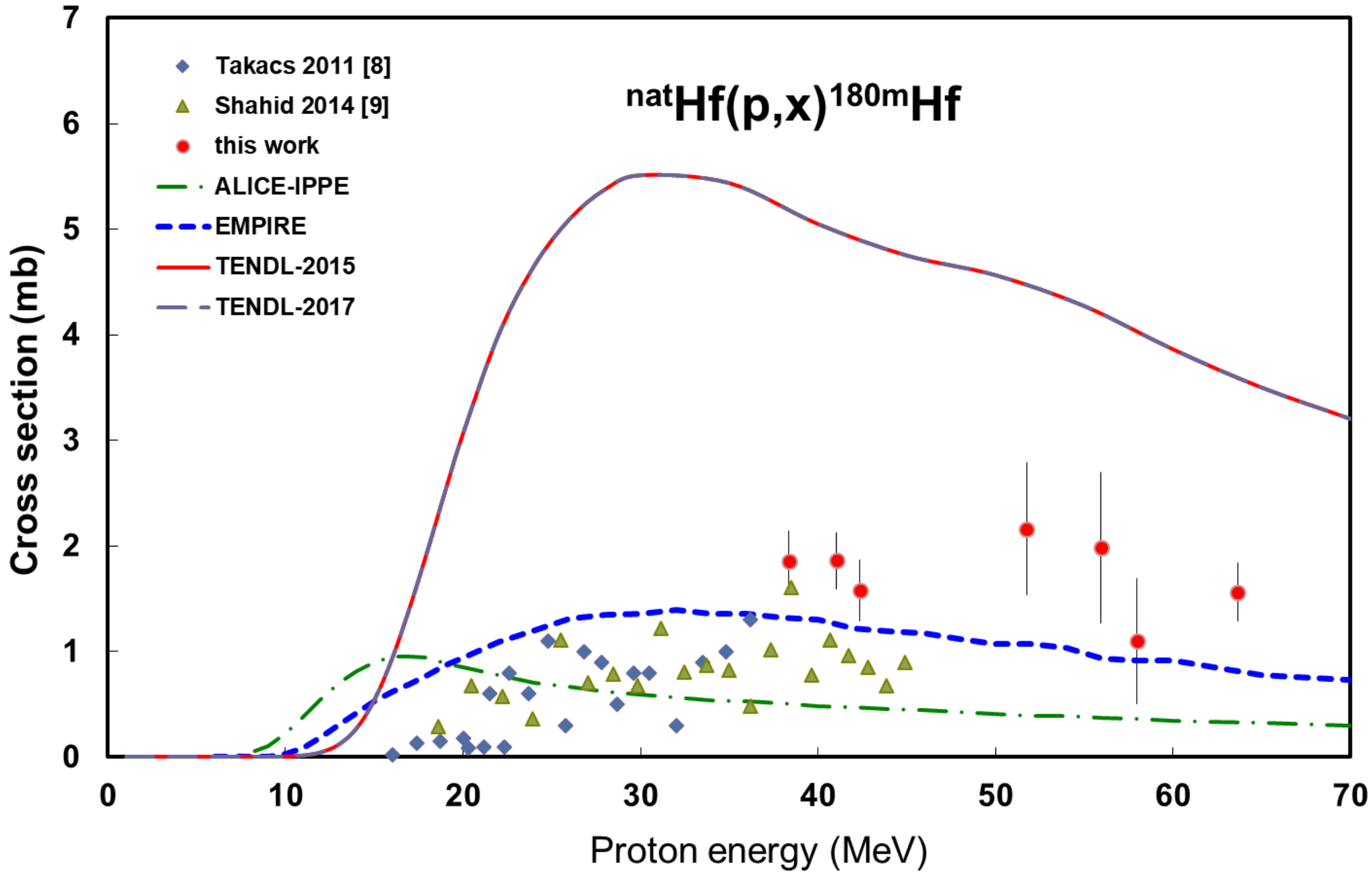

Fig. 8. Experimental and theoretical cross sections for the formation of $^{180m}Hf$ by the proton bombardment of hafnium.

### 4.1.2.2 $^{nat}Hf(p,x)^{179m}Hf$ reaction

The $^{179m}Hf$ (25.05 d) is produced directly (Fig. 9) as no contribution from the isobars $^{179}Lu$ (4.59 h) and $^{179}Ta$ (1.82 y) occurs. The TENDL predictions significantly overestimate the experimental data and differ significantly from predictions of the other two theoretical models (probably due to the too high branching). The best prediction is given by the EMPIRE code.

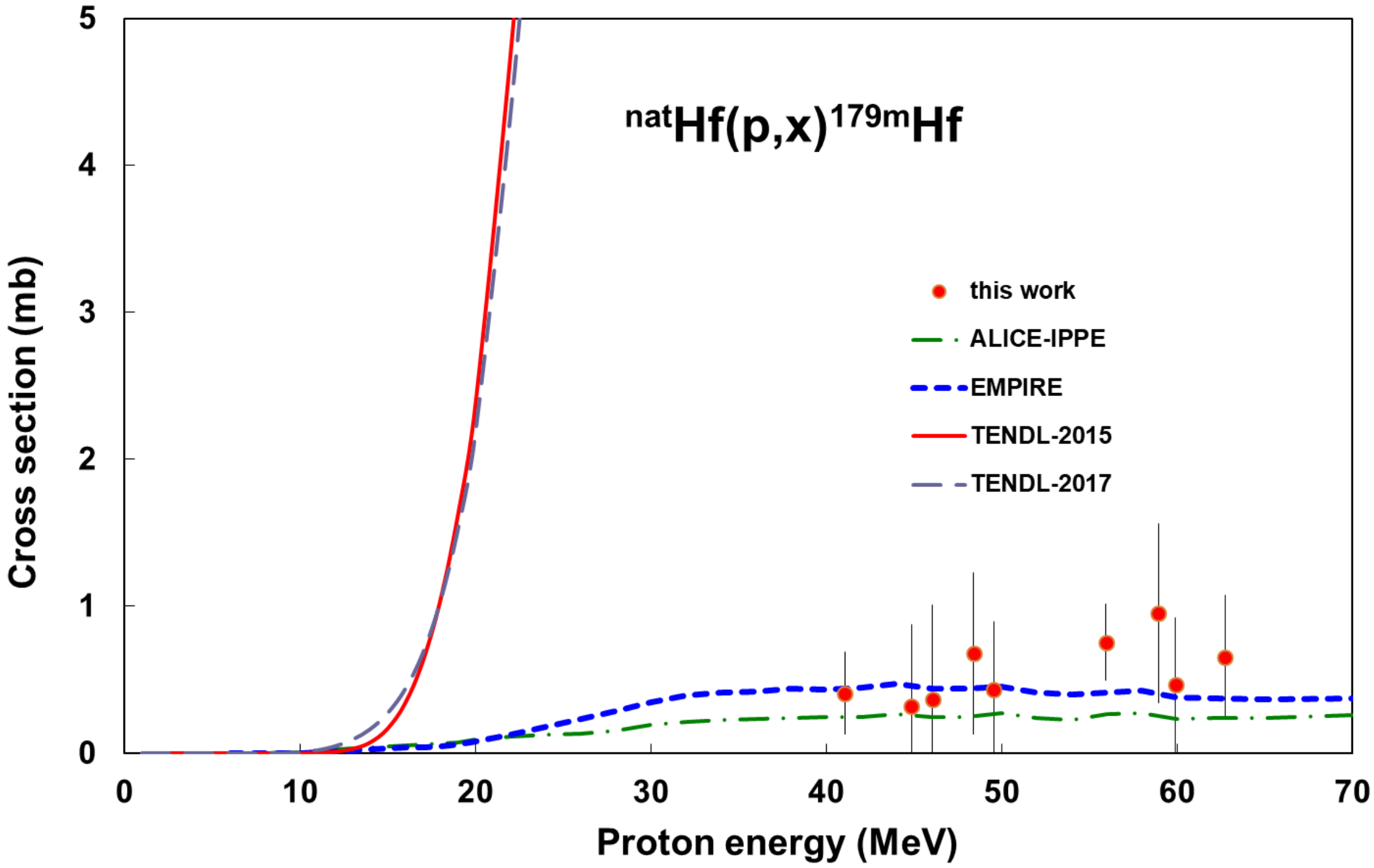

Fig. 9. Experimental and theoretical cross sections for the formation of $^{179m}Hf$ by the proton bombardment of hafnium.

### 4.1.2.3 *$^{nat}Hf(p,x)^{175}Hf$ reaction*

The measured cross section of $^{175}Hf$ (70 d) are cumulative, including the complete decay of $^{175}Ta$ (10.5 h) (Fig. 10). Our data are in acceptable agreement with the previous lower energy experimental data in the overlapping energy range and with the predictions of ALICE and EMPIRE. The TENDL results are lower and energy shifted.

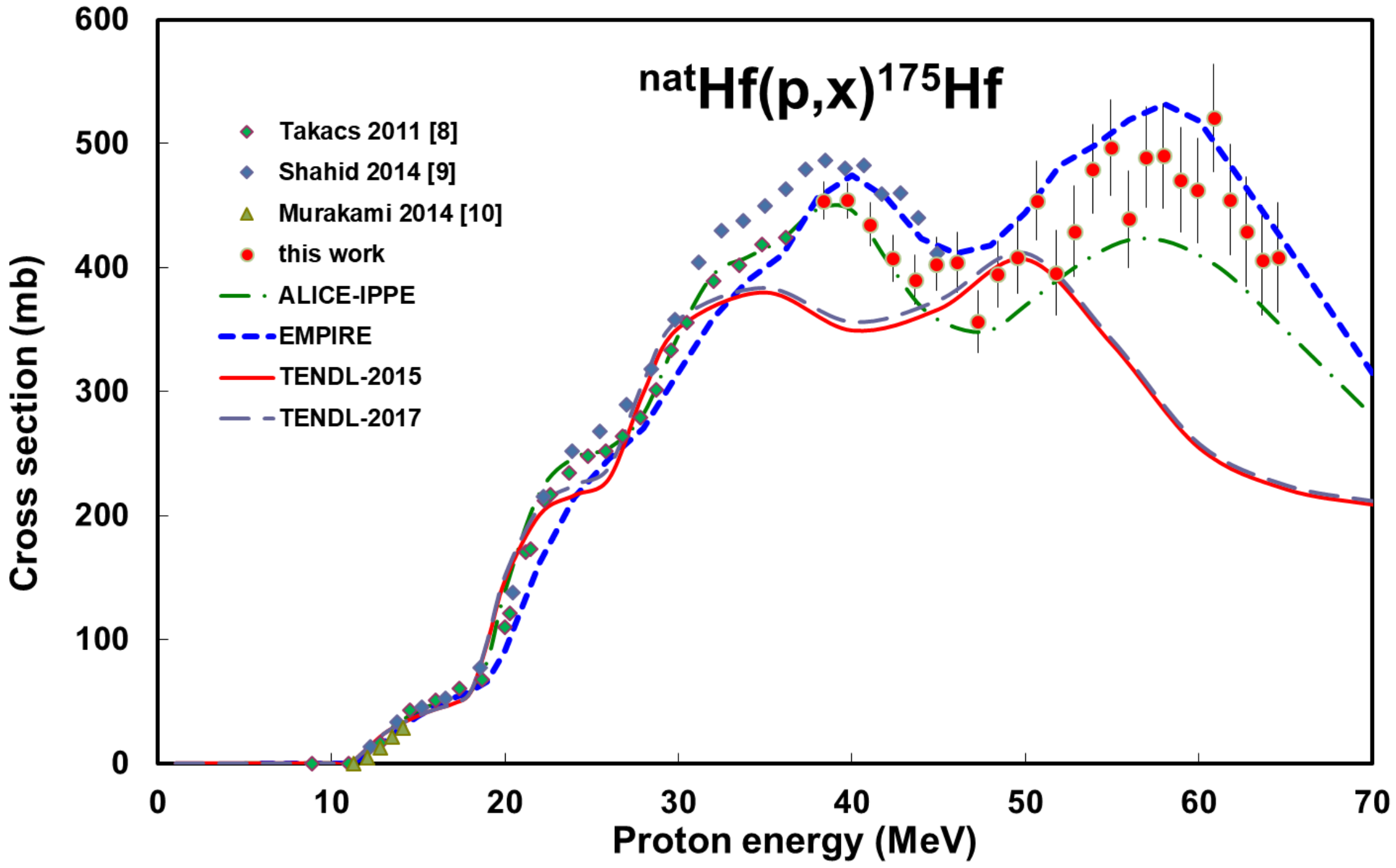

Fig. 10. Experimental and theoretical cross sections for the formation of $^{175}Hf$ by the proton bombardment of hafnium.

### *4.1.2.4 $^{nat}$Hf(p,x)$^{173}$Hf reaction*

The measured cross sections of $^{173}$Hf (23.6 h) are cumulative as they include the decay of $^{173}$Ta (3.14 h) parent (Fig. 11). The trend of our new results is comparable to our earlier results [8], and the magnitude is also in good agreement with the results of Shahid 2014 [9] in the overlapping energy range. The TENDL libraries significantly underestimate the experiment at high energies, while the EMPIRE predictions are acceptable good.

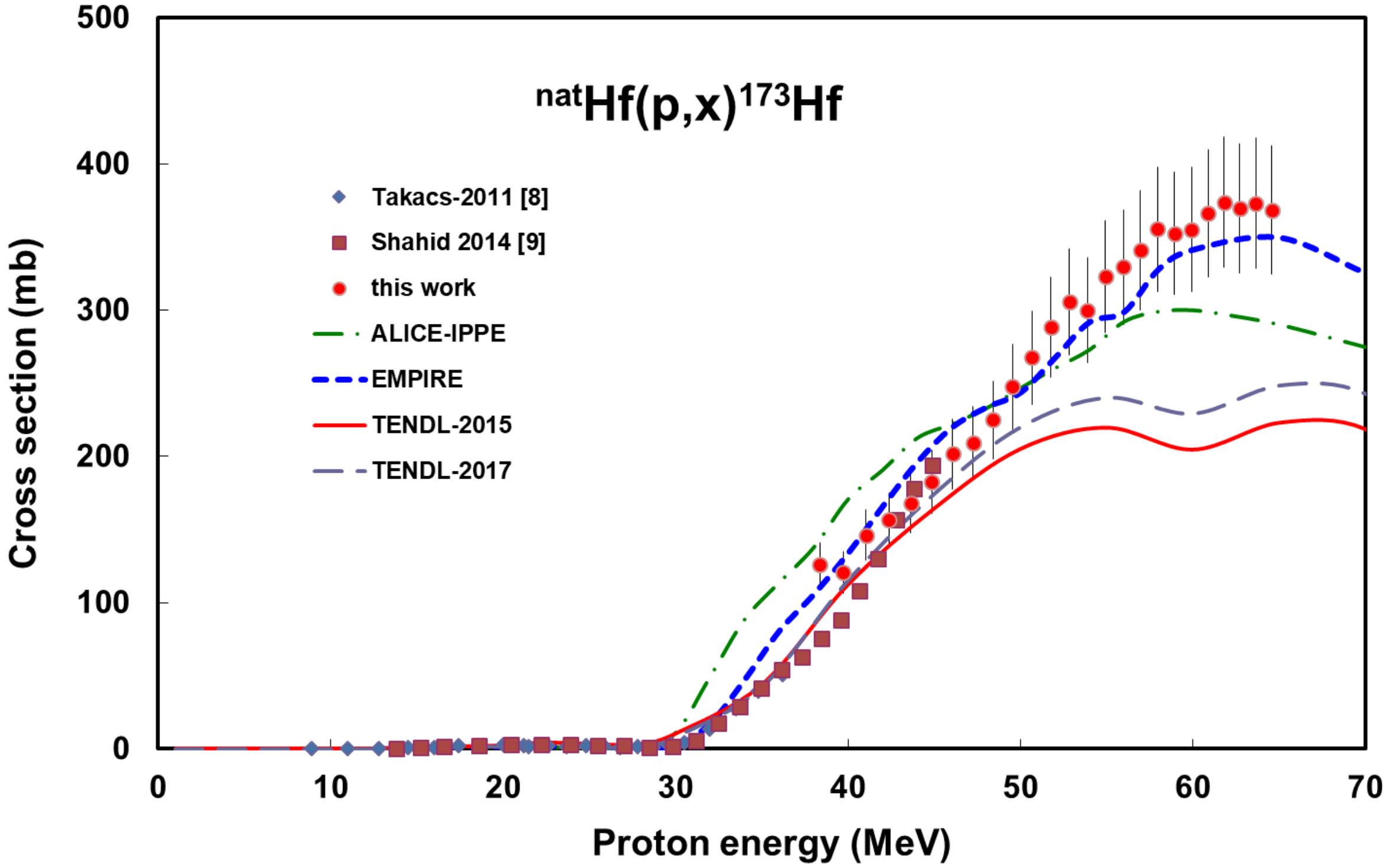

Fig. 11 Experimental and theoretical cross sections for the formation of $^{173}$Hf by the proton bombardment of hafnium.

### *4.1.2.5 $^{nat}$Hf(p,x)$^{172}$Hf reaction*

The measured cross sections for production of $^{172}$Hf (1.87 a) are cumulative, including the decay of $^{172}$Ta (36.8 min) parent (Fig.12), No earlier experimental data were found in the literature. All calculated excitation functions are systematically higher than our new experimental data.

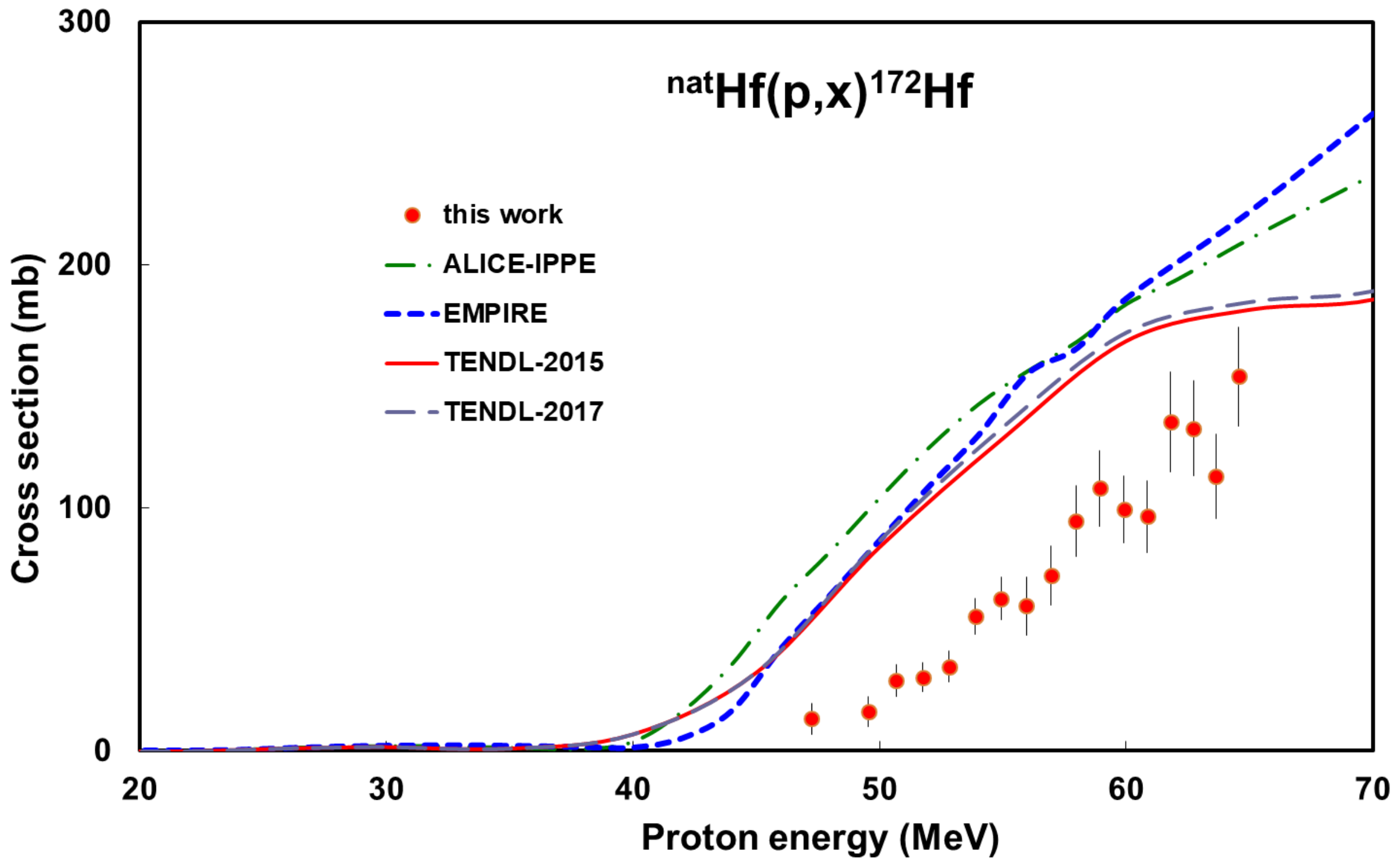

Fig. 12. Experimental and theoretical cross sections for the formation of $^{172}$Hf by the proton bombardment of hafnium.

### *4.1.2.6 $^{nat}Hf(p,x)^{171}Hf$ reaction*

The cumulative cross sections of $^{171}$Hf (12.2 h) include the contributions from the $^{171}$Ta parent (23.3 min) decay (Fig. 13). The agreement with the ALICE, EMPIRE and especially with TENDL predictions is acceptable.

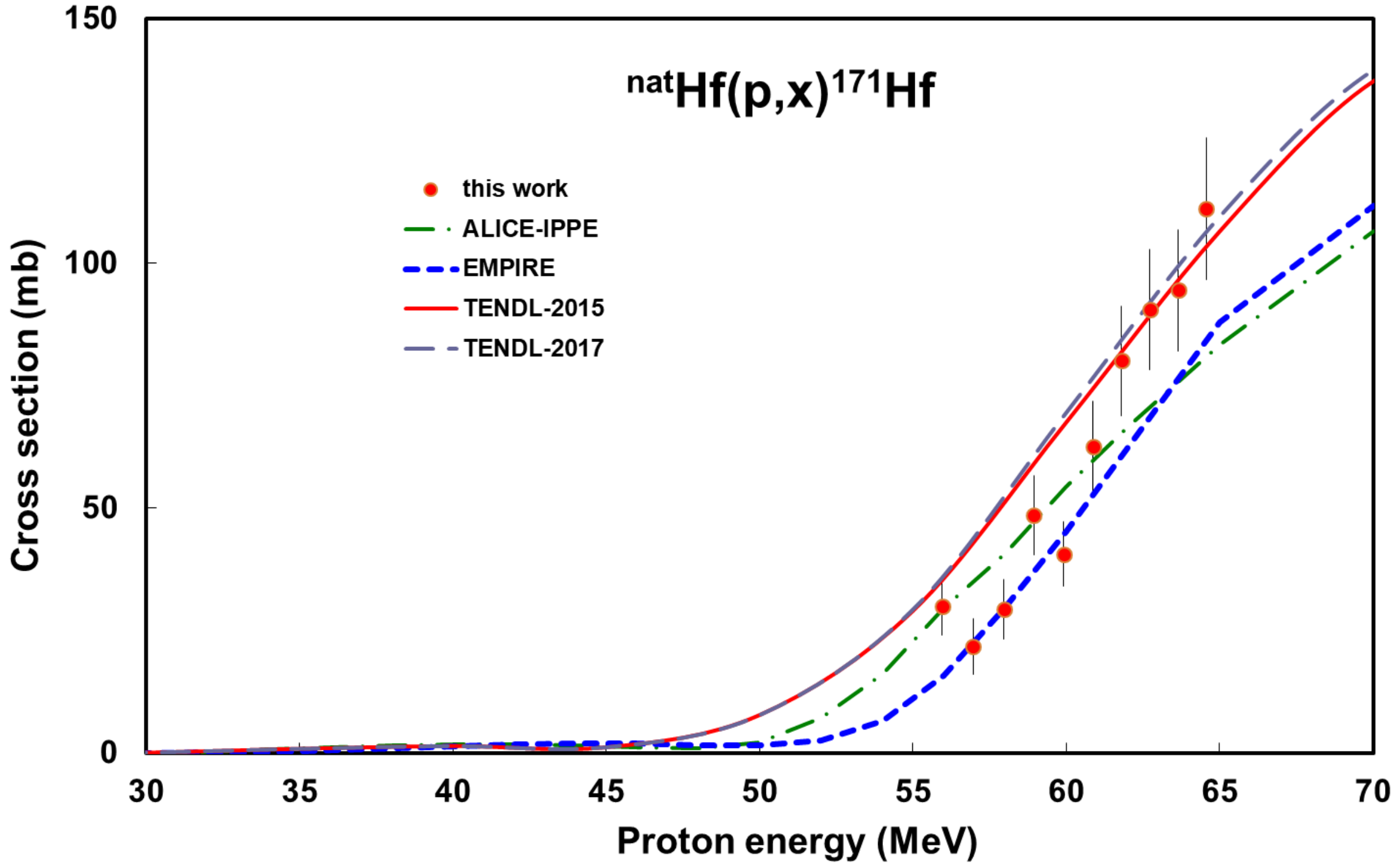

Fig. 13. Experimental and theoretical cross sections for the formation of $^{171}$Hf by the proton bombardment of hafnium.

### *4.1.3 Radioisotopes of lutetium*

The radioisotopes of lutetium are produced directly via reactions where two protons and one or multiple neutrons are emitted (possible clustered in α-particle) complemented with the ($\varepsilon+\beta^+$) decay of isobaric Ta and Hf parents.

#### *4.1.3.1 $^{nat}$Hf(p,x)$^{179}$Lu reaction*

The $^{179}$Lu (4.59 h) isotope is a candidate for radionuclide therapy and in this work is only produced via a direct $^{180}$Hf(p,2p) reaction. We only obtained two scattered experimental data points after decomposition of 215 keV complex gamma-line (Fig. 14). There are large differences between the theoretical predictions.

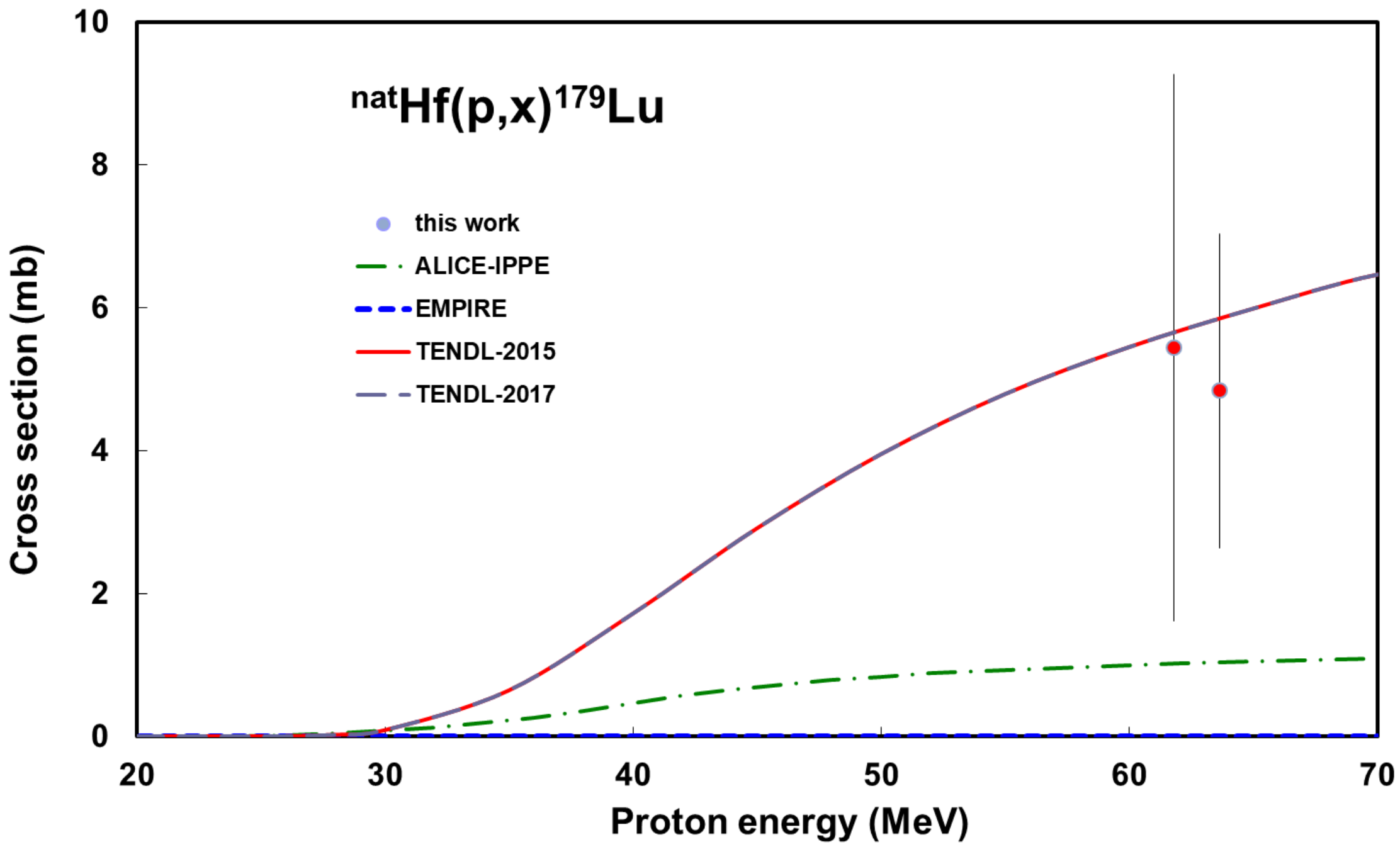

Fig. 14. Experimental and theoretical cross sections for the formation of $^{179}$Lu by the proton bombardment of hafnium.

### 4.1.3.2 $^{nat}Hf(p,x)^{177g}Lu$ reaction

The measured cross section of $^{177g}$Lu (6.647 d) does not include the decay of the much longer-lived $^{177m}$Lu isomeric state (160.44 d, IT: 21.4 %) (Fig. 15). Earlier experimental data at low energy are contradictory. Our scattered data are higher compared to the data of Shahid 2014 [9] in the overlapping energy range. (Fig. 15).

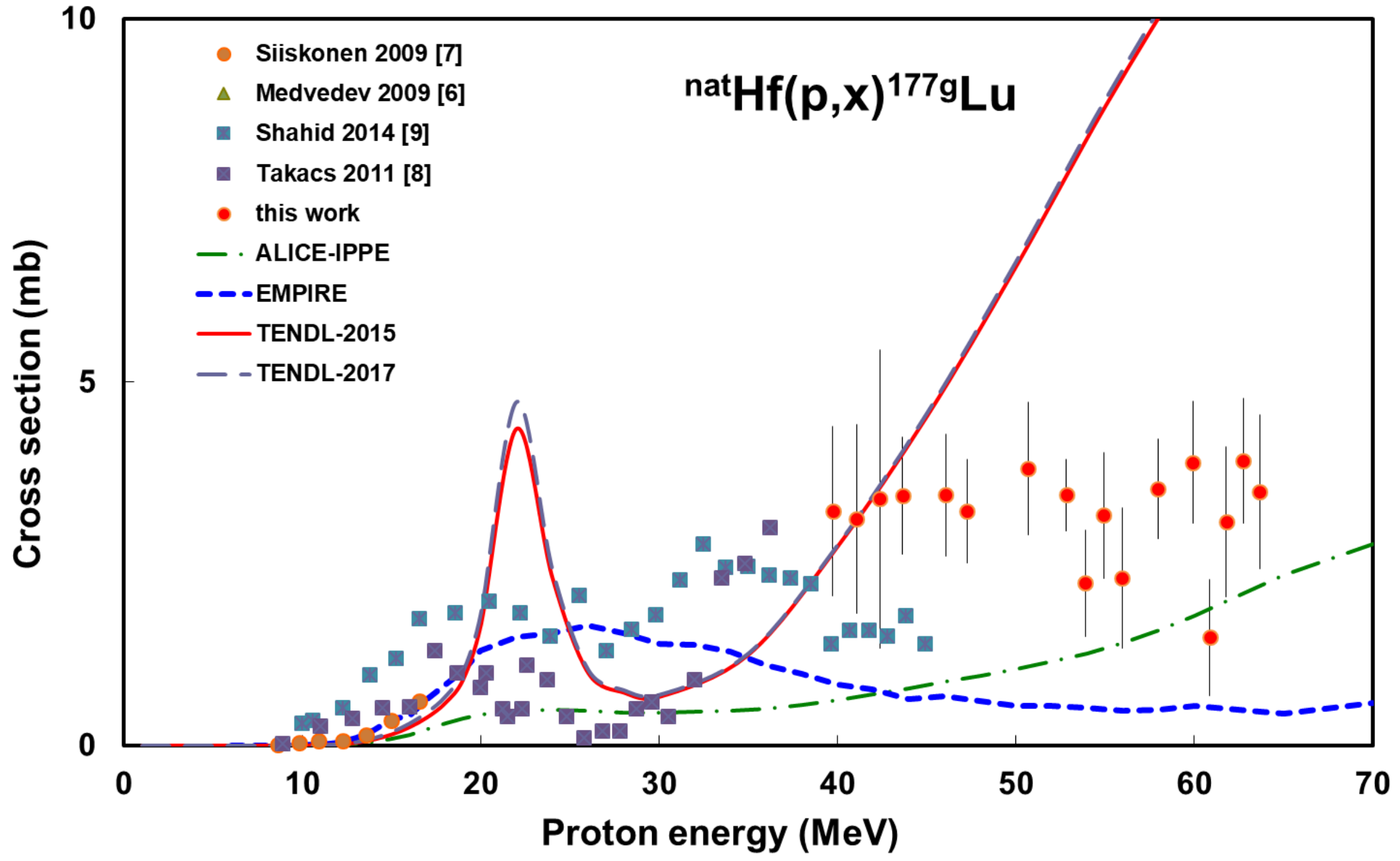

Fig. 15. Experimental and theoretical cross sections for the formation of $^{177g}$Lu by the proton bombardment of hafnium.

### 4.1.3.3 $^{nat}Hf(p,x)^{173}Lu$ reaction

The cumulative cross sections of $^{173}$Lu (1.37 a) include the decay of the $^{173}$Ta (3.14 h) →$^{173}$Hf (23.6 h) parent chain. (Fig. 16). The agreement with the earlier experimental data and with the theoretical results is acceptable.

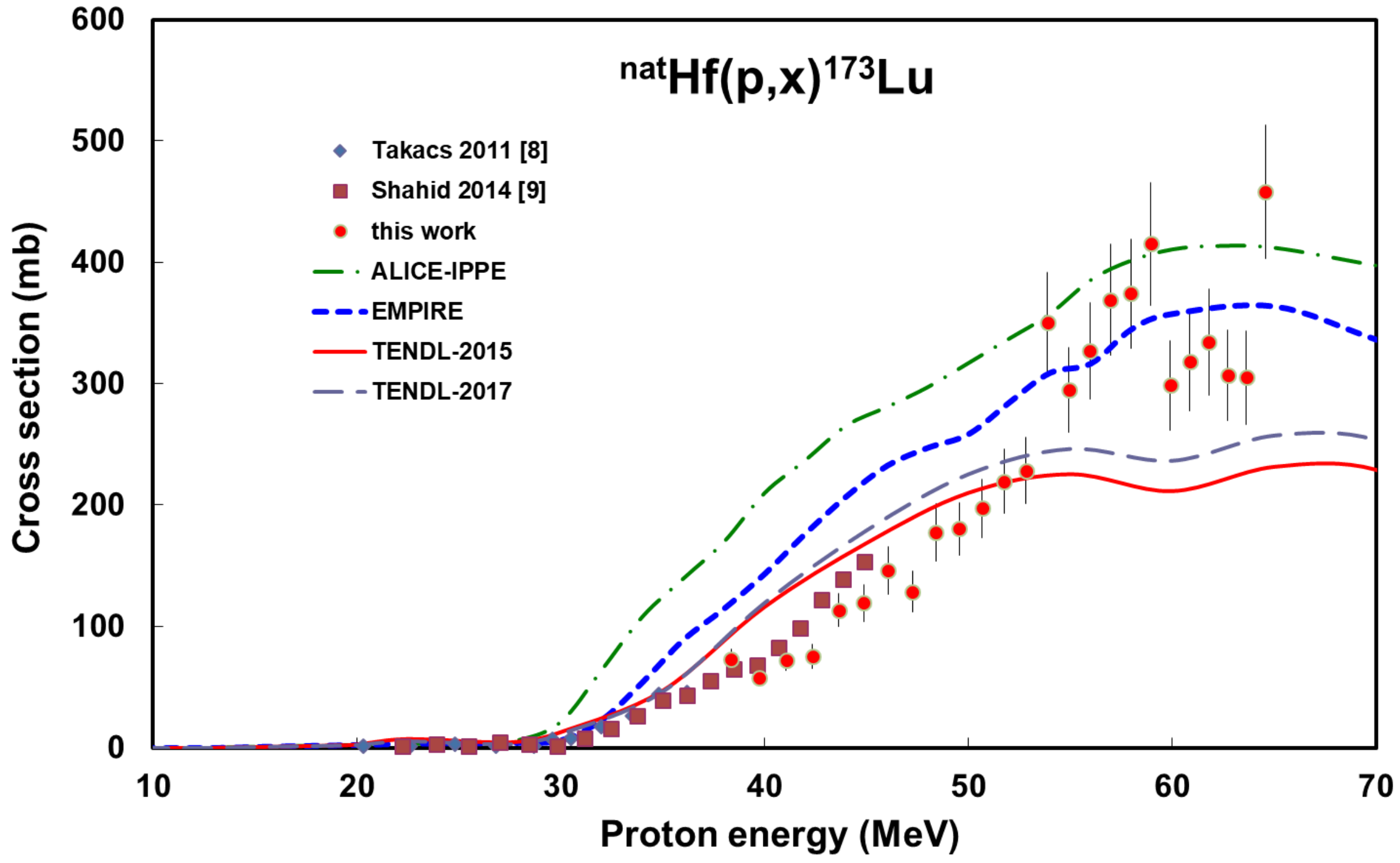

Fig. 16. Experimental and theoretical cross sections for the formation of $^{173}$Lu by the proton bombardment of hafnium.

### 4.1.3.4 $^{nat}Hf(p,x)^{172g}Lu$ reaction

The cumulative activation cross section of $^{172}Lu$ (6.70 d) includes the decay of the short-lived $^{172m}Lu$ isomeric state (3.7 min, IT 100%) (Fig. 17). A possible contribution of $^{172}Hf$ (1.87 y) decay was neglected due to the low (not observable) contribution. No overlapping earlier experimental data were found. There are large differences between the theoretical results.

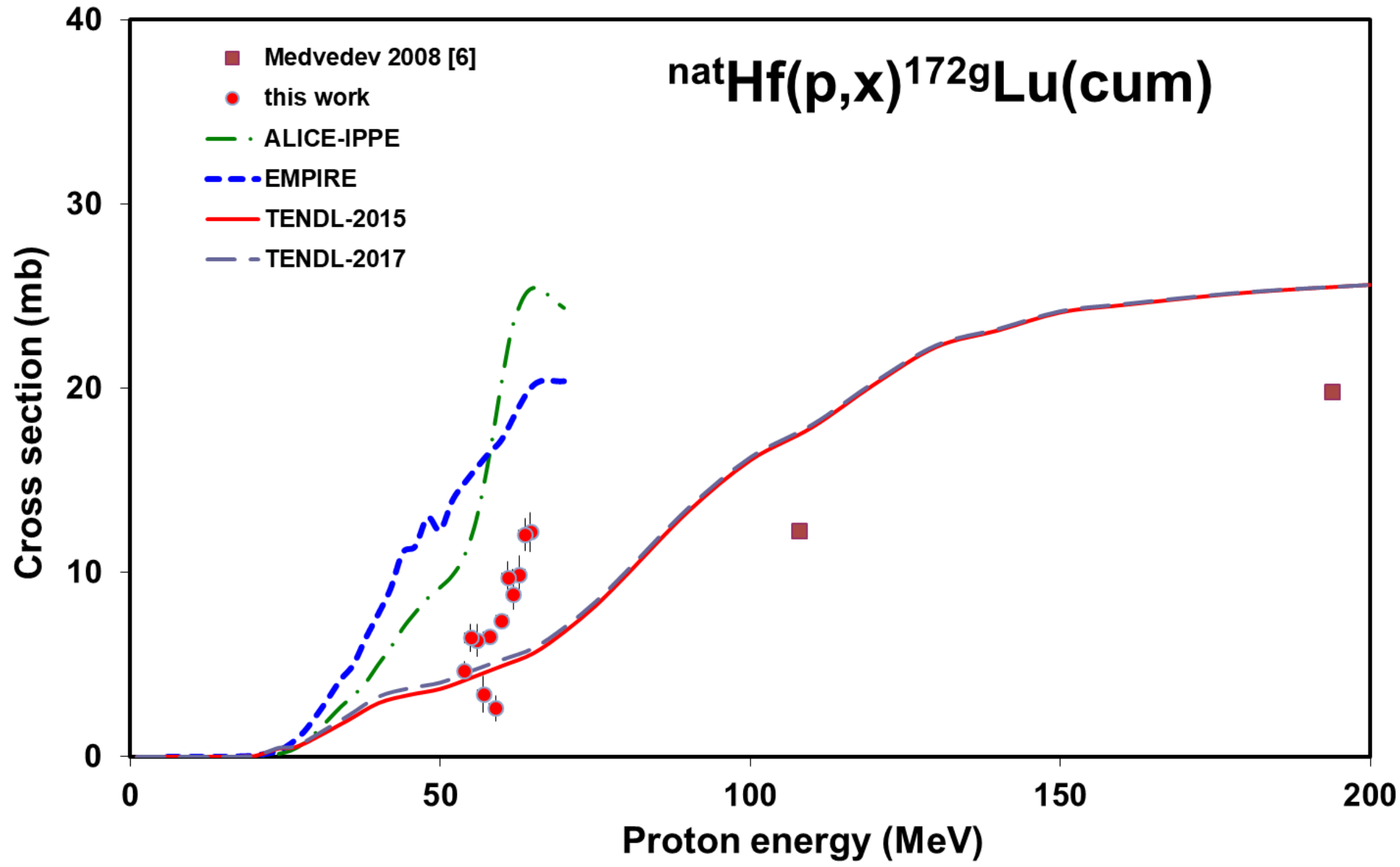

Fig. 17**.** Experimental and theoretical cross sections for the formation of $^{172}Lu$ by the proton bombardment of hafnium.

### 4.1.3.5 $^{nat}Hf(p,x)^{171}Lu$ reaction

The measured cumulative cross sections of $^{171}Lu$ (8.24 d) are shown in Fig. 18. They include the decay of the $^{171}Ta$ (23.3 min) → $^{171}Hf$ (12.2 h) parent decay chain. No overlapping earlier experimental data are available. Our experimental data follow the shape of the theory, but are systematically lower.

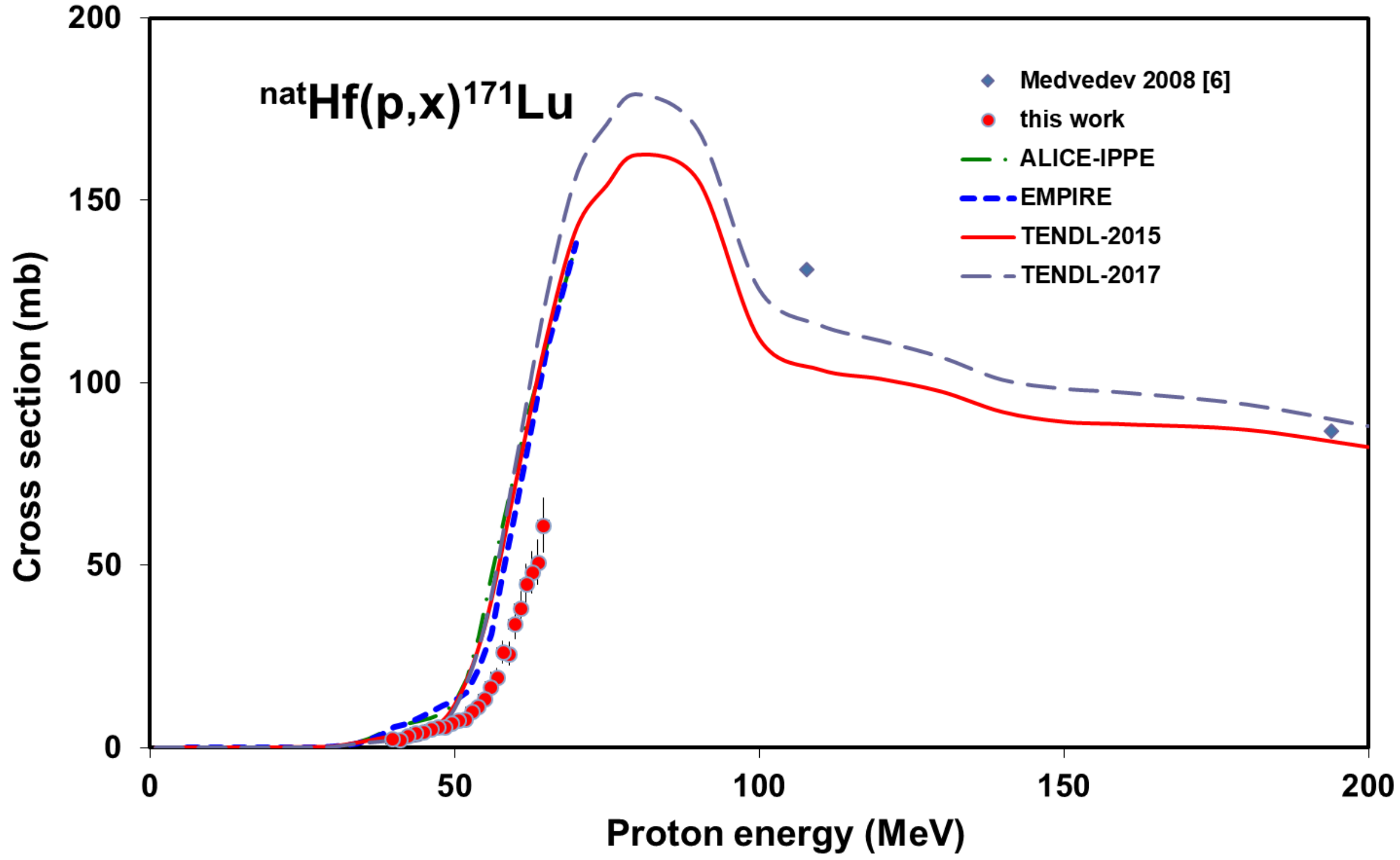

Fig. 18. Experimental and theoretical cross sections for the formation of $^{171}Lu$ by the proton bombardment of hafnium.

### 4.1.3.6 *$^{nat}$Hf(p,x)$^{170}$Lu reaction*

The cumulative cross section of $^{170}$Lu (2.012 d) (Fig. 19) were obtained after the complete decay of the $^{170}$Ta (6.76 min) →$^{170}$Hf (16 h) parent radioisotopes. The agreement between the experimental data and the theory is acceptable.

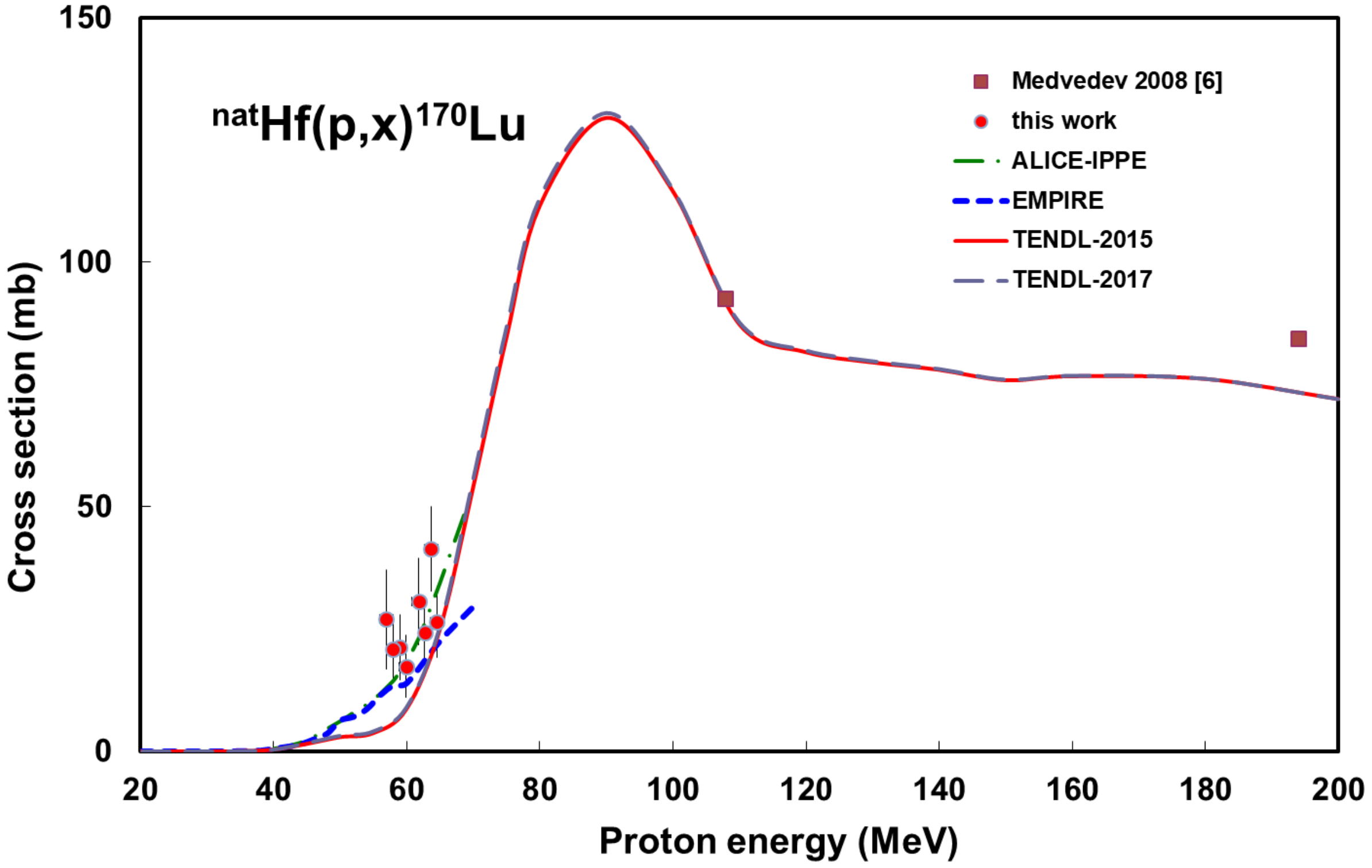

Fig. 19**.** Experimental and theoretical cross sections for the formation of $^{170}$Lu by the proton bombardment of hafnium.

### 4.1.3.7 $^{nat}Hf(p,x)^{169}Lu$ reaction

The few cross sections points of $^{169}$Lu (34.06 h), measured near the effective threshold, are cumulative and include the decay of the $^{169}$Ta (3.25 min) → $^{169}$Hf (5.0 min) decay chain (Fig. 20). Our results are comparable to the predictions of the theoretical codes.

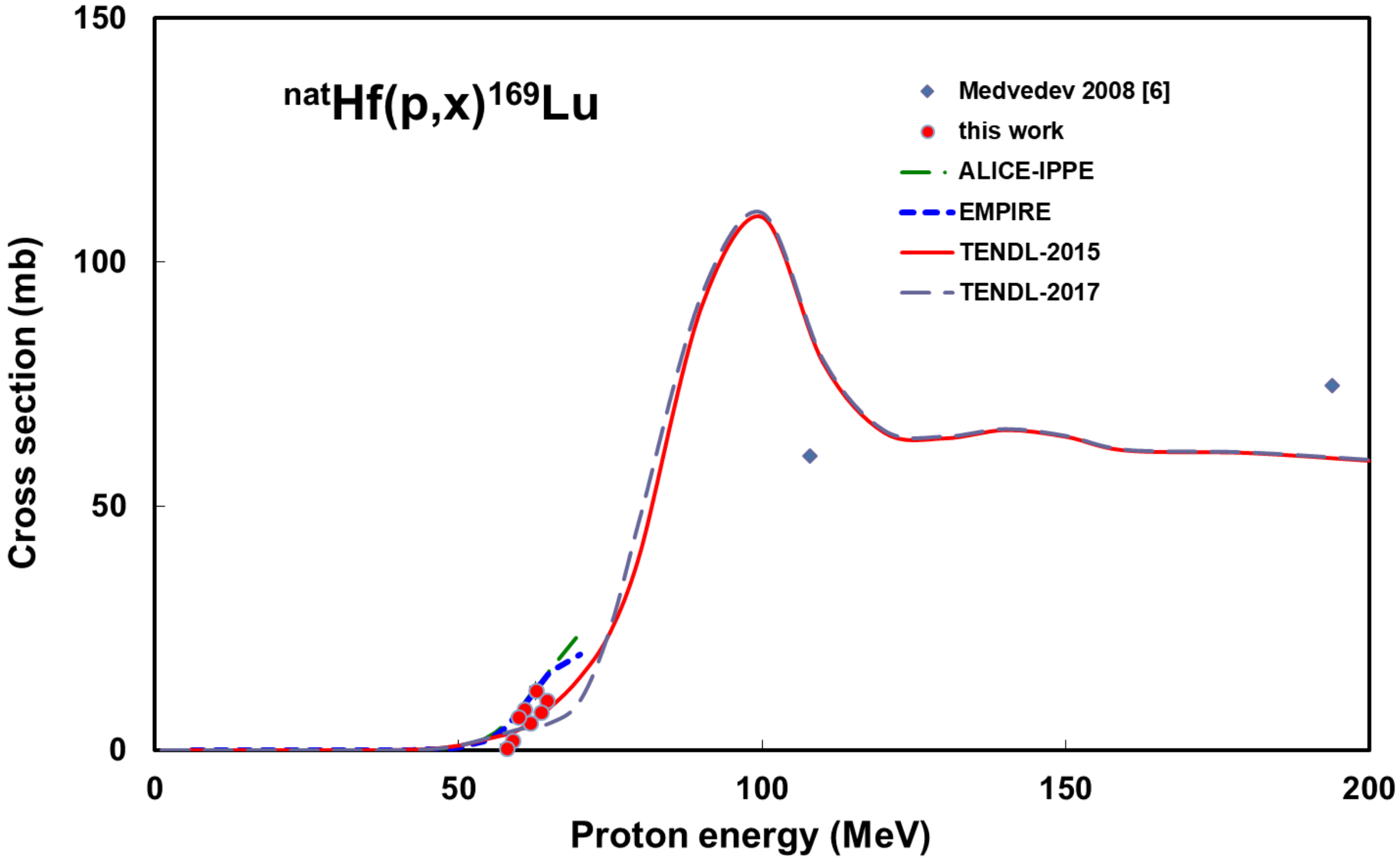

Fig. 20. Experimental and theoretical cross sections for the formation of $^{169}$Lu by the proton bombardment of hafnium.

### *4.1.4 Radioisotopes of ytterbium*

The radioisotopes of ytterbium can be produced directly via reactions where three protons and one or multiple neutrons but are more probably formed in ($\varepsilon+\beta^+$) chain decay of isobaric Ta-Hf-Lu parents.

#### *4.1.4.1 $^{nat}$Hf(p,x)$^{169}$Yb reaction*

The cumulative cross sections for production of $^{169}$Yb (32.018 d) were obtained after complete decay of the $^{169}$Ta (3.25 min) → $^{169}$Hf (5.0 min) → $^{169}$Lu (34.06 h) parent radioisotopes (Fig. 21). The best theoretical prediction is given by the TENDL-2017.

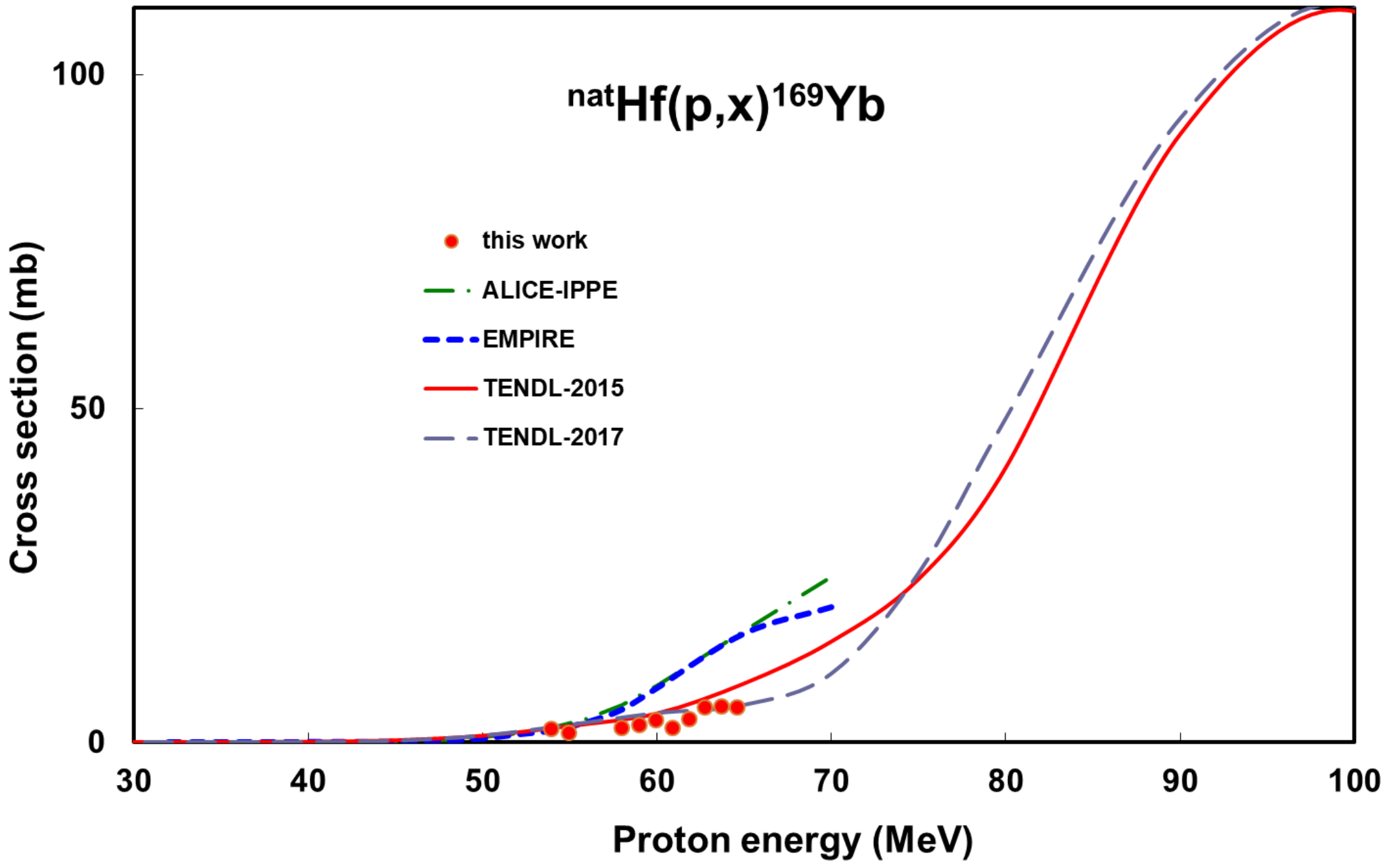

Fig. 21. Experimental and theoretical cross sections for the formation of $^{169}$Yb by the proton bombardment of hafnium.

## 4.2. Integral yields

The integral yields, calculated from spline fits to our experimental excitation functions (complemented with data of Takács et al. 2011 [8], Shahid et al. 2014 [9], Batij et al. 1986 [5], Murakami et al. 2014 [10], Medvedev et al. 2008 [6]), are shown in Figs. 22 (Ta-isotopes), 23 (Hf-isotopes) and 24 (Lu-isotopes). The integral yields represent so called physical yields i.e. activity instantaneous production rates [21, 22]. No experimental thick target yield data were found in the literature for comparison.

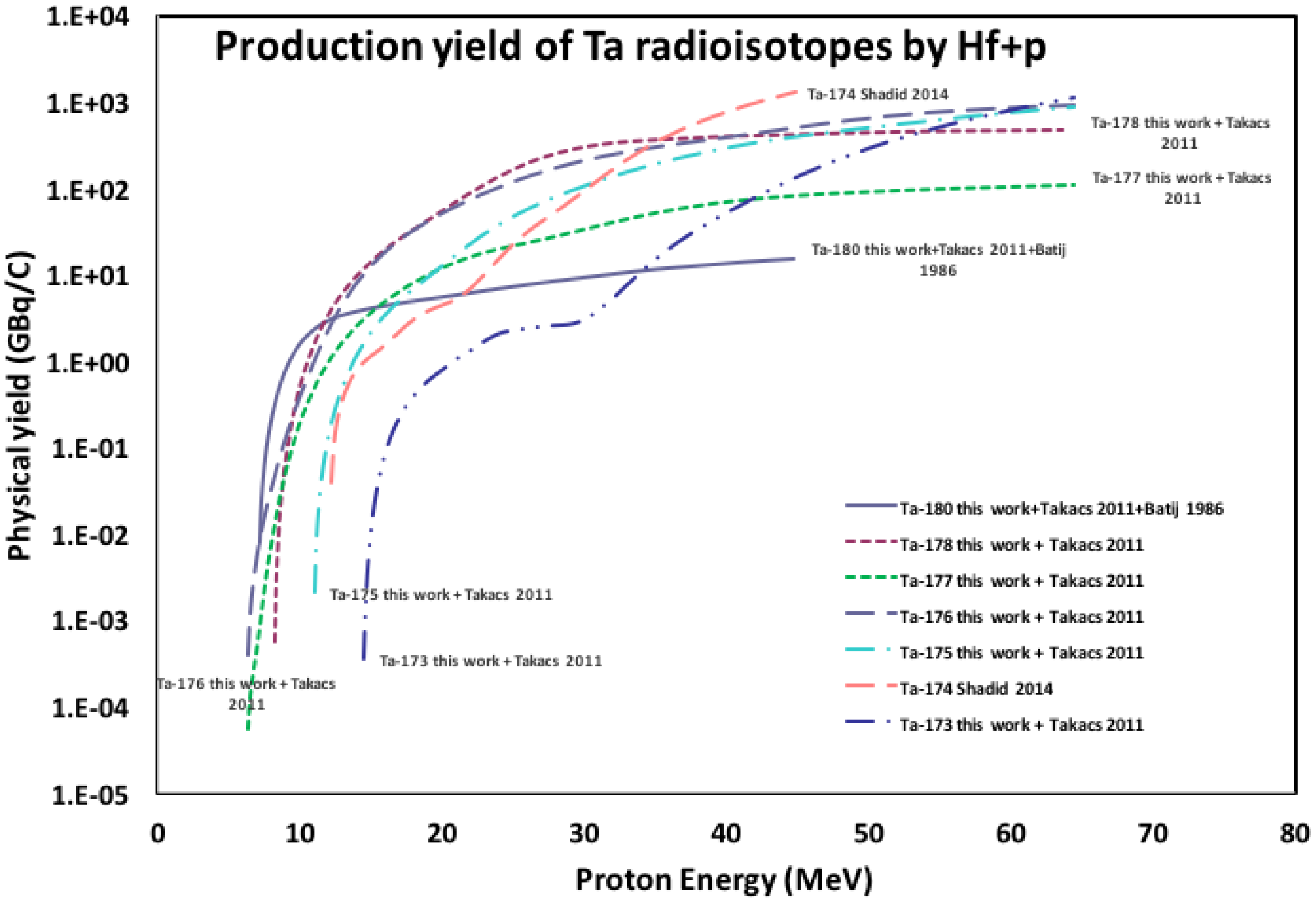

Fig. 22. Integral thick target yields for the formation of the investigated radioisotopes of tantalum as a function of the energy.

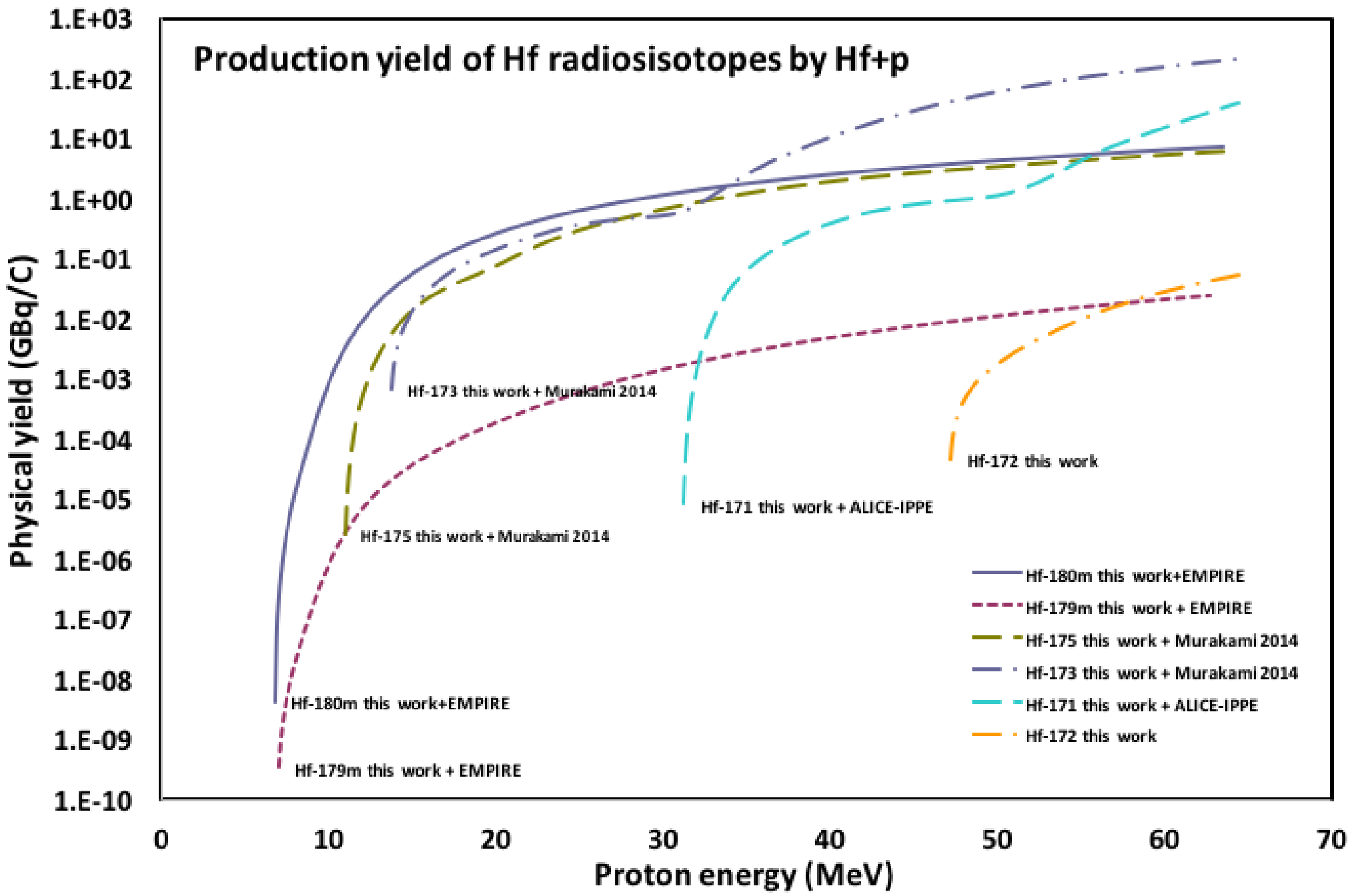

Fig. 23. Integral thick target yields for the formation of the investigated radioisotopes of hafnium as a function of the energy.

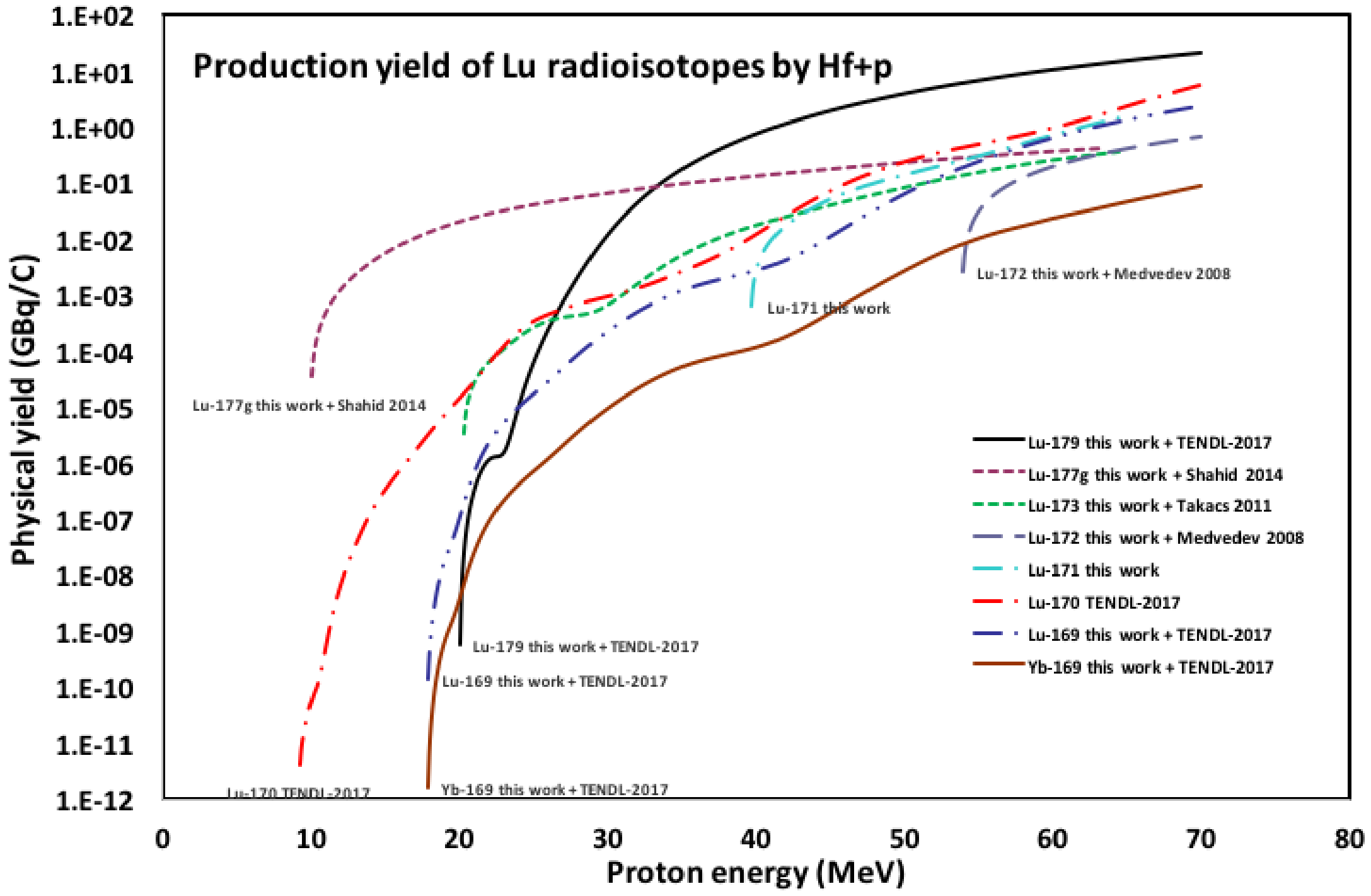

Fig. 24. Integral thick target yields for the formation of the investigated radioisotopes of lutetium and ytterbium as a function of the energy.

# 5. Review of production routes of $^{172}$Lu and $^{172}$Hf

$^{172}$Lu decays to stable $^{172}$Yb with a half-life of 6.70 days. It can be produced directly and by the decay of its long-living ($T_{1/2}$= 1.87 y) $^{172}$Hf parent. $^{172}$Lu appears to be suitable as a long-lived rare-earth tracer for compound labeling for animal bio-distribution studies [28] and for industrial applications. It has abundant low and high energy gamma-lines. The gamma-energies allow industrial transmission studies [29].

The isotope $^{172}$Lu can be produced directly via $^{172}$Yb(p,n)$^{172}$Lu, $^{nat}$Yb(p,xn)$^{172}$Lu, $^{171}$Yb(d,n)$^{172}$Lu, $^{nat}$Yb(d,xn)$^{172}$Lu, $^{169}$Tm ($\alpha$,n),$^{172}$Lu and $^{nat}$Yb($\alpha$,x)$^{172}$Lu charged particle nuclear reactions.

## 5.1 Direct production routes

### *5.1.1 $^{172}$Yb(p,n)$^{172}$Lu and $^{nat}$Yb(p,xn)$^{172}$Lu reactions*

Natural ytterbium is multi-isotopic. No experimental data are available for the $^{172}$Yb(p,n)$^{172}$Lu reaction**.** The theoretical data for $^{172}$Yb(p,n)$^{172}$Lu are shown in Fig. 25 together with the experimental [30] and theoretical data of the $^{nat}$Yb(p,xn)$^{172}$Lu reaction. On $^{nat}$Yb there are a lot of contaminants of Lu that have longer half-life than $^{172}$Lu, so $^{nat}$Yb can not be used in practice for $^{172}$Lu production.

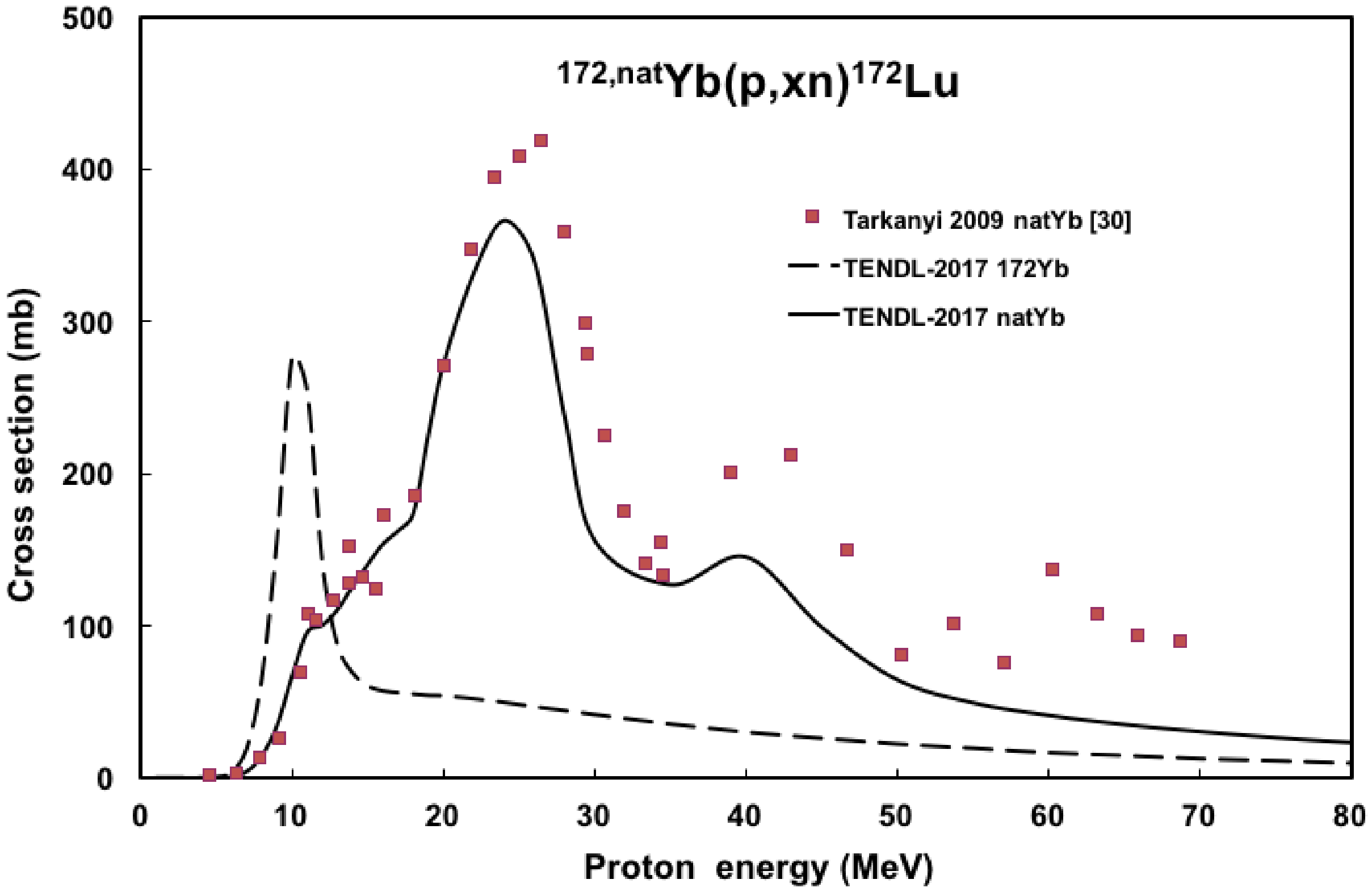

Fig. 25. Excitation functions of the $^{172}Yb(p,n)^{172}Lu$ and $^{nat}Yb(p,xn)^{172}Lu$ reactions.

### 5.1.2 *The $^{171}Yb(d,n)^{172}Lu$ and $^{nat}Yb(d,xn)^{172}Lu$ reactions*

The theoretical data for $^{171}Yb(d,n)^{172}Lu$ together with the experimental [31-35] and theoretical cross sections of $^{nat}Yb(d,xn)^{172}Lu$ reactions are shown in Fig. 26. The experimental data are consistent and show acceptable agreement with the TENDL-2017 calculations. $^{nat}Yb$ cannot be used for medically acceptable $^{172}Lu$ production as the proton induced reaction.

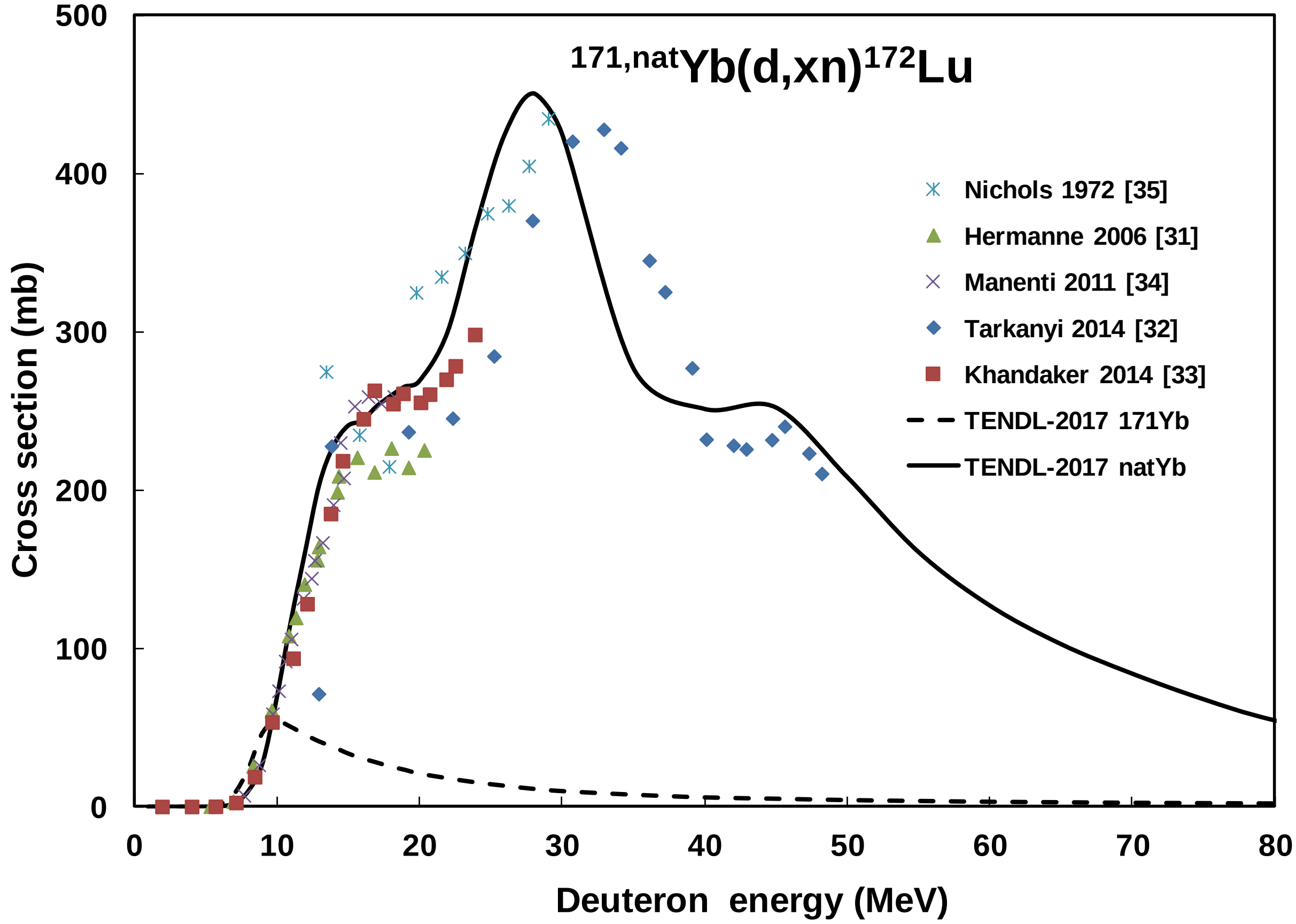

Fig. 26. Excitation functions of the $^{171}Yb(d,n)^{172}Lu$ and $^{nat}Yb(d,xn)^{172}Lu$ reactions.

### 5.1.3 $^{169}Tm(\alpha,n)^{172}Lu$

Natural thulium is monoisotopic. The excitation function of $^{169}Tm(\alpha,n)^{172}Lu$ reaction according to Fig. 27 is well measured [36-42]. This reaction can produce rather pure $^{172}Lu$ if the contribution of the $^{169}Tm(\alpha,2n)^{171}Lu$ is controlled by limiting incident energy.

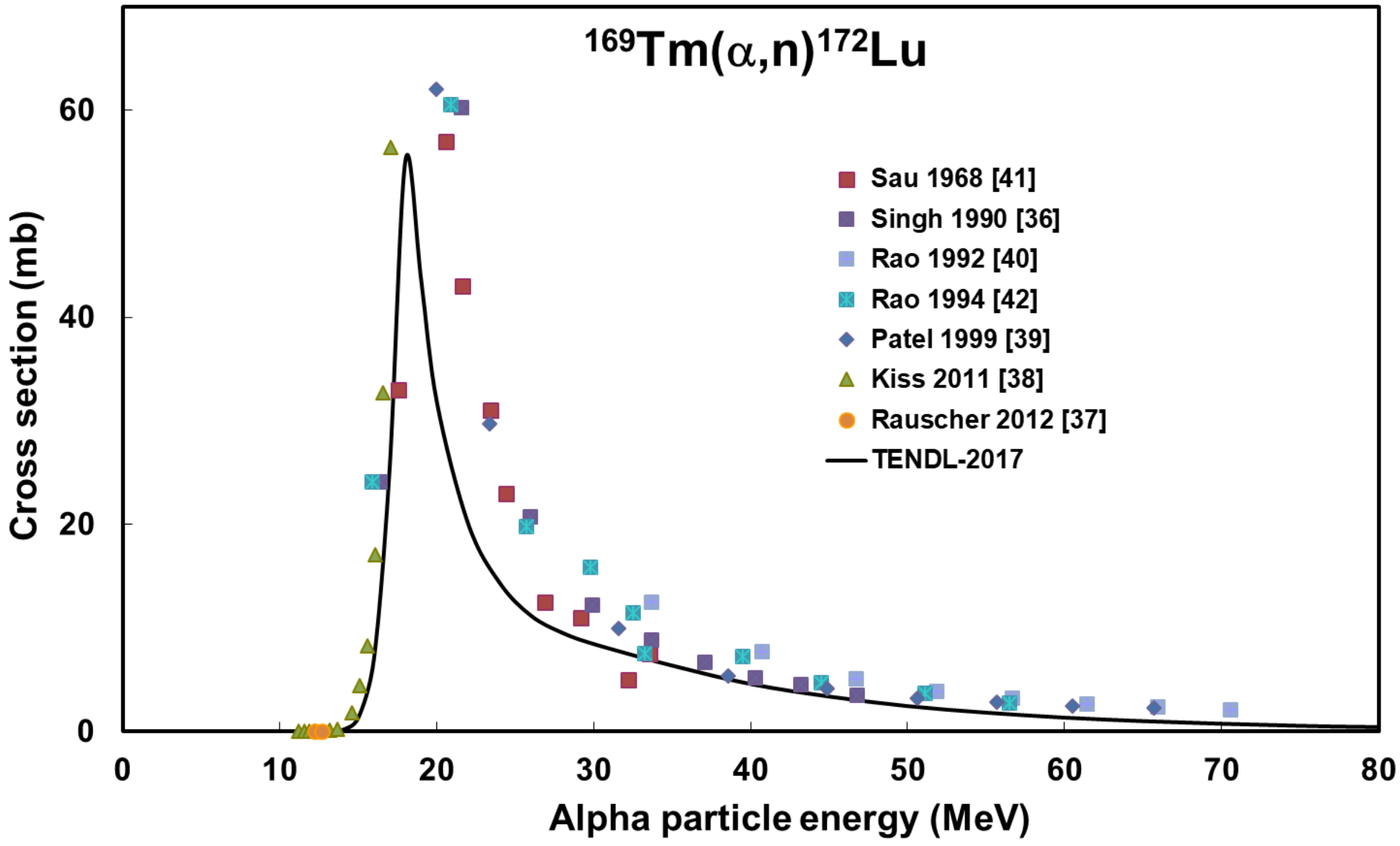

Fig. 27. Excitation functions of the $^{169}Tm(\alpha,n)^{172}Lu$ reaction.

### 5.1.4 $^{nat}Yb(\alpha,x)^{172}Lu$ reaction

The two experimental data sets by Romo 1992 [43] and Kiraly 2008 [44] show good agreement. The TENDL-2017 prediction is significantly lower (Fig. 28). Like for the previously reviewed reactions on $^{nat}Yb$ this route will result in multiple contaminations with longer-lived Hf and Lu radionuclides.

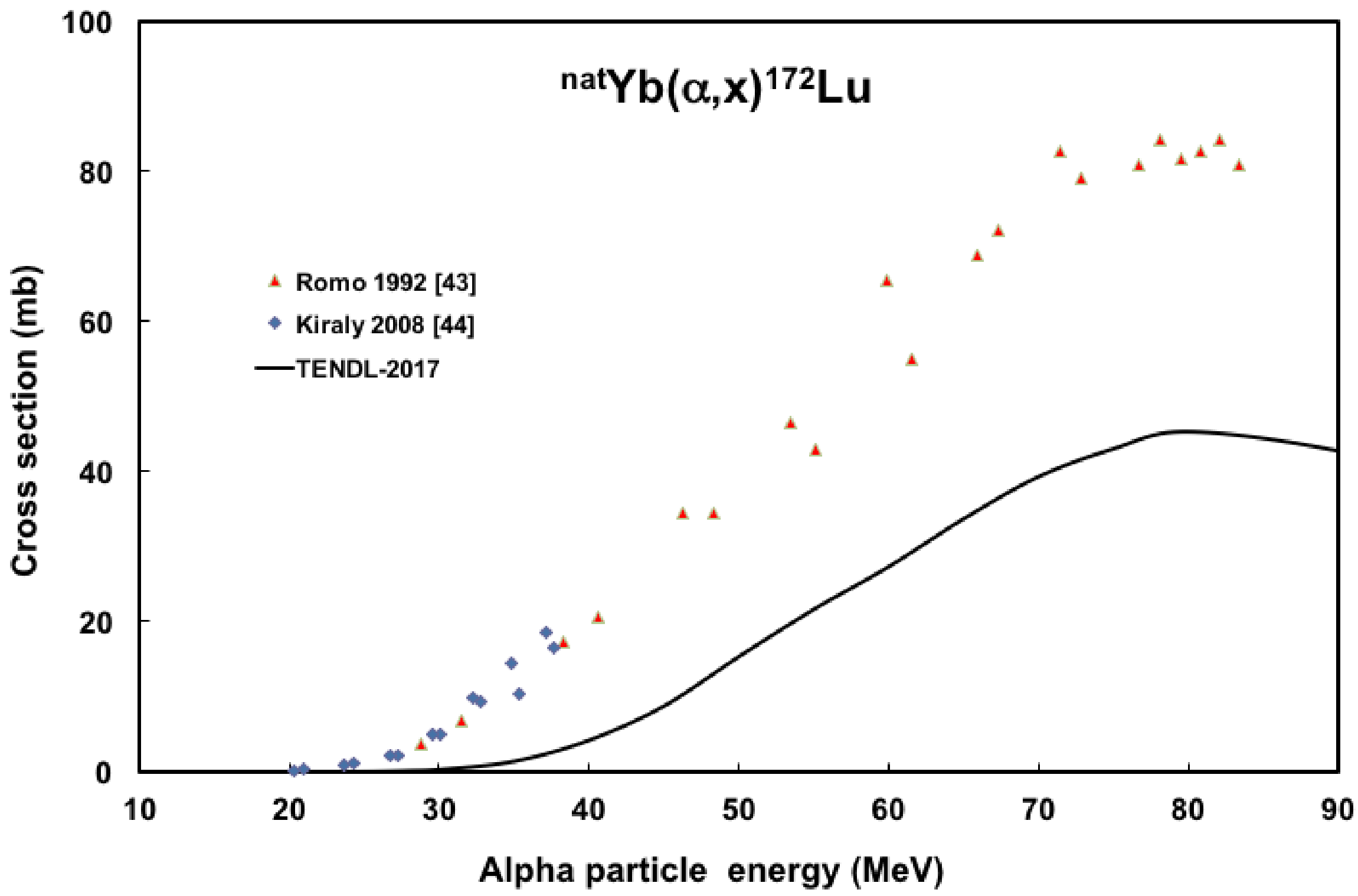

Fig. 28. Excitation functions of the $^{nat}Yb(\alpha,x)^{172}Lu$ reaction.

### 5.2 Indirect production routes

The $^{172}$Hf parent isotope (1.87 y) can be produced at low and middle energy charged particles via $^{nat}$Hf(p,x)$^{172}$Hf, $^{nat}$Lu(p,xn)$^{172}$Hf, $^{nat}$Lu(d,xn)$^{172}$Hf, $^{nat}$Ta(p,x)$^{172}$Hf and $^{nat}$W(p,x)$^{172}$Hf nuclear reactions.

#### *5.2.1 $^{nat}$Hf(p,x)$^{172}$Hf reaction*

The excitation function of the recently measured $^{nat}$Hf(p,x)$^{172}$Hf reaction is shown in Fig. 12.

#### *5.2.2 $^{nat}$Lu(p,xn)$^{172}$Hf reaction*

No experimental data were found for the $^{nat}$Lu(p,xn)$^{172}$Hf reaction. The theoretical prediction is shown in Fig. 29.

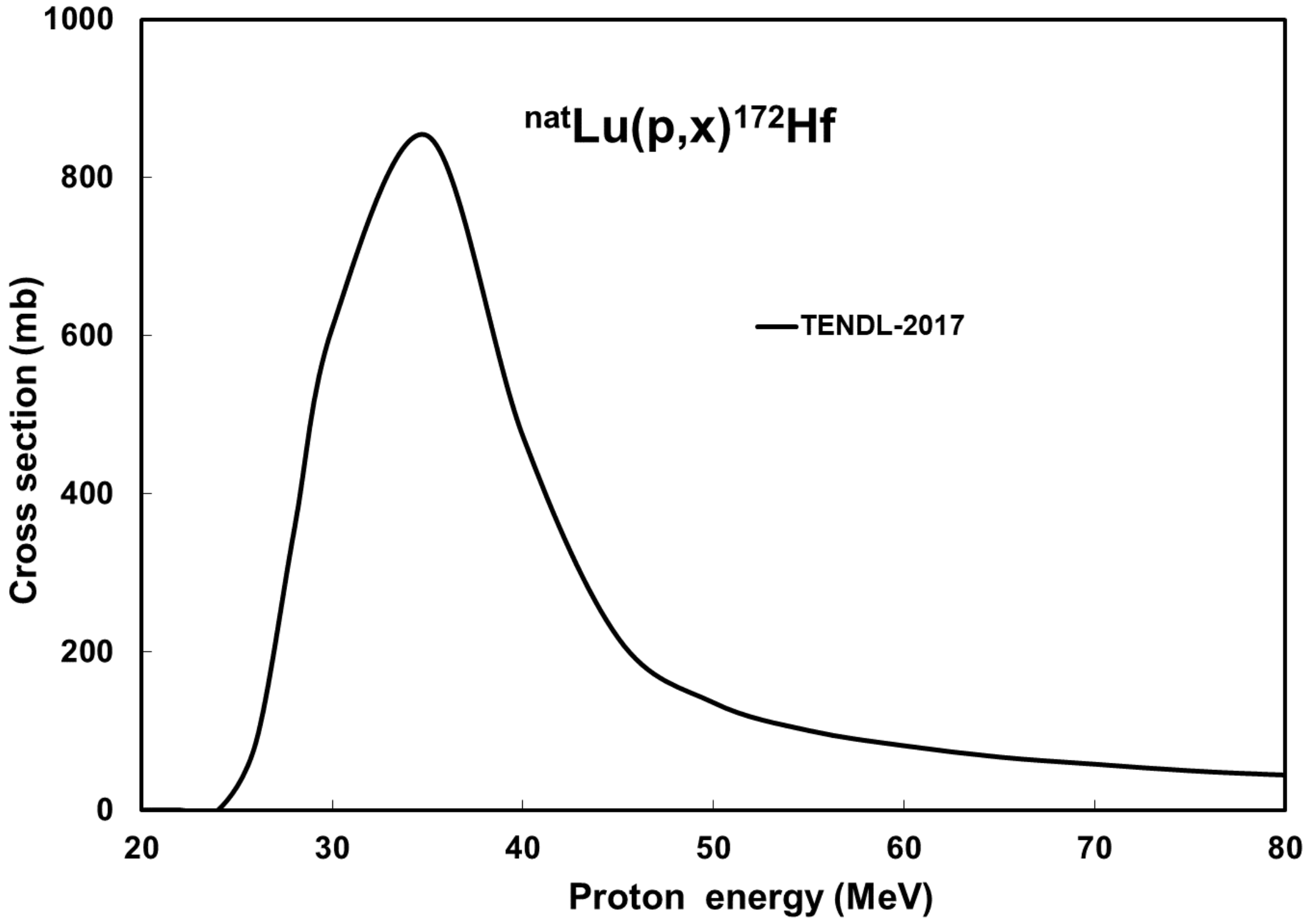

Fig. 29. Excitation functions of the $^{nat}$Lu(p,xn)$^{172}$Hf reaction.

### 5.2.3 $^{nat}Lu(d,xn)^{172}Hf$ reaction

Our new experimental results (presented preliminarily at the RANC (2016) Conference in Budapest [45]) are reproduced in Fig. 30 and are shown in Table 7. The irradiation was performed at the same accelerator and with the same overall experimental conditions as for the present Hf+p measurements, using $^{27}Al(d,x)^{22,24}Na$ monitor reactions and high resolution gamma-ray spectroscopy.

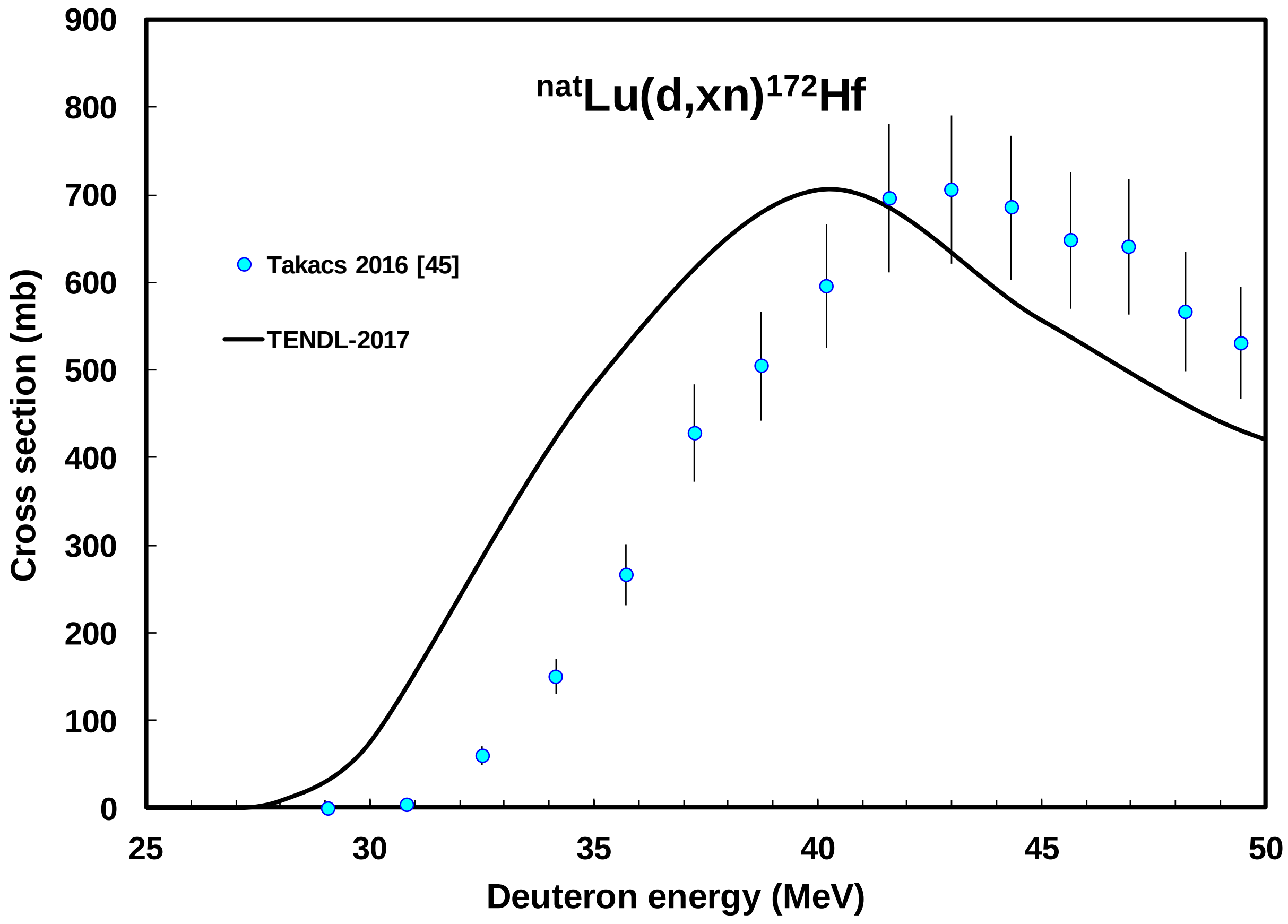

Fig. 30. Excitation functions of the $^{nat}Lu(d,x)^{172}Hf$ reaction.

Table 7 Cross section data of the $^{nat}Lu(d,x)^{172}Hf$ reaction [45].

| | | $^{172}$Hf | |
|---|---|---|---|
| E | ΔE | σ | Δ σ |
| MeV | | mb | |
| 49.4 | 0.3 | 531 | 64 |
| 48.2 | 0.3 | 567 | 68 |
| 46.9 | 0.3 | 641 | 77 |
| 45.6 | 0.4 | 649 | 78 |
| 44.3 | 0.4 | 687 | 82 |
| 43.0 | 0.4 | 707 | 85 |
| 41.6 | 0.4 | 697 | 85 |
| 40.2 | 0.5 | 596 | 71 |
| 38.7 | 0.5 | 506 | 62 |
| 37.2 | 0.5 | 429 | 55 |
| 35.7 | 0.5 | 267 | 35 |
| 34.1 | 0.6 | 150 | 20 |
| 32.5 | 0.6 | 60 | 11 |
| 30.8 | 0.6 | 4 | 2 |
| 29.0 | 0.7 | 0 | 0 |

### 5.2.4 $^{nat}Ta(p,x)^{172}Hf$ reaction

Two experimental data sets, measured by Michel [46] and Titarenko [47] ($^{nat}$Ta consists for 99.988% of $^{181}$Ta) are compared with TENDL-2017 in Fig. 31. The TENDL results significantly differ from the experimental data both in shape and magnitude.

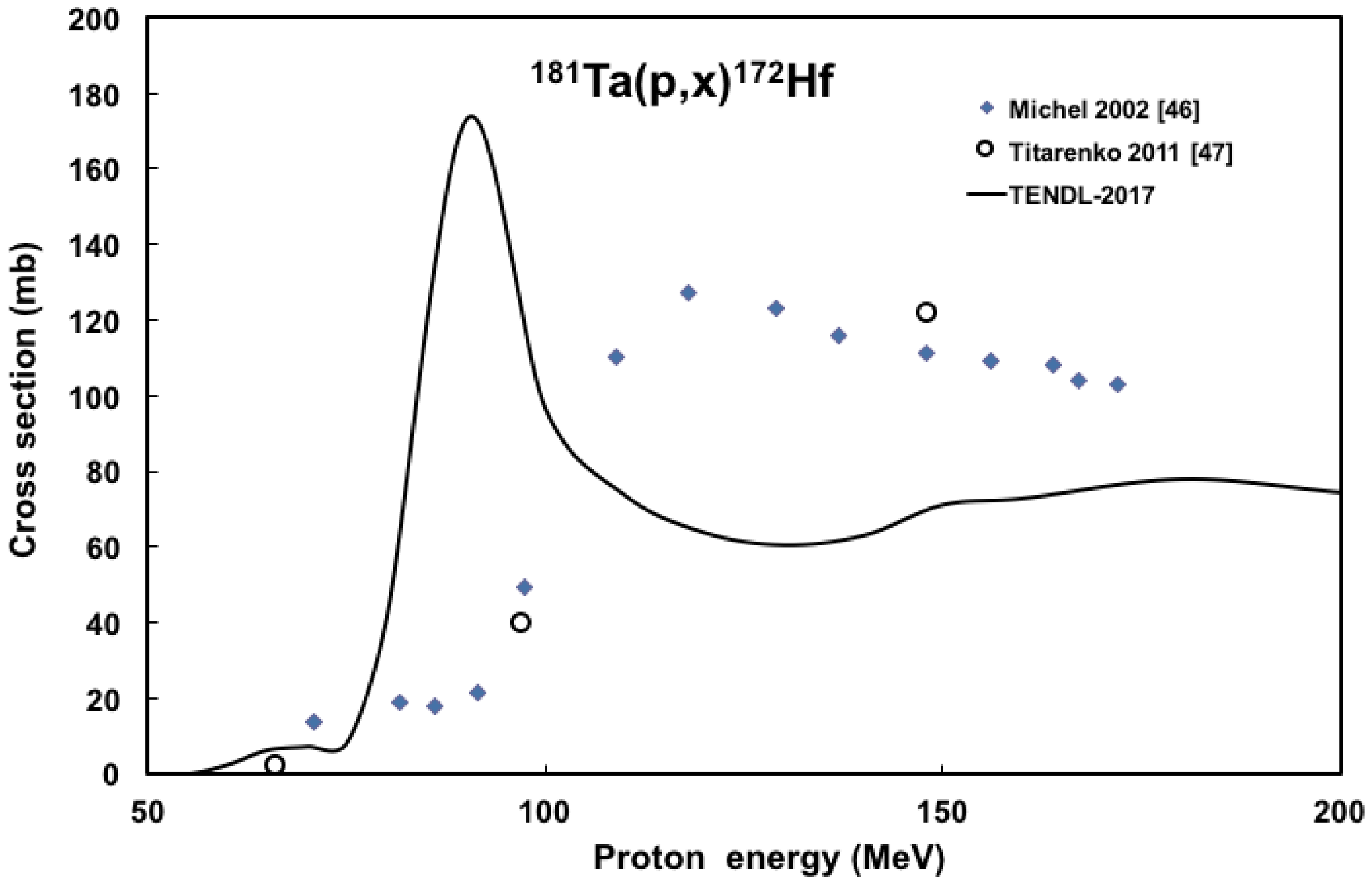

Fig. 31. Excitation functions of the $^{nat}Ta(p,x)^{172}Hf$ reaction.

### *5.2.5 $^{nat}W(p,x)^{172}Hf$ reaction*

The situation for the $^{nat}W(p,x)^{172}Hf$ reaction is the same as for $^{nat}Ta$+p. The experimental data sets of Michel [46] and Titerenko [47] are completely different from the TENDL prediction (Fig. 32). The experimental data are cumulative containing the whole decay chain.

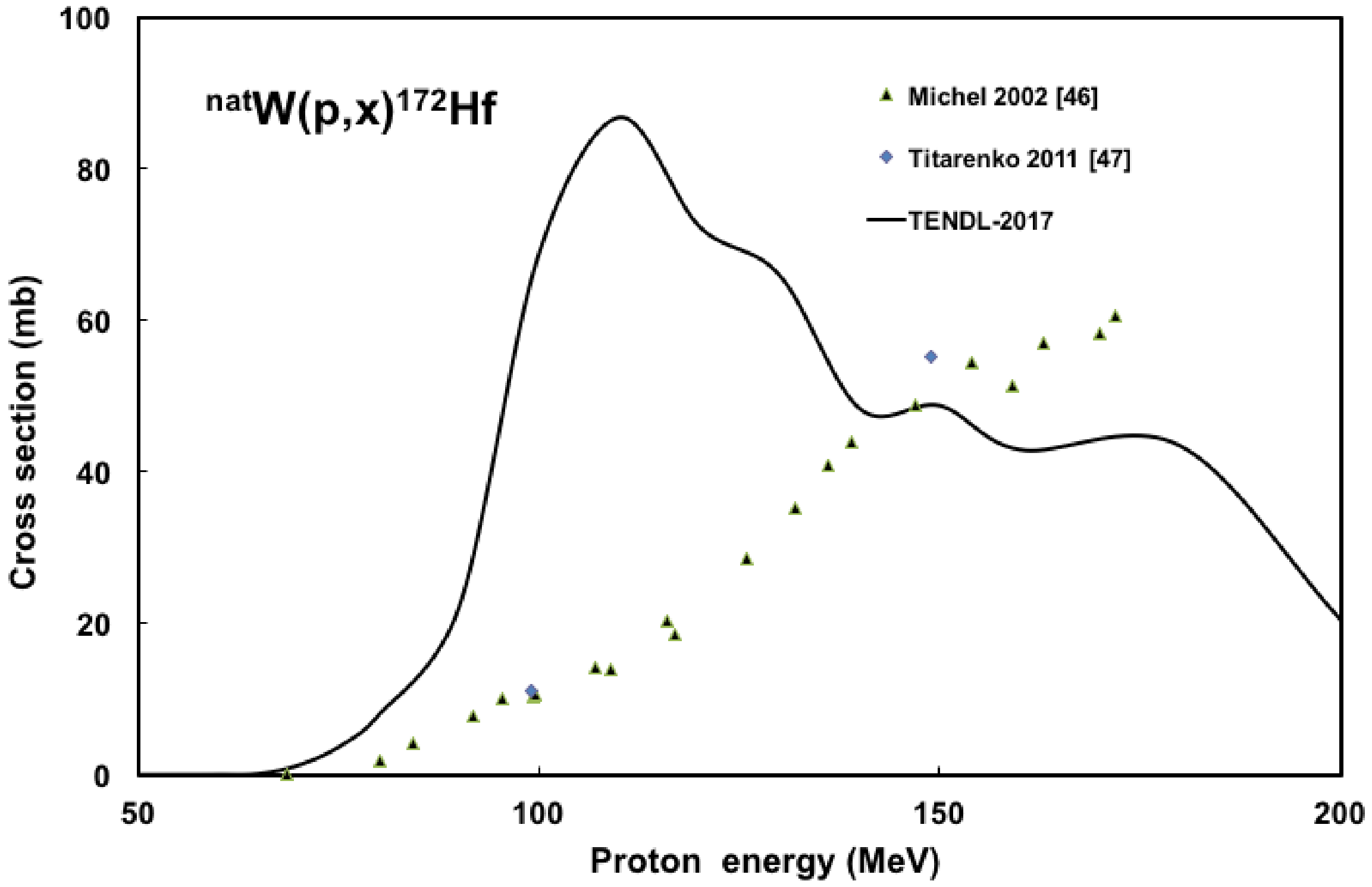

Fig. 32. Excitation functions of the $^{nat}W(p,x)^{172}Hf$ reaction.

### *5.2.6 $^{nat}Yb(\alpha,x)^{172}Hf$ reaction*

The experimental data of Romo 1992 [43] and Tarkanyi 2017 [48] are in good agreement with the TENDL prediction. (Fig. 33).

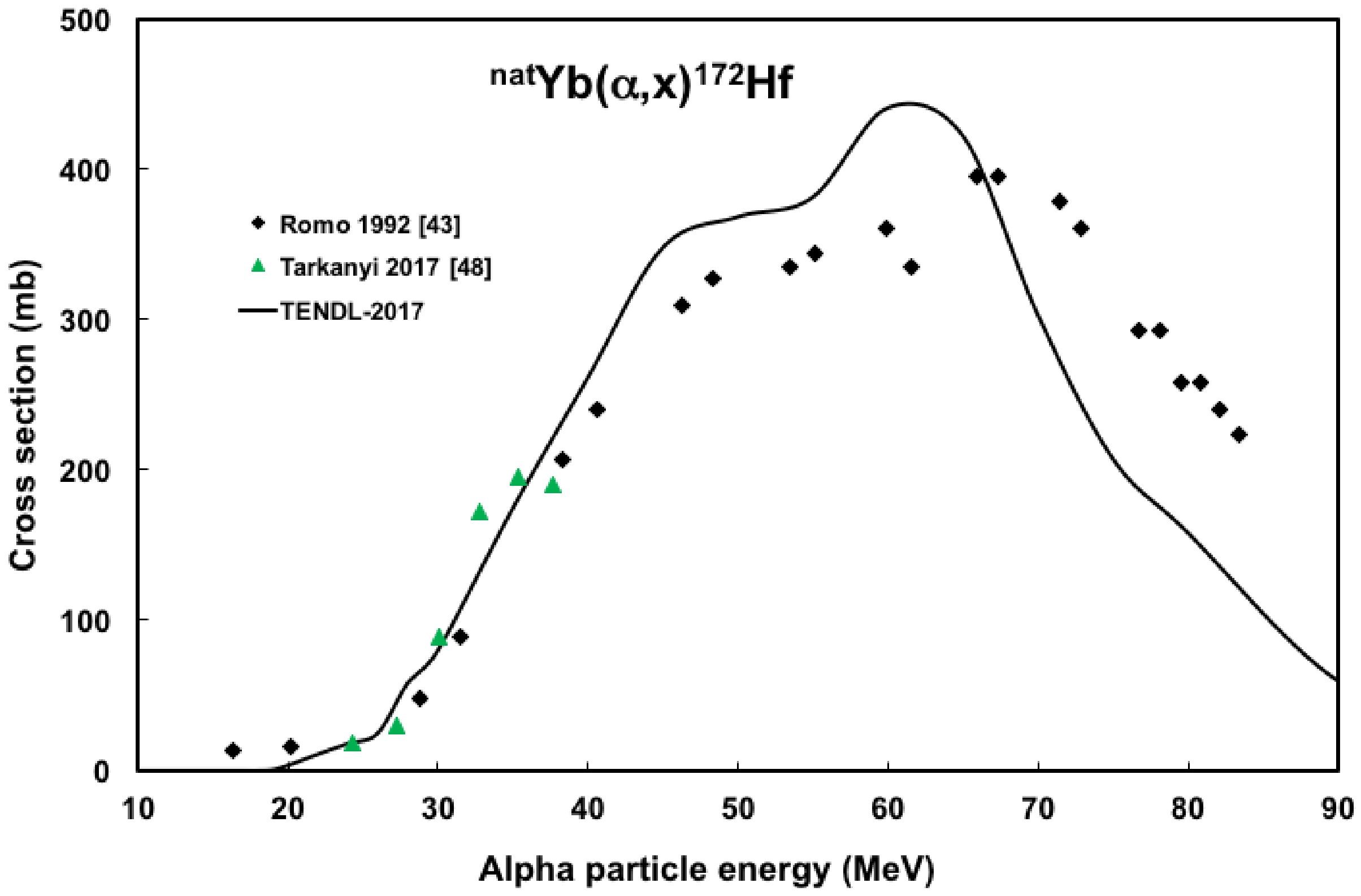

Fig. 33. Excitation functions of the $^{nat}Yb(\alpha,x)^{172}Hf$ reaction.

## 6. Comparison of the production routes of $^{169}Yb$

Due to its decay properties, the $^{169}Yb$ (32.01 d) can be used both in brachytherapy and in medical diagnostics. The production routes were evaluated by Sadeghi 2010 [49], but only part of the reactions were taken into account, and the evaluation is based on theoretical data. The present review deals with the following direct reactions: $^{168}Yb(n,\gamma)^{169}Yb$, $^{169}Tm(p,n)^{169}Yb$, $^{169}Tm(d,2n)^{169}Yb$, $^{nat}Er(\alpha,xn)^{169}Yb$, $^{nat}Er(^{3}He,xn)^{169}Yb$ and formation of parent $^{169}Lu$ through $^{nat}Yb(p,xn)^{169}Lu$, $^{nat}Yb(d,xn)^{169}Lu$, $^{169}Tm(\alpha,x)^{169}Lu$, $^{nat}Lu(p,x)^{169}Lu$ and $^{nat}Hf(p,x)^{169}Lu$.

## 6.1 $^{169}Tm(p,n)^{169}Yb$ reaction,

Three experimental data sets are available measured by Spahn [50], Sonnabend [51] and Tarkanyi [52]. The TENDL-2017 significantly underestimates the experimental data (Fig. 34).

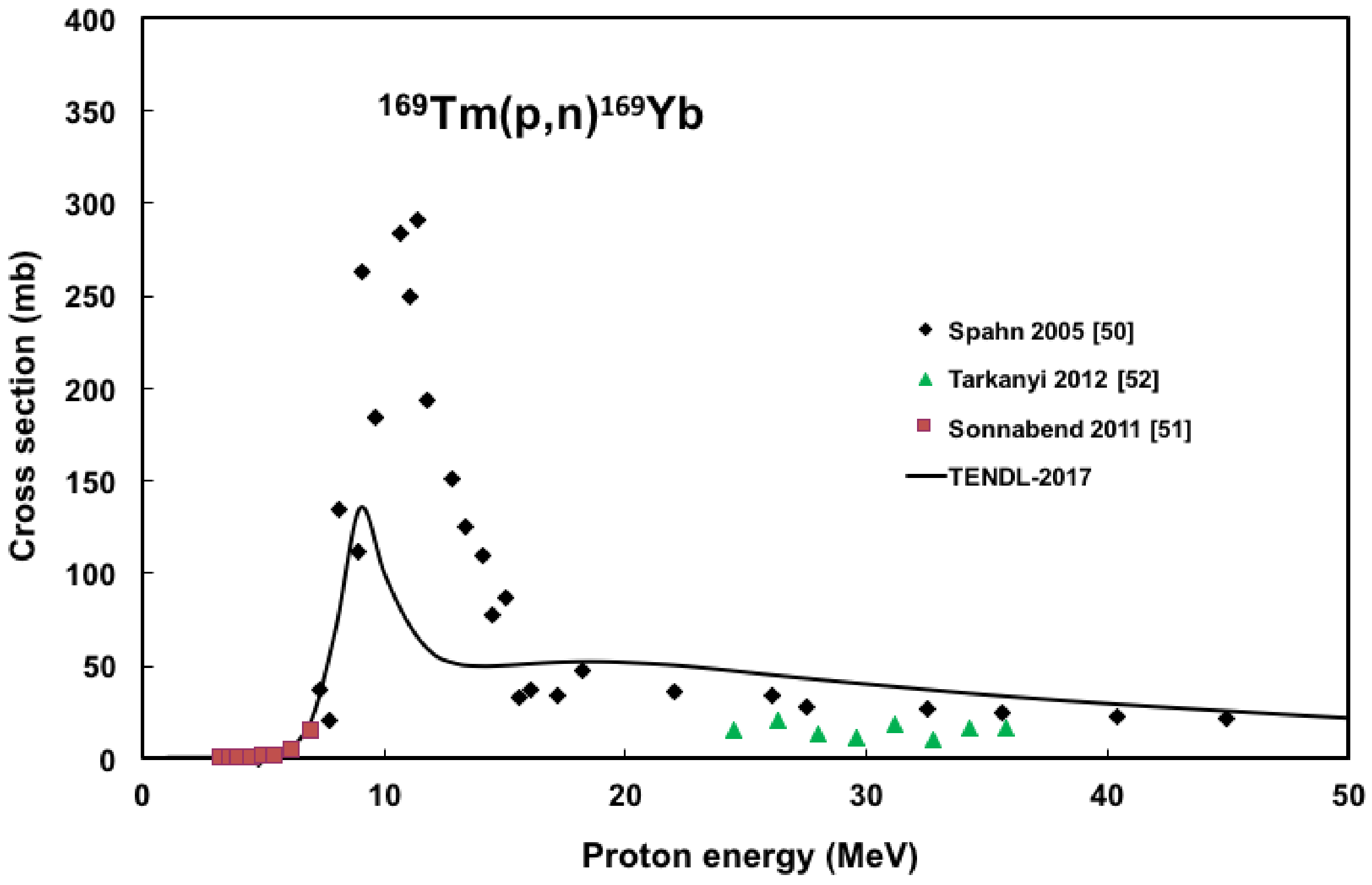

Fig. 34. Excitation functions of the $^{169}Tm(p,n)^{169}Yb$ reaction.

## *6.2 $^{169}Tm(d,2n)^{169}Yb$ reaction*

The three experimental data sets, Tarkanyi 2007 [53], Hermanne 2006 [31], 2009 [54] are in good agreement. The theoretical description is also acceptably good (Fig. 35).

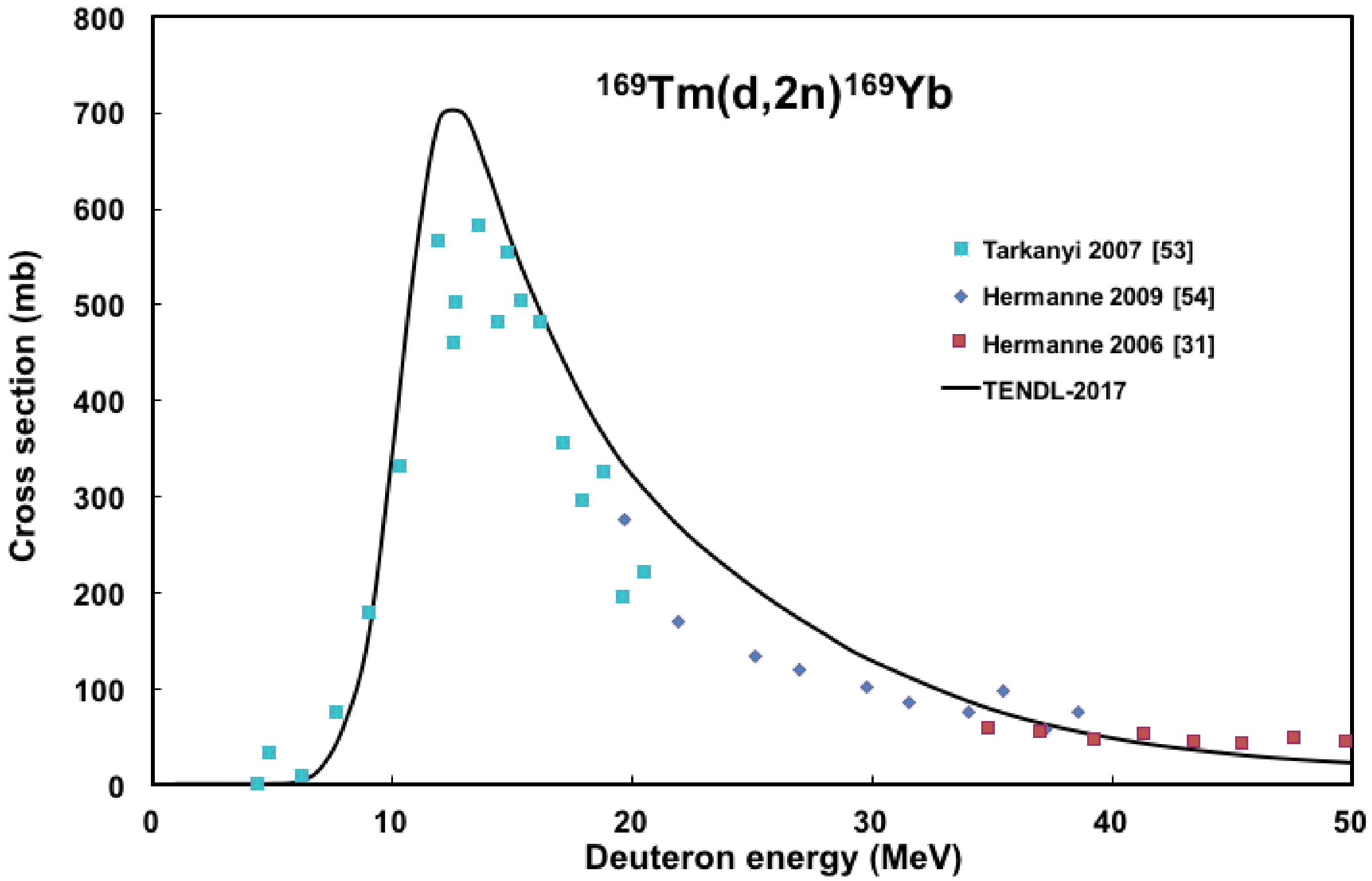

Fig. 35. Excitation functions of the $^{169}$Tm(d,2n) reaction.

### *6.3 $^{nat}$Er($\alpha$,xn)$^{169}$Yb reaction*

There is large disagreement between experimental data of Homma 1980 [55], Archenti [56], Sonzogni 1992 [57] and Kiraly 2007 [58] (Fig. 36).

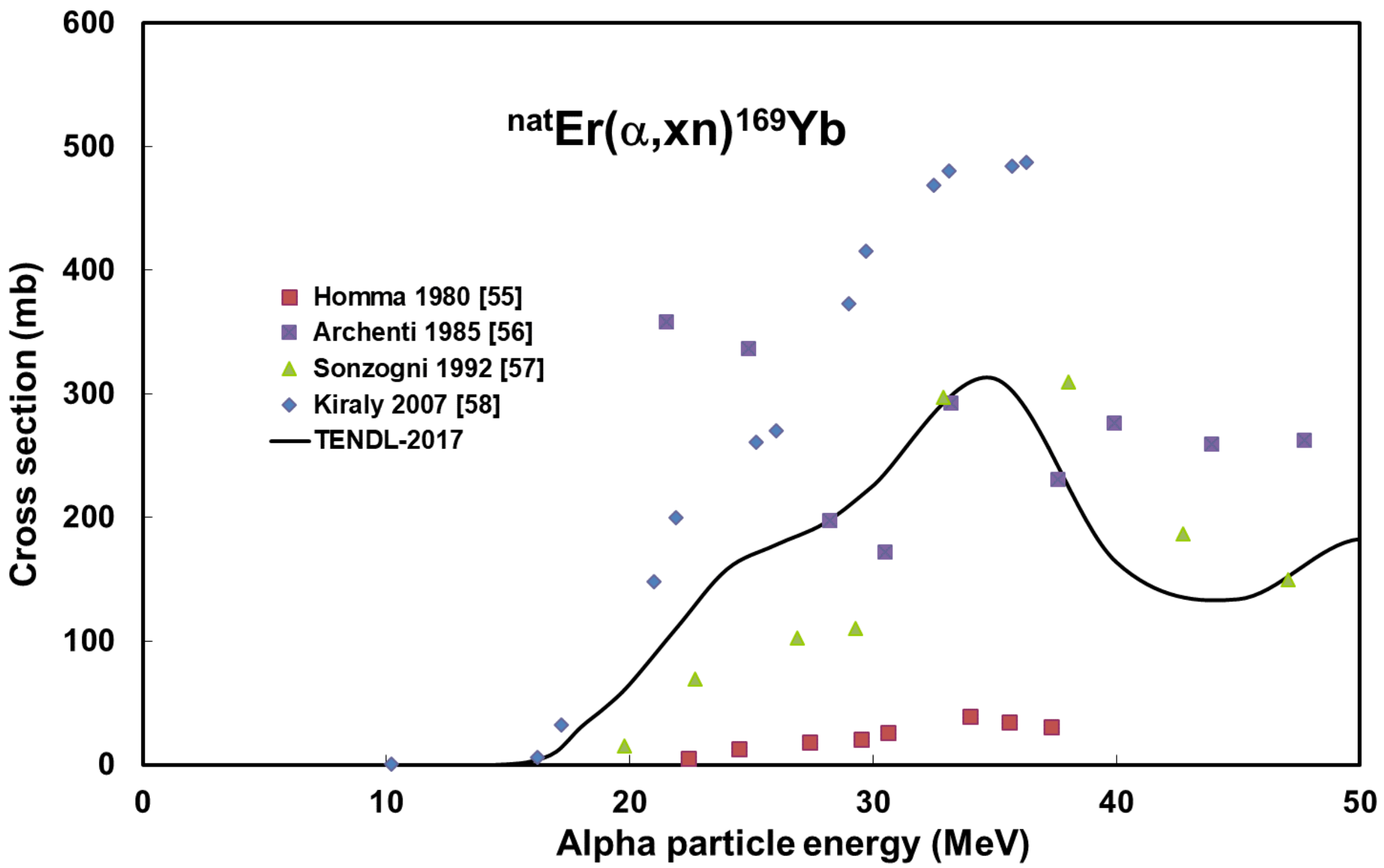

Fig. 36. Excitation functions of the $^{nat}Er(\alpha,xn)^{169}Yb$ reaction.

### 6.4 $^{nat}Er(^{3}He,xn)^{169}Yb$ reaction

The experimental data measured by Homma 1980 [55] differ for nearly a magnitude from values in TENDL-2017 (Fig. 37).

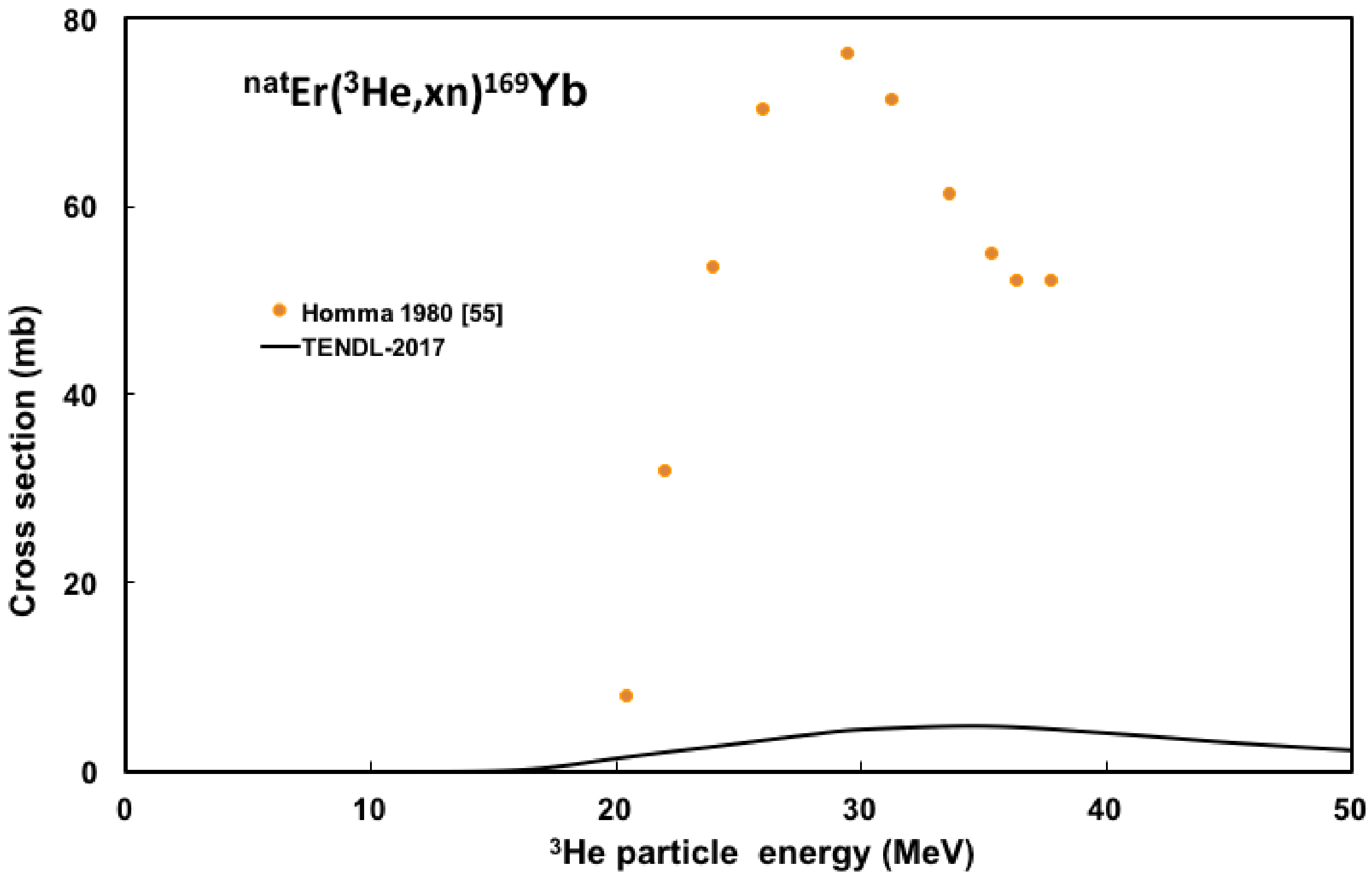

Fig. 37. Excitation functions of the $^{nat}Er(^{3}He,xn)^{169}Yb$ reaction.

### *6.5 $^{nat}Yb(p,xn)^{169}Lu$*

Only one experimental data set is available, measured by Tarkanyi 2009 [59]. At high energies the TENDL-2017 underestimates the experimental data (Fig. 38).

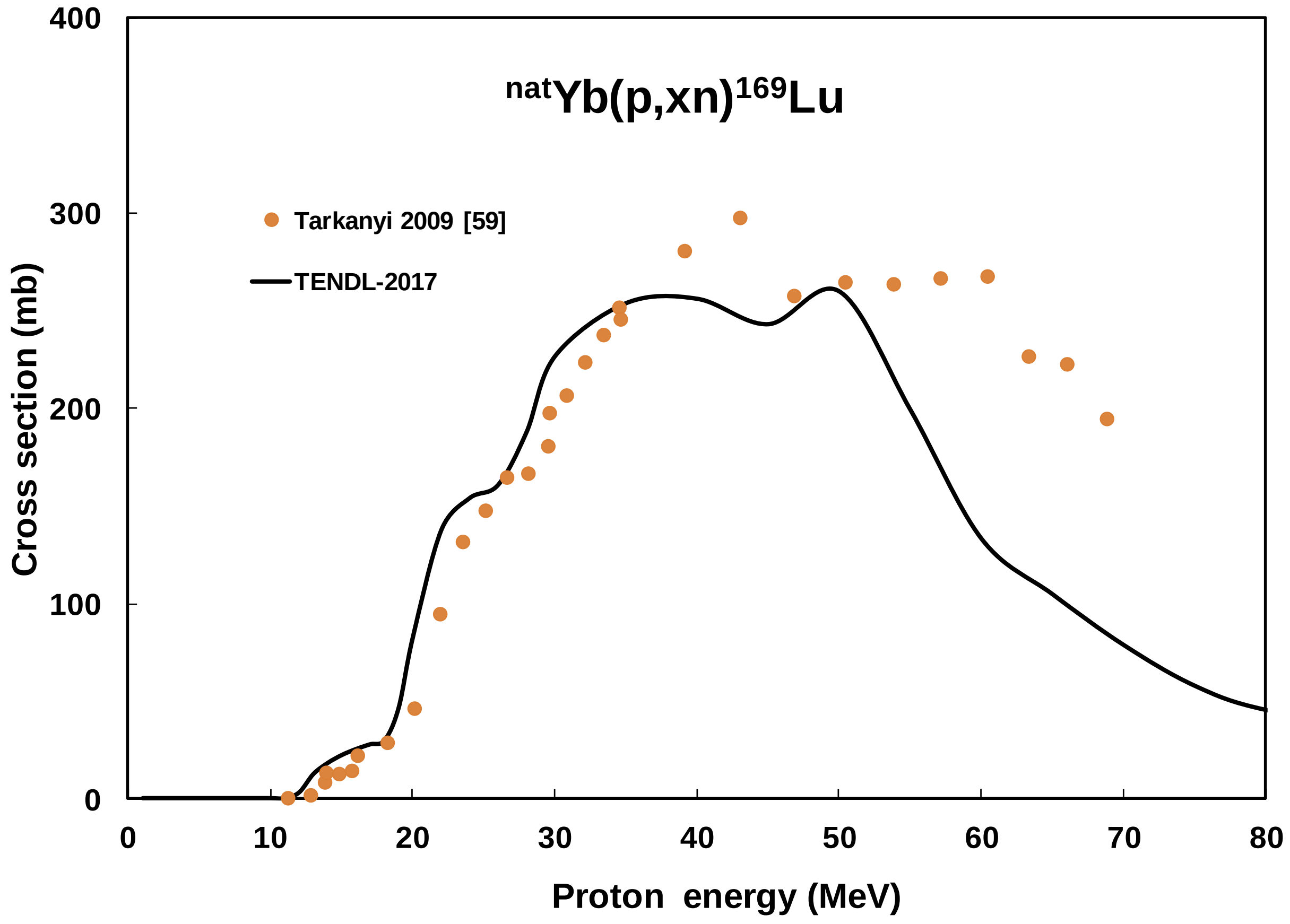

Fig. 38. Excitation functions of the $^{nat}$Yb(p,xn)$^{169}$Lu reaction.

### *6.6 $^{nat}$Yb(d,xn)$^{169}$Lu reaction*

The reaction cross sections were measured earlier by Manenti 2011 [34], Tarkanyi 2013 [60], Tarkanyi 2014 [32] and Khandaker 2014 [33]. The experimental data are in good agreement and, not considering the small energy shift, the TENDL description is also acceptably good (Fig. 39).

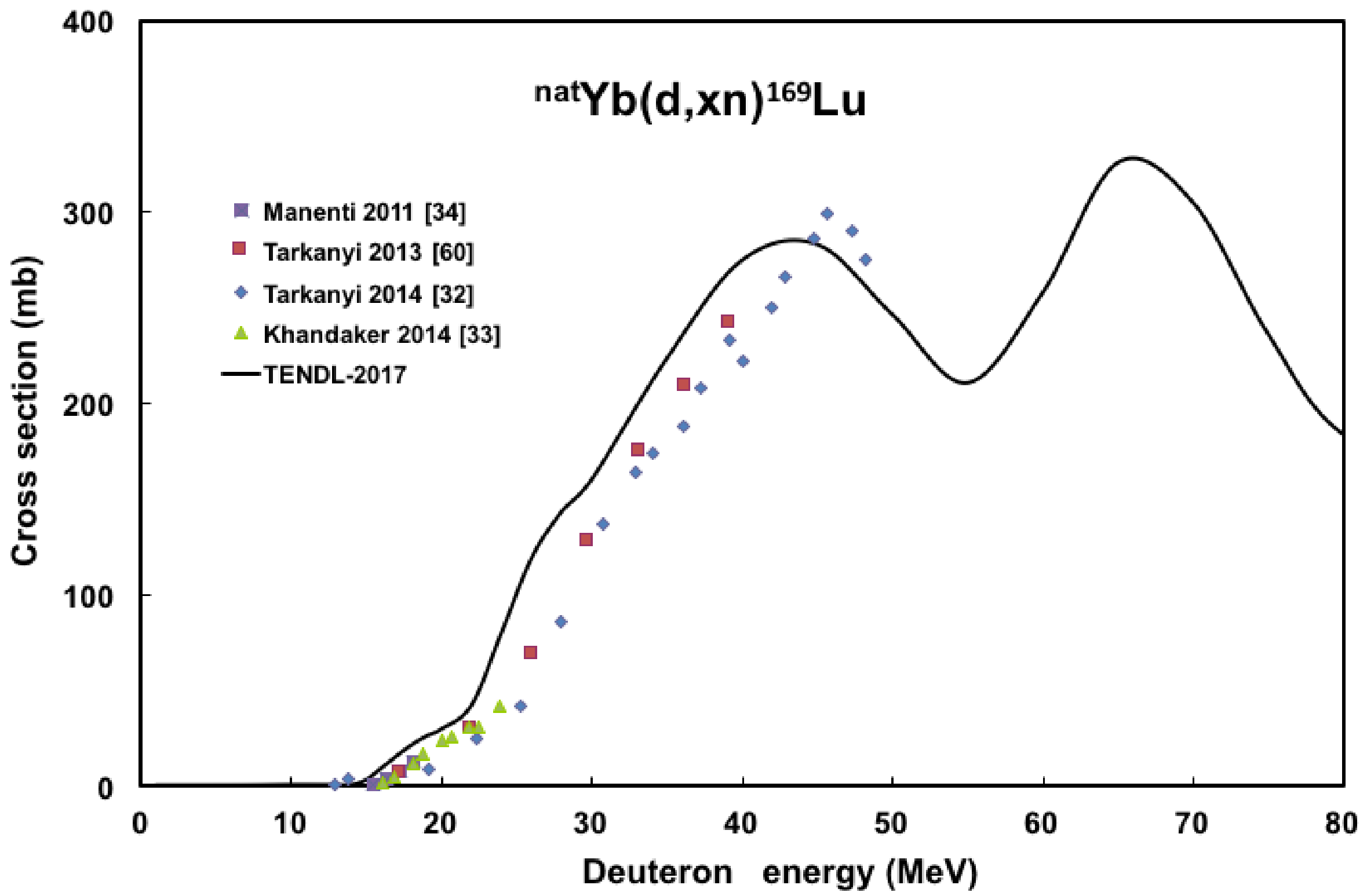

Fig. 39. Excitation functions of the $^{nat}Yb(d,xn)^{169}Lu$ reaction.

### *6.7 $^{169}Tm(\alpha,xn)^{169}Lu$ reaction*

Four experimental data sets were found measured by Sau 1968 [41], Singh 1990 [36], Mahon Rao 1994 [42] and Patel 1999 [39]. A good agreement for the experimental data, but significant disagreement with the TENDL-2017 is seen (Fig. 40).

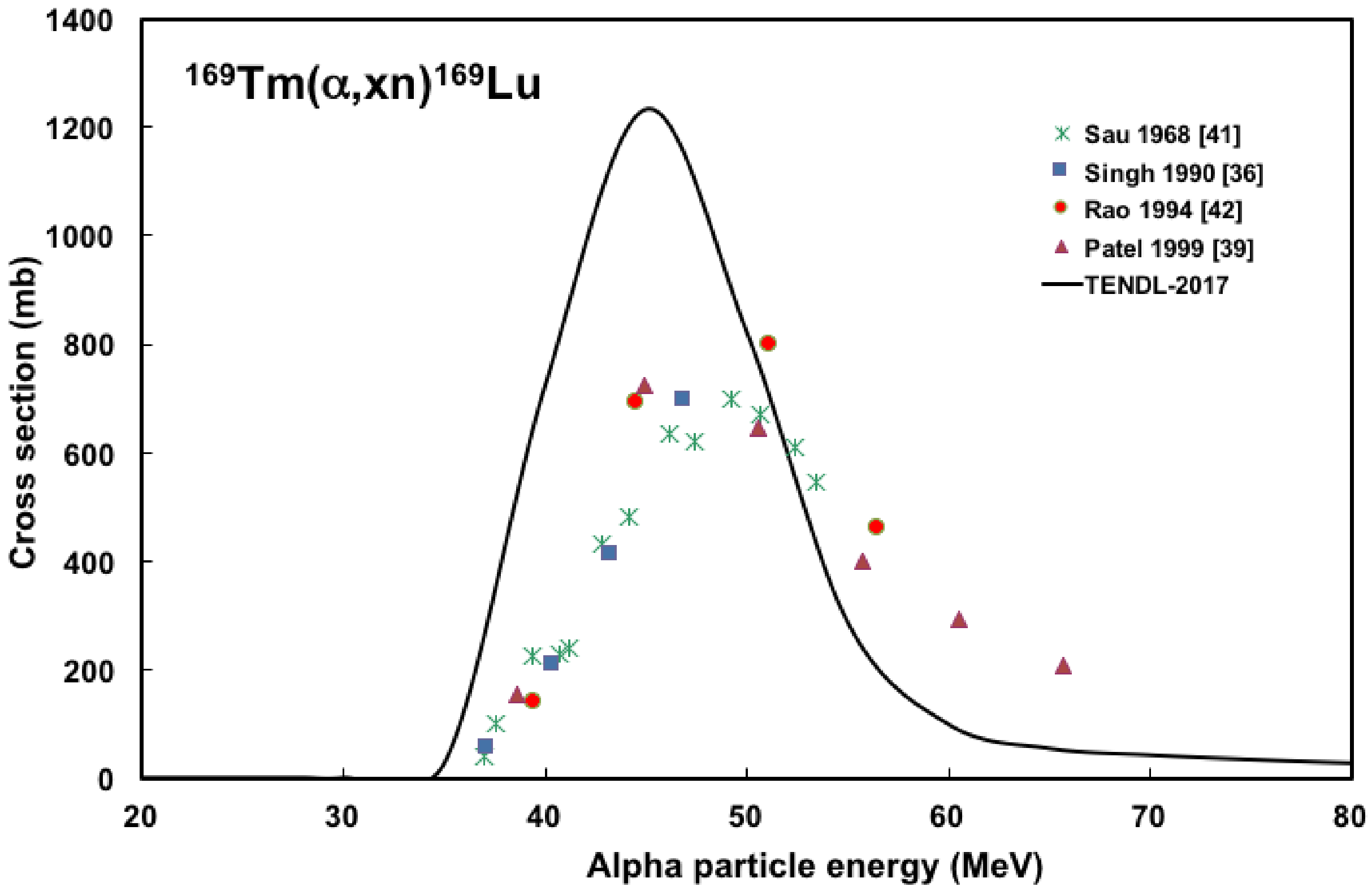

Fig. 40. Excitation functions of the $^{169}Tm(\alpha,xn)^{169}Lu$ reaction.

### *6.8 $^{169}Tm(\alpha,x)^{169}Yb$ reaction*

Only one experimental data set by Rao 1992 [40], measuring independently the production cross section for the metastable state and cumulative production of the ground state (Fig. 41) is available. In accordance with Fig. 41, the $^{169}Yb$ direct production is small and this is confirmed by the TENDL-2017 prediction. The main production is through decay of $^{169}Lu$ parent.

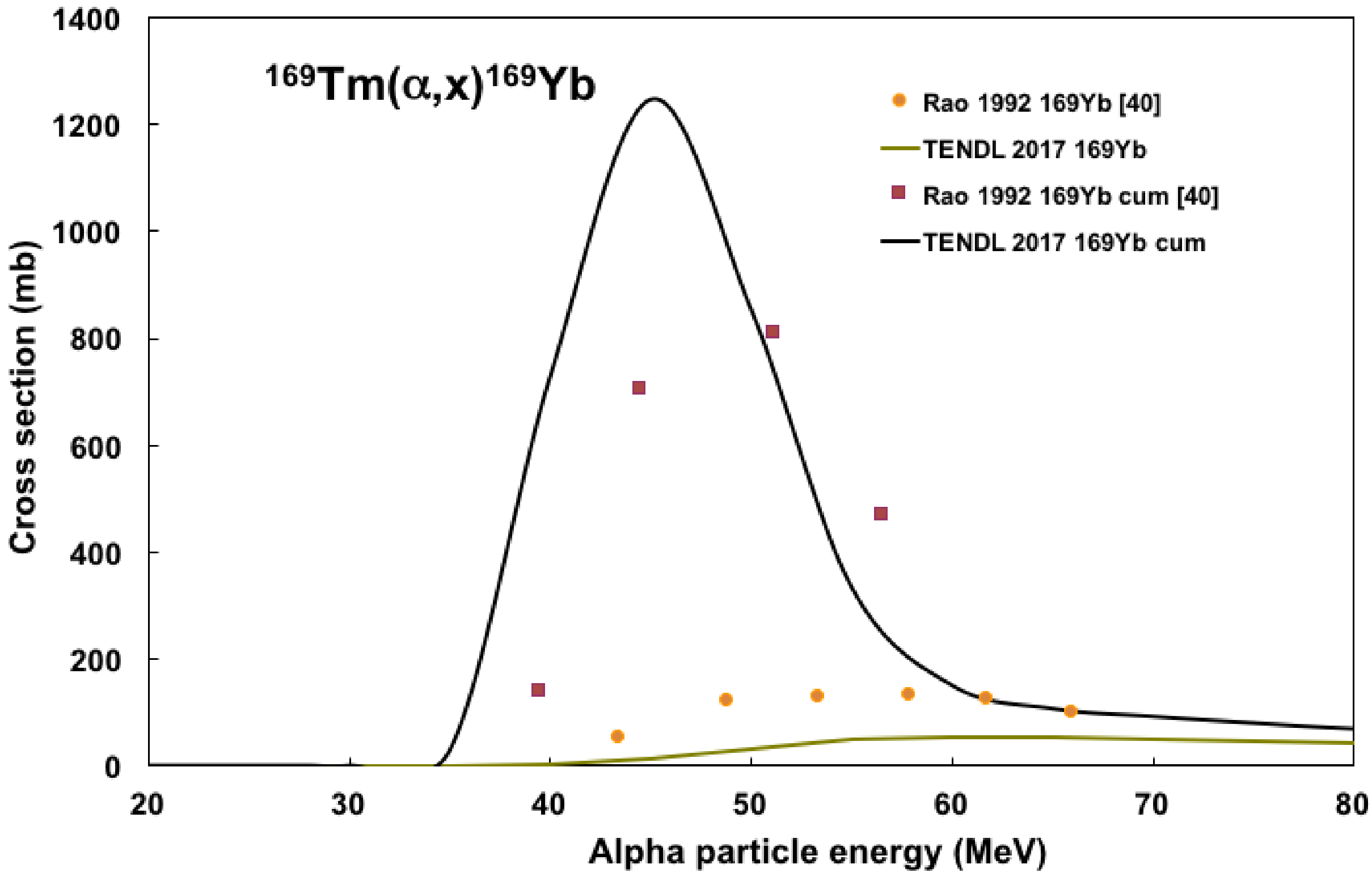

Fig. 41. Excitation functions of the $^{169}Tm(\alpha,x)^{169}Yb$ reaction.

### *6.9 $^{nat}Lu(p,x)^{169}Yb$ reaction*

No experimental data were found for this reaction. The theoretical cross sections are shown in Fig. 42, which includes only the direct reactions on lutetium isotopes.

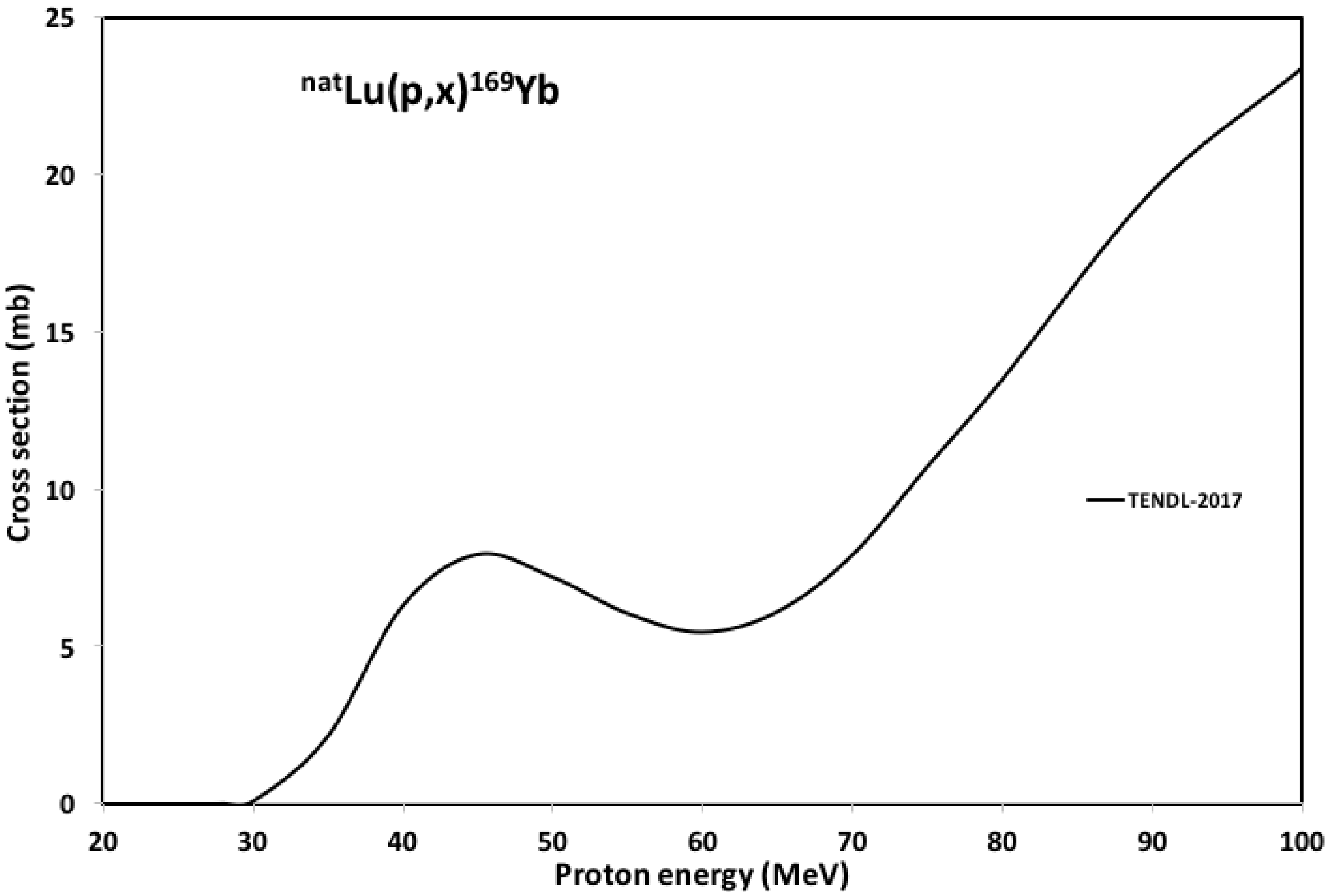

Fig. 42. Excitation functions of the $^{nat}Lu(p,x)^{169}Yb$ reaction.

### *6.10 $^{nat}Hf(p,x)^{169}Yb$ reaction*

Only our present experimental data are available near the effective threshold. The experimental and the theoretical data are shown in Fig. 21 for cumulative cross sections obtained after complete decay of the $^{169}Ta$ (3.25 min) → $^{169}Hf$ (5.0 min) → $^{169}Lu$ (34.06 h) parent radioisotopes.

## 7. Integral yields of charged particle induced reactions for production of $^{172}Lu$, $^{172}Hf$, $^{169}Yb$ and $^{169}Lu$

The integral yields calculated on the basis of the fitted available experimental data, and in case of missing experimental data using the TENDL prediction, are shown in Figs. 43-46.

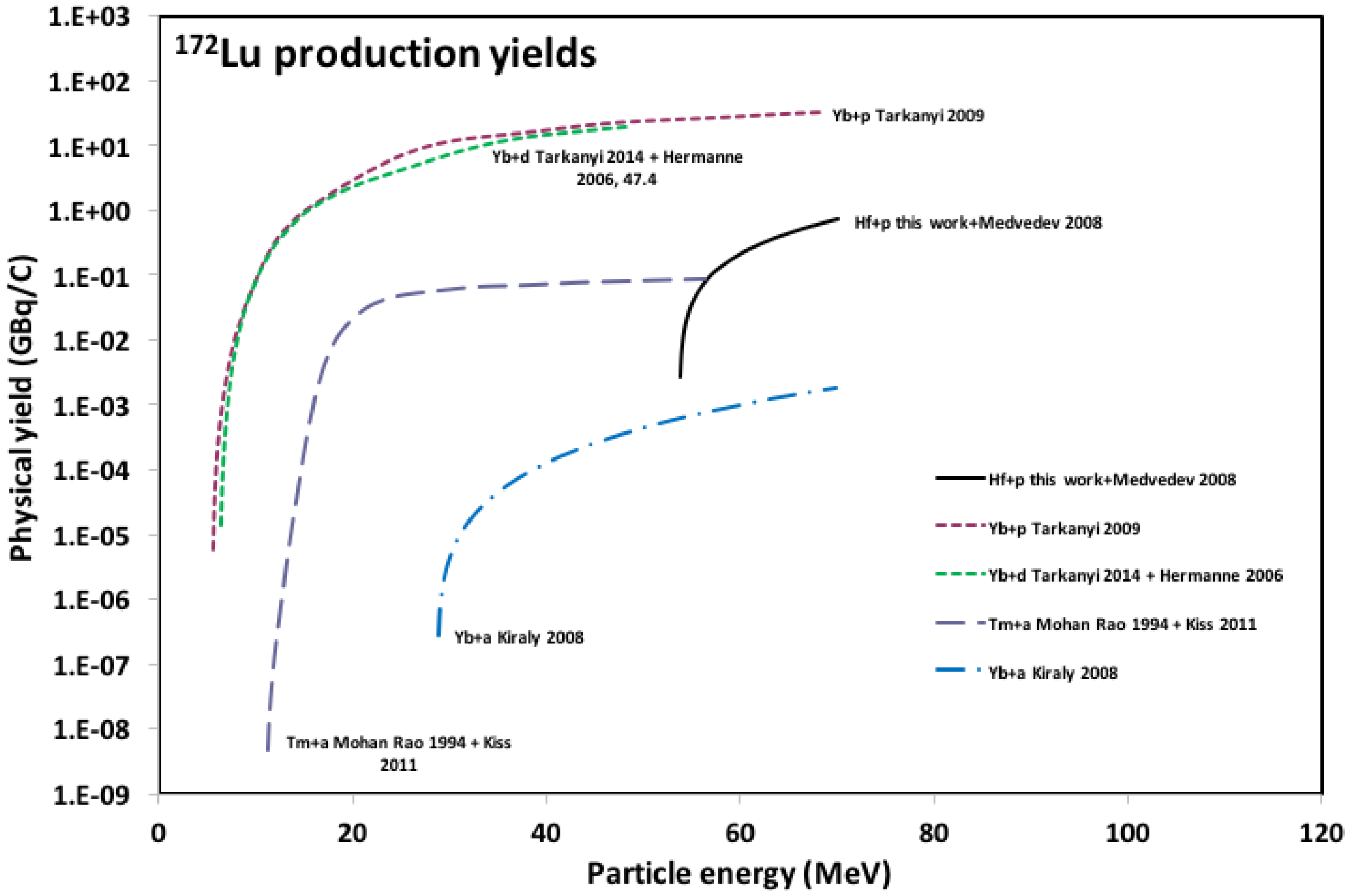

Fig. 43 Integral yields for production of $^{172}$Lu.

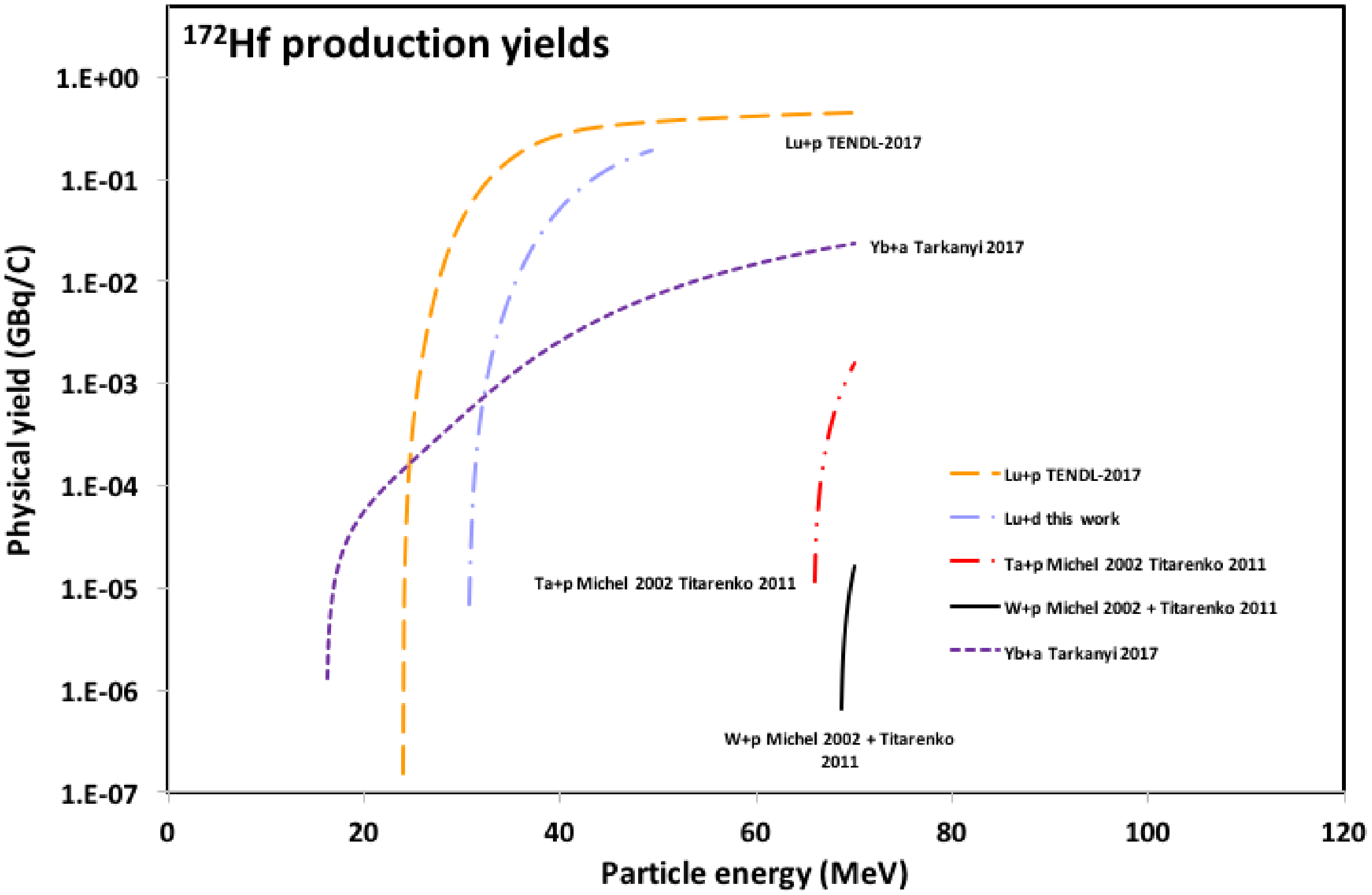

Fig. 44. Integral yields for production of $^{172}$Hf.

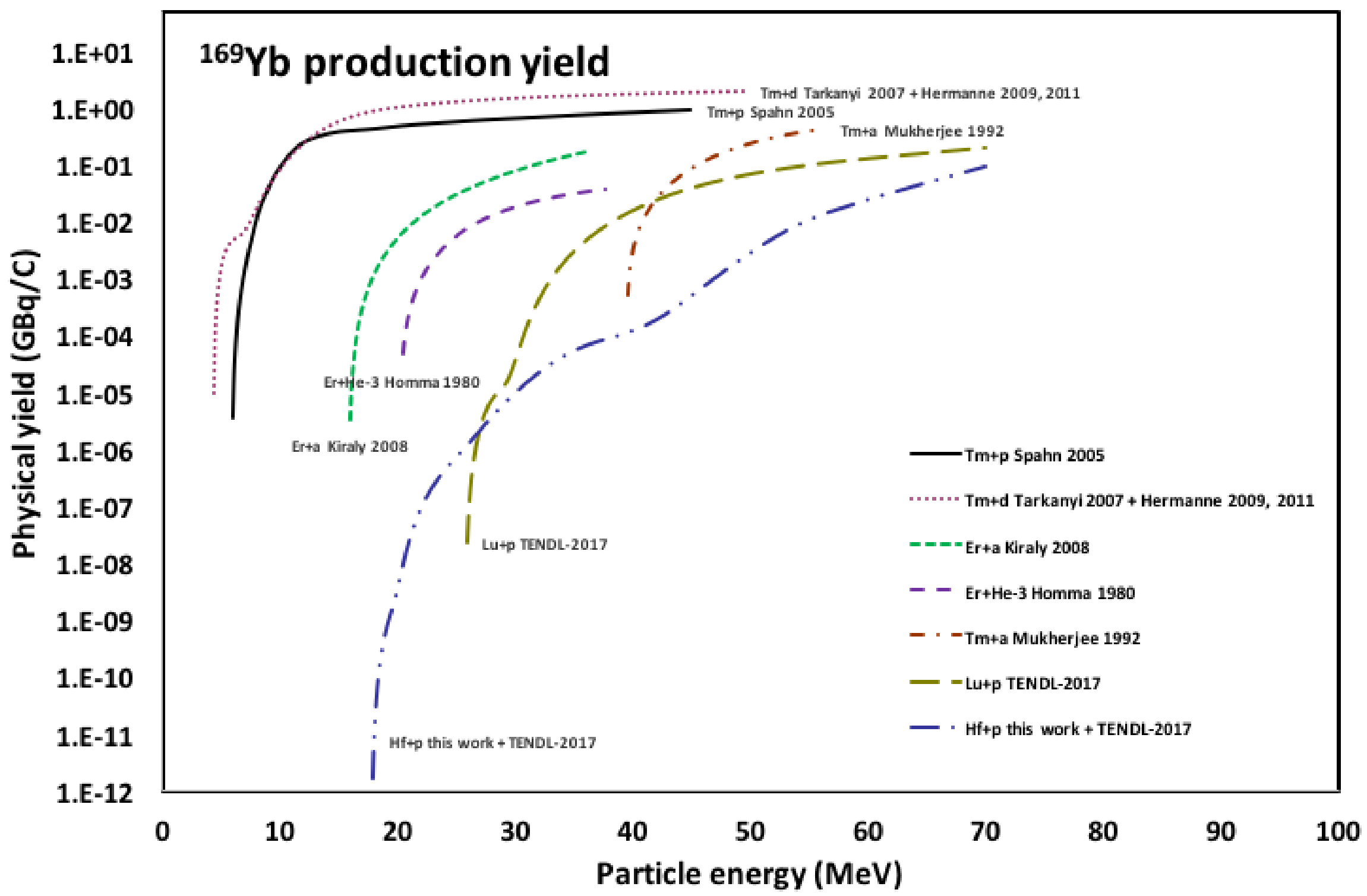

Fig. 45 Integral yield for production of $^{169}$Yb.

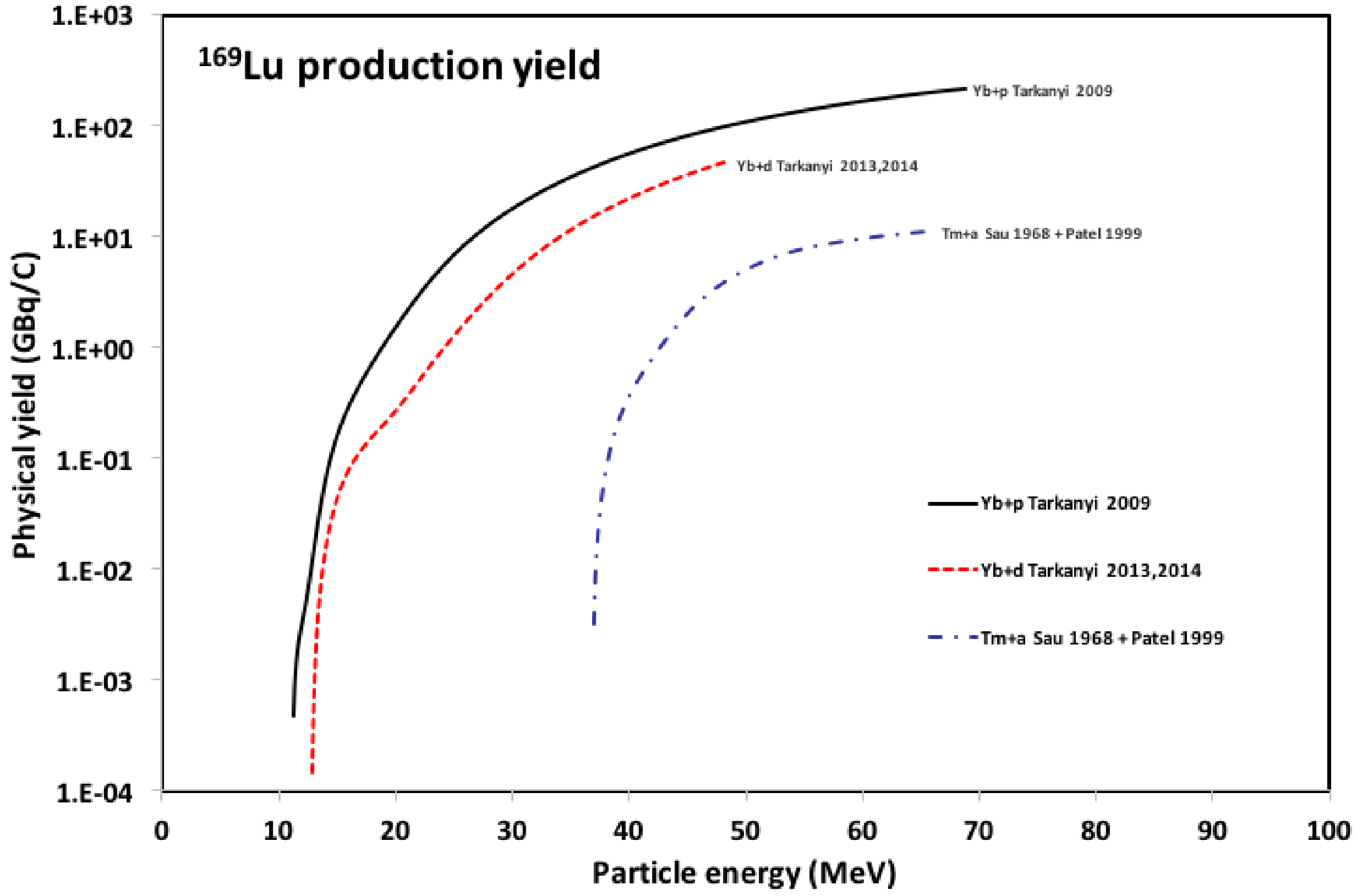

Fig. 46 Integral yield for production of $^{169}$Lu.

# 8. Comparison of the production routes of $^{172}$Lu and $^{169}$Yb

The comparison of the different production routes is not a simple task. It is well known that the selection of the optimal reactions for each medically relevant radionuclide depends on many parameters:

- the available bombarding particles and their energy ranges,
- the amount of the required activity,
- the simultaneously produced radioactive waste,
- the application of the radioisotope (direct or generator method),
- the possibility of the parallel use of the accelerator (tandem targets),
- the cost, preparation and recovery of the used targets,
- the required chemical separation process,
- the possible radionuclidic impurities and their limits imposed for practical use,
- the required specific activity,
- etc.

The following comparison deals only with a few of these parameters.

## 8.1 Production of $^{172}$Lu

### 8.1.1 Direct production of $^{172}$Lu (6.70 d)

In case of $^{nat}$Yb+p and $^{nat}$Yb+d the production yield is high (Fig 43), but important contamination with long-lived Lu radionuclides will always be present:$^{169}$Lu (34.06 h), $^{171}$Lu (8.24 d), $^{172}$Lu (6.70 d), $^{173}$Lu (1.37 y), $^{174}$Lu (142 d). Only use of highly enriched $^{172}$Yb or $^{171}$Yb targets and limitation of incident energy can result in rather pure end-product after separation from the target material. By using the $^{169}$Tm($\alpha$,n) reaction the product is carrier free and has high specific activity, without any radionuclidic impurities.

### 8.1.2 Indirect production through $^{172}$Hf (1.87 y) parent

The $^{172}$Hf is a long-lived radioisotope, other hafnium radioisotope by-products formed during the discussed reactions have significantly shorter half-life (except $^{175}$Hf, 70 days), which, however decays to stable $^{175}$Lu) and hence rather pure (no radioisotopic contaminants) $^{172}$Lu can be obtained by decay of the parent.

In all cases discussed above some stable Hf isotopes are also produced. The $^{nat}$Hf(p,x)$^{172}$Hf reaction has high yield but low (Fig. 44) specific activity (taking into account that the target is from the same element), but that does not influence the specific activity of the daughter.

In case of $^{nat}$Yb($\alpha$,x)$^{172}$Hf, $^{nat}$Lu(p,x)$^{172}$Hf, $^{nat}$Ta(p,x)$^{172}$Hf, $^{nat}$W(p,x)$^{172}$Hf reactions the $^{172}$Hf specific activity is much higher using one or more chemical separations and observing adapted waiting times for establishing equilibrium in some decay chains involved. The production yields in case of Lu+p, Ta+p and W+p are high when comparing to the yield of the $^{nat}$Yb($\alpha$,x)$^{172}$Hf reaction (Fig. 44). In practice the long-lived $^{172}$Hf is produced mainly through high energy spallation reactions on tantalum and tungsten, with simultaneous production of other radionuclides suitable for applications.

Taking into account the possible applications of $^{172}$Lu (industrial, preclinical animal studies), the requirements from point of view of presence of carriers, specific activity and of the radionuclide purity are rather low, compared to requirements for in-vivo medical use.

If carrier free $^{172}$Lu is needed, it can be produced only via the $^{172}$Yb(p,n), $^{170}$Yb($^{3}$He,n)$^{172}$Hf $\rightarrow$ $^{172}$Lu reactions. These production routes however require highly enriched targets, expensive $^{3}$He beam, long irradiation times as the production yields are low.

## 8.2 Production of $^{169}$Yb

### 8.2.1 Direct route

Using the $^{169}$Tm(p,n) reaction the product is carrier free. A low energy accelerator can be used, but the yield is low (Fig. 45). By using $^{169}$Tm(d,2n) reaction the stable $^{170}$Lu by-product is formed (lowering the specific activity) but the yield is significantly higher.

The Hf(p,x)$^{169}$Yb and Lu(p,x)$^{169}$Yb reactions need higher energy accelerators, the yields are low up to 70 MeV compared to yields of the above reactions on $^{169}$Tm.

In case of the $^{nat}$Er($^{3}$He,x)$^{169}$Yb and $^{nat}$Er($\alpha$,x)$^{169}$Yb reactions the yields are low and accompanied by other stable Yb by-products. To get a carrier free, high specific activity product, highly enriched erbium targets are required, using the $^{166}$Er($\alpha$,n) and $^{165}$Er($^{3}$He,n) reactions.

### 8.2.2 Indirect route through $^{169}$Lu parent.

In case of the discussed indirect production routes of $^{169}$Yb through decay of parent $^{169}$Lu all other lutetium contaminating radionuclides decay to stable Yb isotopes or shorter-lived isotopes away from the stability line. Among the investigated reactions the $^{nat}$Yb(p,x) and $^{nat}$Yb(d,x) reactions have the highest yields (Fig. 46) but result in many simultaneously produced Lu radioisotopes decaying to stable Yb isotopes. By using enriched $^{168}$Yb targets (low abundance: 0.13%) the $^{168}$Yb(d,n)$^{169}$Lu $\rightarrow$ $^{169}$Yb route gives carrier free $^{169}$Yb product after chemical separation of the parent from the Yb target material.

In case of $^{169}$Tm($\alpha$,4n)$^{169}$Lu (or ($^{3}$He,3n)$^{169}$Lu) we have only $^{170,171,172}$Lu by-products, which means that the $^{169}$Yb will contain stable $^{170,171,172}$Yb carriers.

## 9. Summary and conclusions

Excitation functions were measured in the 38 - 65 MeV energy range for the $^{nat}Hf(p,xn)^{180g,177,176,175,173}Ta$, $^{nat}Hf(p,xn)^{180m,179m,175,173,172,171}Hf$, $^{179,177g,173,172,171,170,169}Lu$ and $^{nat}Hf(p,x)^{169}Yb$ reactions by using an activation method. The measured excitation functions are in acceptable agreement with the earlier measured low energy data. Comparison with the TENDL predictions shows in most cases significant energy shifts of TENDL cross sections towards lower energies at higher energies. In more than in the half of the cases ALICE and EMPIRE give better prediction both in trend and in value than TALYS/TENDL, and in most cases the predictions of EMPIRE and ALICE are similar.

The review of the nuclear data on direct and indirect production routes of the $^{172}Lu$, and $^{169}Yb$ from point of view of production yield show the advantage of proton induced reactions, and the high yield of the deuteron induced reactions.

The review of the production routes from point of view of database and rates and of the production routes of $^{172}Lu$ and $^{169}Yb$ shows:

- The lack of experimental data on $^{3}He$ induced reactions;
- The low production rate of the alpha particle induced reactions;
- The high production rate of the (d,2n) reaction compared to (p,n) in this mass region;
- High production capability of the high energy reactions;
- The poor prediction capability of the TENDL in case of $^{3}He$ induced reactions, in some cases for alpha particle induced reactions and for high energy proton induced reactions.

## Acknowledgement

The authors acknowledge the support of the respective institutions and the accelerator staffs for providing the beam time at the experimental facilities.

## Figure captions

Fig. 1. Experimental and theoretical cross sections for the formation of $^{180g}$Ta by the proton bombardment of hafnium.

Fig. 2. Experimental and theoretical cross sections for the formation of $^{178m}$Ta by the proton bombardment of hafnium.

Fig. 3. Experimental and theoretical cross sections for the formation of $^{177}$Ta by the proton bombardment of hafnium.

Fig. 4. Experimental and theoretical cross sections for the formation of $^{176}$Ta by the proton bombardment of hafnium.

Fig. 5. Experimental and theoretical cross sections for the formation of $^{175}$Ta by the proton bombardment of hafnium.

Fig. 6. Experimental and theoretical cross sections for the formation of $^{174}$Ta by the proton bombardment of hafnium.

Fig. 7. Experimental and theoretical cross sections for the formation of $^{173}$Ta by the proton bombardment of hafnium.

Fig. 8. Experimental and theoretical cross sections for the formation of $^{180m}$Hf by the proton bombardment of hafnium.

FIG. 9. Experimental and theoretical cross sections for the formation of $^{179m}$Hf by the proton bombardment of hafnium.

Fig. 10. Experimental and theoretical cross sections for the formation of $^{175}$Hf by the proton bombardment of hafnium.

Fig. 11. Experimental and theoretical cross sections for the formation of $^{173}$Hf by the proton bombardment of hafnium.

Fig. 12. Experimental and theoretical cross sections for the formation of $^{172}$Hf by the proton bombardment of hafnium.

Fig. 13. Experimental and theoretical cross sections for the formation of $^{171}$Hf by the proton bombardment of hafnium.

Fig. 14. Experimental and theoretical cross sections for the formation of $^{179}$Lu by the proton bombardment of hafnium.

Fig. 15. Experimental and theoretical cross sections for the formation of $^{177g}Lu$ by the proton bombardment of hafnium.

Fig. 16. Experimental and theoretical cross sections for the formation of $^{173}Lu$ by the proton bombardment of hafnium.

Fig. 17**.** Experimental and theoretical cross sections for the formation of $^{172}Lu$ by the proton bombardment of hafnium.

Fig. 18. Experimental and theoretical cross sections for the formation of $^{171}Lu$ by the proton bombardment of hafnium.

FIG. 19**.** Experimental and theoretical cross sections for the formation of $^{170}Lu$ by the proton bombardment of hafnium.

Fig. 20. Experimental and theoretical cross sections for the formation of $^{169}Lu$ by the proton bombardment of hafnium.

Fig. 21. Experimental and theoretical cross sections for the formation of $^{169}Yb$ by the proton bombardment of hafnium.

Fig. 22. Integral thick target yields for the formation of the investigated radioisotopes of tantalum as a function of the energy.

Fig. 23. Integral thick target yields for the formation of the investigated radioisotopes of hafnium as a function of the energy.

Fig. 24. Integral thick target yields for the formation of the investigated radioisotopes of lutetium and ytterbium as a function of the energy.

Fig. 25. Excitation functions of the $^{172}Yb(p,n)^{172}Lu$ and $^{nat}Yb(p,xn)^{172}Lu$ reactions.

Fig. 26. Excitation functions of the $^{171}Yb(d,n)^{172}Lu$ and $^{nat}Yb(d,xn)^{172}Lu$ reactions.

Fig. 27. Excitation functions of the $^{169}Tm(\alpha,n)^{172}Lu$ reaction.

Fig. 28. Excitation functions of the $^{nat}Yb(\alpha,x)^{172}Lu$ reaction.

Fig. 29. Excitation functions of the $^{nat}Lu(p,xn)^{172}Hf$ reaction.

Fig. 30. Excitation functions of the $^{nat}Lu(d,x)^{172}Hf$ reaction.

Fig. 31. Excitation functions of the $^{nat}Ta(p,x)^{172}Hf$ reaction.

Fig. 32. Excitation functions of the $^{nat}W(p,x)^{172}Hf$ reaction.

Fig. 33. Excitation functions of the $^{nat}Yb(\alpha,x)^{172}Hf$ reaction.

Fig. 34. Excitation functions of the $^{169}Tm(p,n)^{169}Yb$ reaction.

Fig. 35. Excitation functions of the $^{169}Tm(d,2n)^{169}Yb$ reaction.

Fig. 36. Excitation functions of the $^{nat}Er(\alpha,xn)^{169}Yb$ reaction.

Fig. 37. Excitation functions of the $^{nat}Er(^{3}He,xn)^{169}Yb$ reaction.

Fig. 38. Excitation functions of the $^{nat}Yb(p,xn)^{169}Lu$ reaction.

Fig. 39. Excitation functions of the $^{nat}Yb(d,xn)^{169}Lu$ reaction.

Fig. 40. Excitation functions of the $^{169}Tm(\alpha,xn)^{169}Lu$ reaction.

Fig. 41. Excitation functions of the $^{169}Tm(\alpha,x)^{169}Yb$ reaction.

Fig. 42. Excitation functions of the $^{nat}Lu(p,x)^{169}Yb$ reaction.

Fig. 43. Integral yields for production of $^{172}Lu$.

Fig. 44. Integral yields for production of $^{172}Hf$.

Fig. 45. Integral yield for production of $^{169}Yb$.

Fig. 46. Integral yield for production of $^{169}Lu$.

## Tables

Table 1. Main parameters of the experiment and the methods of data evaluations.

Table 2. Decay characteristics of the investigated reaction products and Q-values of reactions for their productions.

Table 3. Experimental cross sections for the $^{nat}Hf(p,xn)^{180g,177,176,175}Ta$ reactions.

Table 4. Experimental cross sections for the $^{nat}Hf(p,xn)^{173}Ta$,$^{180m,179m,175,173}Hf$ reactions.

Table 5. Experimental cross sections for $^{172,171}Hf$, $^{179,177g,173}Lu$ reactions.

Table 6. Experimental cross sections for $^{172,171,170,169}Lu$,$^{169}Yb$ reactions.

Table 7. Cross section data of the $^{nat}Lu(d,x)^{172}Hf$ reaction [45].